\documentclass{IEEEtaes}
\usepackage{color,array,amsthm,amsmath,amssymb,gensymb}
\usepackage{graphicx}
\usepackage{bm}
\usepackage{epstopdf}
\usepackage{subfigure}
\usepackage{epstopdf}
\usepackage{booktabs}
\usepackage{stfloats}
\usepackage{multirow,array}
\usepackage{makecell}

\newcounter{TempEqCnt}
\begin{document}

\title{FDA Jamming Against Airborne Phased-MIMO Radar-Part I: Matched Filtering and Spatial Filtering} 

\author{Yan Sun}

\affil{University of Electronic Science and Technology of China, Chengdu, China} 

\author{Wen-qin Wang}
\member{Senior Member, IEEE}
\affil{University of Electronic Science and Technology of China, Chengdu, China}

\author{Zhou He}

\affil{Southwest Jiaotong University, Chengdu, China} 

\author{Shunsheng Zhang}

\affil{University of Electronic Science and Technology of China, Chengdu, China} 

\receiveddate{~~~~This work was supported by the National Natural Science Foundation of China under Grant 62171092. }

\corresp{{\itshape (Corresponding author: Wen-qin Wang)}.}

\authoraddress{~~~~Y. Sun, W. Wang and S. Zhang are with the School of Information and Communication Engineering, University of Electronic Science and Technology of China, Chengdu 611731, China 
(e-mail: \href{mailto:sunyan_1995@163.com}{sunyan\_1995@163.com}; \href{mailto:wqwang@uestc.edu.cn}{wqwang@uestc.edu.cn}; \href{mailto:zhangss@uestc.edu.cn}{ zhangss@uestc.edu.cn}). Z. He is with the School of Mathematics, Southwest Jiaotong University, Chengdu 611756, China, (e-mail: \href{mailto:zhou.he@swjtu.edu.cn}{zhou.he@swjtu.edu.cn}).}

\maketitle

\begin{abstract}


Phased multiple-input multiple-output (Phased-MIMO) radar has received increasing attention for enjoying the advantages of waveform diversity and range-dependency from frequency diverse array MIMO (FDA-MIMO) radar without sacrificing coherent processing gain through partitioning transmit subarray. This two-part series proposes a framework of electronic countermeasures (ECM) inspired by frequency diverse array (FDA) radar, called FDA jamming, evaluating its effectiveness for countering airborne phased-MIMO radar. This part introduces the principles and categories of FDA jammer and proposes the FDA jamming signal model based on the two cases of phased-MIMO radar, phased-array (PA) radar and FDA-MIMO radar. Moreover, the effects of FDA jamming on matched filtering and spatial filtering of PA and FDA-MIMO radar are analyzed. Numerical results verify the theoretical analysis and validate the effectiveness of the proposed FDA jamming in countering phased-MIMO radar.


%
 
\end{abstract}


\section{Introduction}

A{\scshape s} a tradeoff technique between phased-array (PA) and multiple-input multiple-output (MIMO) radar, phased-MIMO radar has been continuously developing by jointly exploiting the advantages of coherent processing gain for PA radar [\ref{cite1}] and waveform diversity for MIMO radar [\ref{cite2}] over the last 10 years [\ref{cite3}]-[\ref{cite6}]. When the waveform diversity between the subarrays is implemented by using a frequency offset that is larger than the signal bandwidth, the phased-MIMO radar can be investigated as a combination of PA and frequency diverse array MIMO (FDA-MIMO) radar [\ref{cite7},\ref{cite8}], a tradeoff between coherency, waveform diversity, and range-dependency. Moreover, it benefits from the PA and FDA-MIMO radar in airborne radar target detection under the challenges of background clutter and hostile electronic countermeasures (ECM) [\ref{cite9}]-[\ref{cite12}]. From the perspective of ECM [\ref{cite13}], this two-part series proposes a new jamming technique to counter the airborne phased-MIMO radar, aiming to deteriorate its anti-jamming and anti-clutter performance in target detection.


Dividing the phased-MIMO radar into two cases, PA radar and FDA-MIMO radar, they take advantages of the coherent processing gain [\ref{cite14}], waveform diversity [\ref{cite15}], and range-dependency [\ref{cite16}] in countering the hostile jamming or background clutter. A higher coherent array gain for airborne radar allows for higher robustness to environment noise, improving the signal-to-noise ratio (SNR) for target detection [\ref{cite17}]. The radar spatial resolution can be enhanced by transmitting different waveforms to form virtual arrays, improving the performance of spatial filtering to anti-jamming [\ref{cite18}, \ref{cite19}]. Furthermore, by adding a frequency offset larger than the signal bandwidth on the transmit elements, FDA-MIMO radar has a range-dependent transmit spatial frequency while inheriting the waveform diversity from MIMO radar [\ref{cite20}], which provides significant advantages in both sidelobe and mainlobe jamming suppression [\ref{cite21}]-[\ref{cite23}]. Overall, through partitioning the transmit elements, PA radar without waveform diversity and FDA-MIMO radar with subarray diversity have covered the main types of airborne radar in the electronic counter-countermeasures (ECCM) [\ref{cite24}, \ref{cite25}]. Therefore, our works take them as the objective for investigating the ECM.


In ECM, active jamming techniques have been increasingly developed with the progress of airborne radar [\ref{cite26}]-[\ref{cite29}], which can be categorized as self-protection jammers and support jammers. Jammers typically deployed into the main beam of victim-radar and required significant prior information of the target are referred to as self-protection jammers [\ref{cite29}], which delays-and-forwards victim-radar signals and load the deceptive information, such as range deceptive jamming [\ref{cite14}], angle deceptive jamming [\ref{cite31}], and velocity deceptive jamming [\ref{cite32}]. Jammers deployed far from the target and directed into the victim-radar sidelobes are referred to as support jammers [\ref{cite28}], such as modulation jamming [\ref{cite33}], typically engineering differences in the pulse signals to confuse radar. However, many anti-jamming algorithms have been extensively studied to distinguish target and jamming and suppress the jamming power from the range dimension, azimuth dimension, and Doppler dimension. Matched filtering (MF) can decrease the jamming and noise power by mismatching [\ref{cite34}]. Spatial filtering can form nulls at the location of interference to decrease the sidelobe or mainlobe jamming power in azimuth or range dimension [\ref{cite35}]. Space-time adaptive processing (STAP) can suppress the clutter by jointly using spatial and temporal information [\ref{cite36}]. In this series of works, the effectiveness of the jamming technique is evaluated from the matched filtering, spatial filtering, and STAP against phased-MIMO radar.

\begin{figure}[t]
\centerline{\includegraphics[width=20pc]{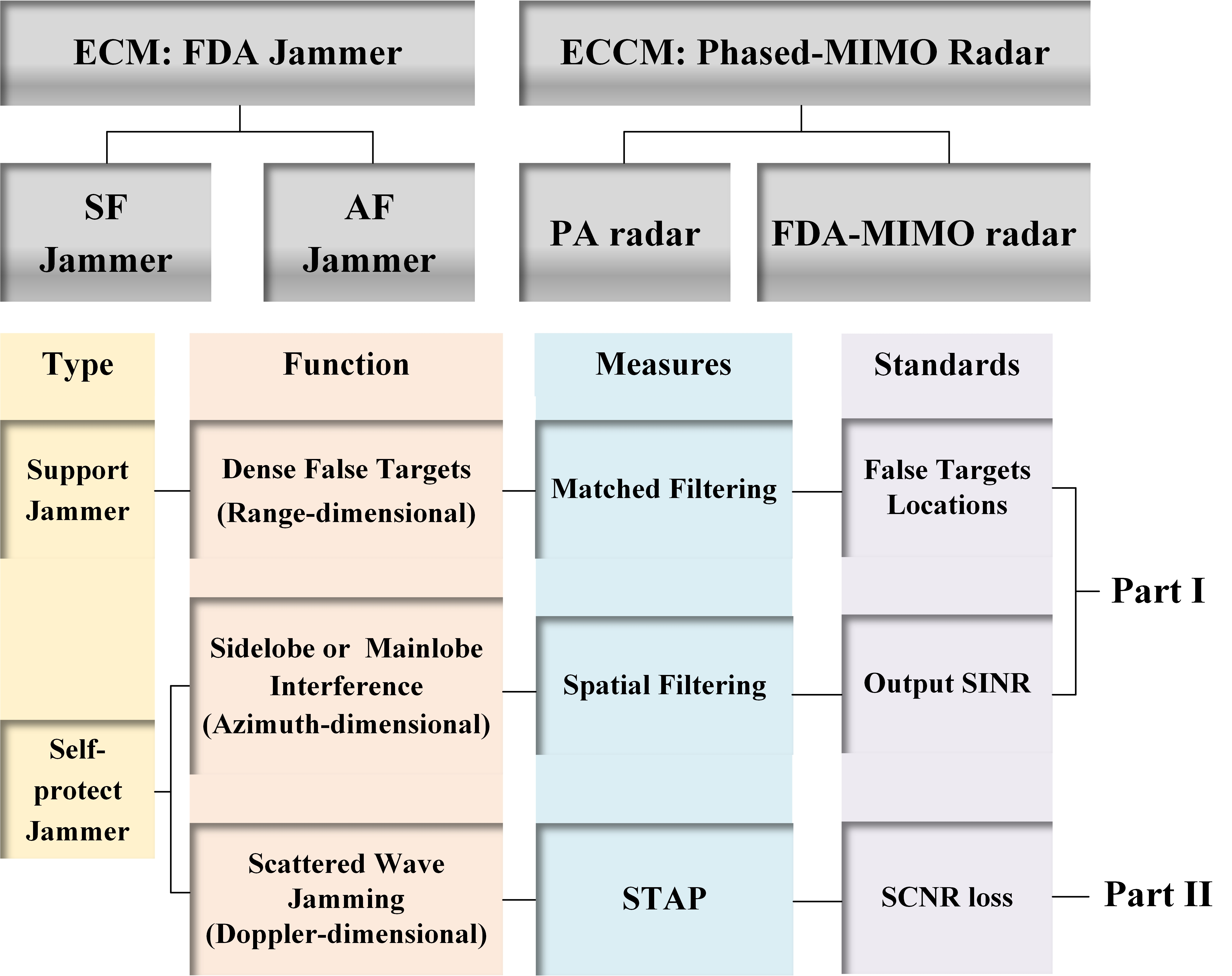}}
\caption{The content structure of this two-part series.}
\label{FIG.1}
\end{figure}

Thanks to the development of the digital radio frequency memory (DFRM) technique [\ref{cite37}], the deceptive jammer can copy, process, and forward the radar signal in a very short time (nanoseconds) after intercepting them [\ref{cite38}]. Inspired by FDA radar [\ref{cite39}], we design a much smaller jamming frequency offset on each jammer antenna, i.e., each transmit jamming signal has a different carrier frequency, which is called \emph{FDA jammer}. Meanwhile, the FDA jammer can estimate the radar parameters through the intercepted radar signal and use the prior target information to improve the jamming effectiveness. Specifically, the FDA jammer can adjust the jamming frequency offset and complete the beamforming by the known locations of the target and radar. In part I of the series, we propose the FDA jamming to counter the MF process and spatial filtering of phased-MIMO radar, disturbing the matched filtering outputs and deteriorating the spatial filtering performance. In part II of the series, we propose the scattered wave FDA jamming to counter the STAP of phased-MIMO radar, increasing the clutter rank and deteriorating the clutter suppression performance. The content structure of this two-part series is shown in Fig.\ref{FIG.1}. The main contributions in this part are briefly summarized as follows.  
\begin{enumerate}
\def\labelenumi{\arabic{enumi})}
\item
  Based on the signal model of two cases of phased-MIMO radar, PA and FDA-MIMO radar, we introduce the principles of FDA jammer and derive the two types of FDA jamming signals, stepped frequency (SF) jamming signal and arrayed frequency (AF) jamming signal.  
\item
  Focusing on matched filtering, we derive the results of SF and AF jamming after the MF process for PA and FDA-MIMO radar, respectively, and prove the relationship between the jamming frequency offset and the location of false targets generated by FDA jamming in the range dimension after the MF process. 
\item
  Focusing on spatial filtering, we propose three measurements to evaluate the effectiveness of FDA jamming in countering the spatial filtering for PA and FDA-MIMO radar and prove the relationship between the jamming frequency offset and the jamming notch depth and the output signal-to-interference-plus-noise (SINR) after spatial filtering.
\end{enumerate}

The remainder of the paper is organized as follows. The next section contains the background presentation of the phased-MIMO radar signal model. Section III introduces the principles of the FDA jammer and the derivation of the FDA jamming signal model. Section IV is devoted to the MF process of SF and AF jamming for PA and FDA-MIMO radar. The spatial filtering process of two types of FDA jammers at different cases of phased-MIMO radar, namely PA radar and FDA-MIMO radar, is provided in Section V. Numerical simulation results are presented to illustrate the effectiveness of FDA jammers in Section VI. In Section VII, we compare the FDA jamming with the existing jamming techniques and discuss the advantages and limitations. Finally, in Section VIII, we draw the conclusions. Proofs and derivations are confined to the Appendices.

\textbf{Notations}: vectors and matrices are denoted by boldface lower-case and upper-case letters, respectively. Symbols $\left( \cdot \right) ^{\mathrm{*}}$, $\left( \cdot \right) ^{\mathrm{T}}$, $\left( \cdot \right) ^{\mathrm{H}}$, $\odot$, $\circledast$, and $\otimes$ denote the conjugate, transpose, conjugate transpose, Hadamard product, convolution operation, and Kronecker product, respectively. $\boldsymbol{I}_M$, $\mathbf{1}_M$ and $\mathbf{0}_M$ stand for the M-dimensional identity matrix, the M-dimensional all 1 vector or matrix and the M-dimensional null vector or matrix of proper size. Operation symbols $\times$ and $\cdot$ represent scalar multiplication and the product of matrices or vectors, respectively. $\mathbb{C}$ is the set of complex numbers, and $\mathbb{C}^{N\times M}$ is the Euclidean space of ($N\times M$)-dimensional complex matrices (or vectors if $M=1$). The superscript `$^{(P)}$' and `$^{(F)}$' correspond to the PA radar and FDA-MIMO radar, respectively. The $(i,j)$th entry of a matrix $A$ is indicated by $\left[ \boldsymbol{A} \right] _{ij}$. $<$, $\leqslant$, $>$, and $\geqslant$ represent the less-than sign, less-than or equal sign, greater-than sign, and greater-than or equal sign, respectively. $\mathcal{F} \left\{ \cdot \right\}$, $\mathrm{E}\left\{ \cdot \right\}$ denote the Fourier transform operation and the expectation operation, respectively.

\section{PHASED-MIMO RADAR SIGNAL MODEL}

In this two-part series, we take the phased-MIMO radar, a tradeoff technique between PA and FDA-MIMO radar, as an objective to study the jamming countermeasures. Apart from enjoying the benefits from MIMO radar as mentioned in [\ref{cite3}, \ref{cite5}], it inherits the range-dependency from FDA-MIMO by using the frequency offset to satisfy orthogonality between the subarrays. Thanks to the range-dependency, FDA-MIMO radar performs better than MIMO radar in mainlobe interference suppression [\ref{cite9}, \ref{cite10}] and range-ambiguous clutter suppression [\ref{cite11}, \ref{cite12}]. From the perspective of ECM, the research on a jamming technology against the combination of PA and FDA-MIMO radar is more meaningful. Therefore, this paper evaluates the effectiveness of the proposed jamming in countering the phased-MIMO radar with two cases, PA radar without subarrays and FDA-MIMO radar with non-overlapping subarrays.

\begin{figure}[t]
\centerline{\includegraphics[width=20pc]{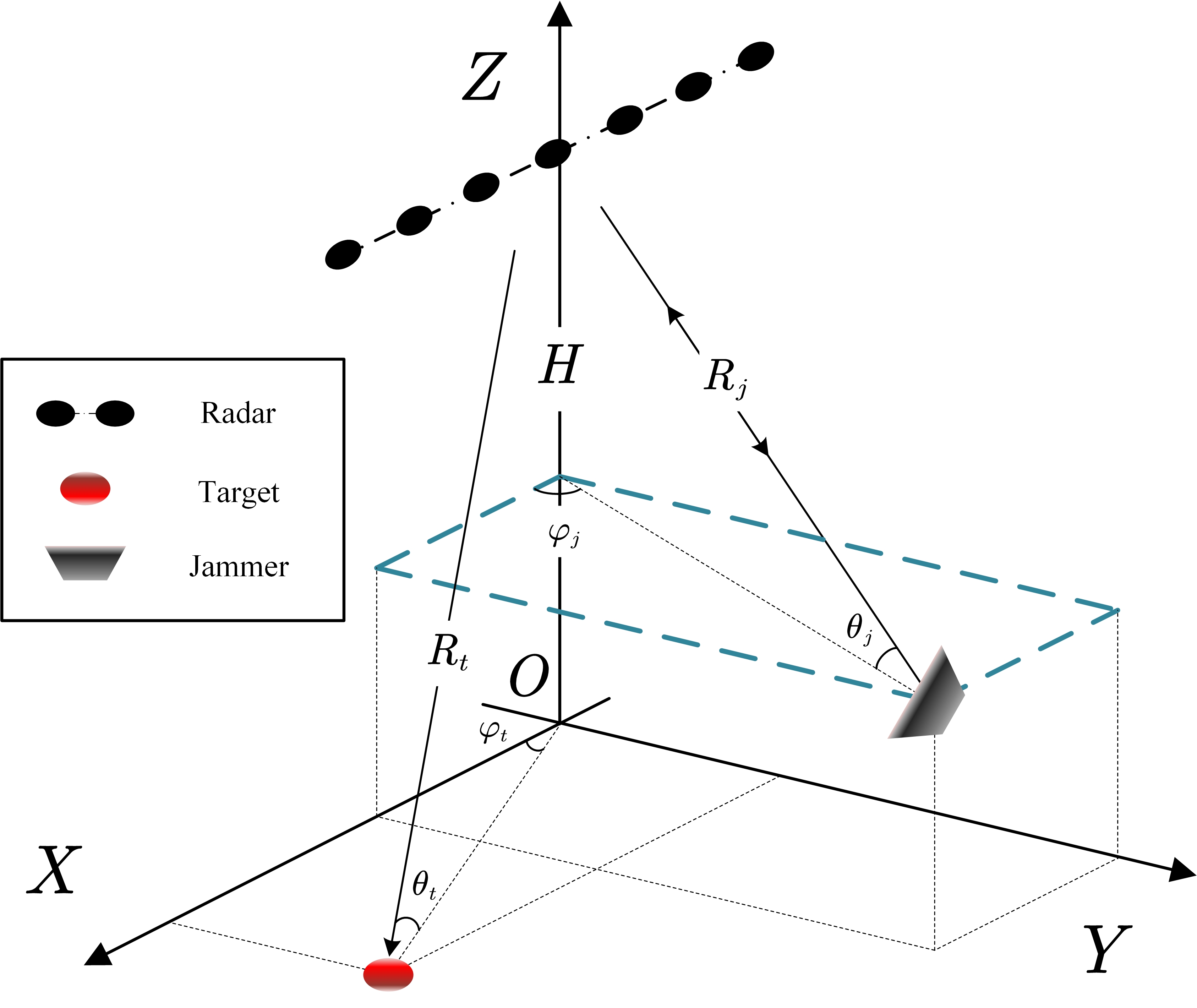}}
\caption{The coordinates scene of FDA jammer.}
\label{FIG.2}
\end{figure}

This section introduces the signal model for a monostatic phased-MIMO radar with $M$ transmit elements and $N$ receive elements. Fig.\ref{FIG.2} illustrates the geometric coordinates of the airborne side-looking array, the jammer, and the target of interest. Assume that the airborne radar array with height $H$ is arranged along the X-axis, and both transmit and receive elements are spaced half wavelength apart, $d=\lambda _0/2$, where $\lambda _0$ denotes the wavelength corresponding to the carrier frequency $f_0=c/\lambda_0$ and $c$ is the speed of light. The radial range, elevation, and azimuth of the target are $R_t$, $\theta_t$ and $\varphi _t$, respectively, while the radial range, elevation, and azimuth of FDA jammer are $R_j$, $\theta_j$ and $\varphi _j$, respectively. The relationship between the radial range and the elevation is $\theta =\mathrm{arc}\sin \left( H/R \right) $.

By dividing $M$ transmit elements into $S$ subarrays without overlapping, the coherent array gain is obtained from $M_S=M/S$ elements within each subarray, and the waveform diversity is provided by $S$ subarrays [\ref{cite3}]. In this paper, the baseband waveform signal transmitted by the s-th subarray can be expressed as
\begin{equation}
u _s\left( t \right) =A\left( t \right) e^{j2\pi \left( s-1 \right) \varDelta ft}
\label{eq.1}
\end{equation}
where $A(t)$ is the unit energy envelope with pulse width $T_p$ and bandwidth $B$, satisfying the narrow-band assumption [\ref{cite17}]. $\varDelta f$ ensures that the transmitted signals of each subarray are orthogonal, which satisfies $\varDelta f\geqslant B$ [\ref{cite3},\ref{cite4}]. 
\begin{align}
\int_{T_p}{u _i\left( t \right) u _{j}^{*}\left( t \right) \mathrm{d}t}=\left\{ \begin{array}{c}
  1, i=j\\
  0, i\ne j\\
\end{array} \right. 
\label{eq.2}
\end{align}
Different from [\ref{cite3}], by using a frequency offset that is larger than the signal bandwidth, this paper considers the phased-MIMO radar as a tradeoff technique between PA radar and FDA-MIMO radar, enjoying the coherency provided by subarray beamforming, and waveform diversity and range-dependency from the frequency offset. Hence, the phased-MIMO radar can be divided into two cases, PA radar with $M$ transmit elements and FDA-MIMO radar with $S$ non-overlapping subarrays, where each subarray has $M_S=M/S$ elements.


\subsection{PA radar}

For the case of PA radar, the baseband waveform signal in \eqref{eq.2} can be modified as $u _m\left( t \right) =A\left( t \right)$ with $\varDelta f=0$, which are transmitted by the m-th element. Define the transmit steering vector $\boldsymbol{a}_t(\varphi,\theta)$ and beamforming vector $\boldsymbol{w}_t(\varphi_0,\theta_0)$ for PA radar, respectively,
\begin{subequations}
\begin{align}
\boldsymbol{a}_t(\varphi,\theta) =&\left[ \begin{matrix}
  1&     \cdots&   e^{j2\pi \frac{d}{\lambda_0}\left( M-1 \right) \cos \varphi \cos \theta}\\
\end{matrix} \right] ^{\mathrm{T}}\label{eq.3a}
\\
\boldsymbol{w}_t(\varphi_0,\theta_0) =&\frac{1}{\sqrt{M}}\left[ \begin{matrix}
  1&     \cdots&   e^{j2\pi \frac{d}{\lambda_0}\left( M-1 \right) \cos \varphi_0 \cos \theta_0}\\
\end{matrix} \right] ^{\mathrm{T}}
\label{eq.3b}
\end{align}
\end{subequations}
where $(\varphi_0, \theta_0)$ is the main beam direction determined by $\boldsymbol{w}_t(\varphi_0, \theta_0)$. $\frac{1}{\sqrt{M}}$ guarantees $\left| \boldsymbol{w}_t(\varphi _0,\theta _0) \right|^2=1$. Focusing on the target of interest located at $(\varphi_t, \theta_t)$ as shown in Fig.\ref{FIG.2}, then the synthetic signal after transmit beamforming can be expressed as 
\begin{align}
z^{\left( \mathrm{P} \right)}\left( t \right)=&\boldsymbol{w}_{t}^{\mathrm{H}}(\varphi_t, \theta_t)\boldsymbol{a}_t(\varphi_t, \theta_t) A\left( t \right) e^{j2\pi f_0t}
\nonumber\\
=&\sqrt{M}\cdot A\left( t \right) e^{j2\pi f_0t}
\label{eq.4}
\end{align}
After reflection by the target, the monopulse signal received by the n-th element can be expressed as 
\begin{equation}
y_{n}^{\left( \mathrm{P} \right)}\left( t \right) =\xi_t\cdot \sqrt{M}\cdot A\left( t \right) e^{j2\pi f_0\left( t-\tau_t \right)}e^{j2\pi \frac{d}{\lambda _0}\left( n-1 \right) \cos \varphi_t \cos \theta_t}
\label{eq.5}
\end{equation}
where $\tau_t=2R_t/c$ denotes the two-way propagation delay and $\xi_t$ is the target reflection coefficient. Note that \eqref{eq.5} uses the narrow-band assumption, $A\left( t \right) \approx A\left( t-\tau _t \right)$. Here we ignore the Doppler frequency shift caused by the platform and target movement since the Doppler information cannot be acknowledged through the monopulse signal when the Doppler frequency shift $f_D$ is less than the signal bandwidth [\ref{cite15}], $f_D=\frac{2\left| v_a-v_t \right|}{\lambda _0}\leqslant B$, where $v_a$ and $v_t$ denote the radial velocities of airborne and target, respectively. The discussion about multi-pulse signals is described in Part II of this series. 

Through the down-conversion $e^{-j2\pi f_0(t-\tau_t)}$ for the distance of $R_t$ and MF by $A(t)$ for PA radar [\ref{cite17}], the fast-time snapshot of the target of interest can be expressed as an N-dimensional vector.
\begin{equation}
 \boldsymbol{x}_{ \mathrm{P}}=\xi _t\cdot \sqrt{M}\cdot \boldsymbol{a}_r(\varphi_t,\theta_t)
\label{eq.6}
\end{equation} 
where $\boldsymbol{a}_r(\varphi ,\theta )$ represents the receive spatial steering vector.
\begin{equation}
\boldsymbol{a}_r(\varphi ,\theta )=\left[ \begin{matrix}
  1&    \cdots&   e^{j2\pi \frac{d}{\lambda _0}\left( N-1 \right) \cos \varphi \cos \theta}\\
\end{matrix} \right] ^{\mathrm{T}}
\label{eq.7}
\end{equation}
Note that \eqref{eq.6} is also considered as a signal model for single-input multiple-output (SIMO) radar [\ref{cite2}].

\begin{figure}[t]
\centerline{\includegraphics[width=20pc]{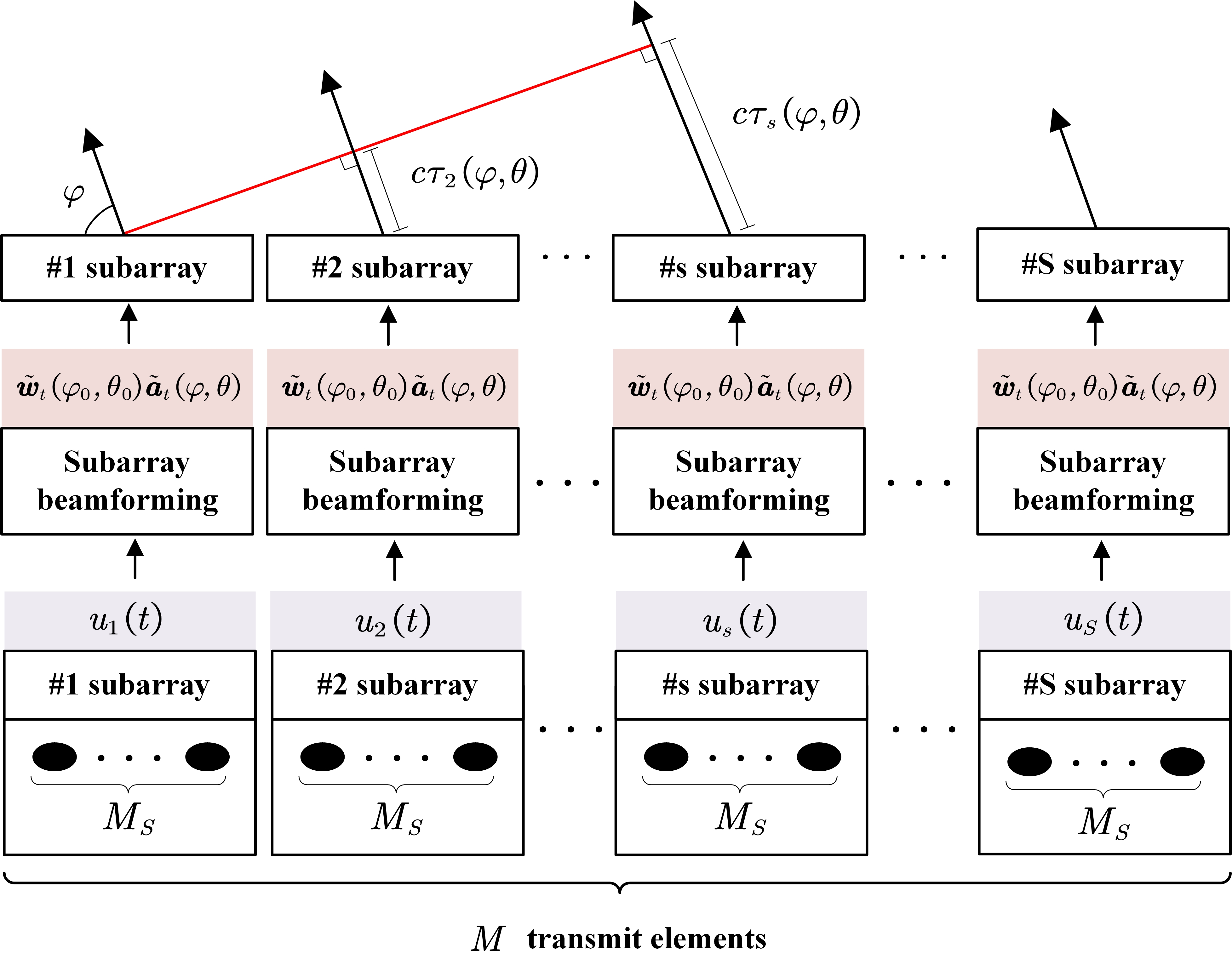}}
\caption{The transmit subarray partitioning and processing for FDA-MIMO radar with $S$ non-overlapping subarrays.}
\label{FIG.3}
\end{figure}

\subsection{FDA-MIMO radar}

For the case of FDA-MIMO radar, each non-overlapping subarray is composed of $M_S$ transmit elements. The transmit array partitioning and processing of FDA-MIMO radar is shown in Fig.\ref{FIG.3}. The spacing between the first elements of neighboring subarrays is $M_Sd$. The transmit beamforming is achieved within each subarray, where the transmit steering vector $\tilde{\boldsymbol{a}}_t(\varphi,\theta)$ and beamforming vector $\tilde{\boldsymbol{w}}_t(\varphi_0,\theta_0)$ within subarray can be expressed as
\begin{subequations}
\begin{align}
\tilde{\boldsymbol{a}}_t(\varphi,\theta) =&\left[ \begin{matrix}
  1&     \cdots&   e^{j2\pi \frac{d}{\lambda_0}\left( M_S-1 \right) \cos \varphi \cos \theta}\\
\end{matrix} \right] ^{\mathrm{T}}\label{eq.8a}
\\
\tilde{\boldsymbol{w}}_t(\varphi_0,\theta_0) =&\frac{1}{\sqrt{M_S}}\left[ \begin{matrix}
  1&     \cdots&   e^{j2\pi \frac{d}{\lambda_0}\left( M_S-1 \right) \cos \varphi_0 \cos \theta_0}\\
\end{matrix} \right] ^{\mathrm{T}}
\label{eq.8b}
\end{align}
\end{subequations}
where $\frac{1}{\sqrt{M_S}}$ guarantees $\left| \tilde{\boldsymbol{w}}_t(\varphi _0,\theta _0) \right|^2=1$. Focusing on the target located at $(\varphi_t, \theta_t)$, let the first subarray as the reference subarray (`1\# subarray' as shown in Fig.\ref{FIG.3}), the transmit signal for the s-th subarray after subarray beamforming can be expressed as 
\begin{align}
\tilde{z}_s\left( t \right)=&\tilde{\boldsymbol{w}}_t^\mathrm{H}(\varphi_t,\theta_t)\tilde{\boldsymbol{a}}_t(\varphi_t,\theta_t) u _s\left( t \right)e^{j2\pi f_0[t+\tau _s\left(  \varphi_t,\theta_t \right)]}
\nonumber\\
=&\sqrt{M_S}\cdot u _s\left( t \right) e^{j2\pi f_0[t+\tau _s\left(  \varphi_t,\theta_t \right)]}
\label{eq.9}
\end{align}
where the subarray time delay $\tau_s(\varphi_t, \theta_t)$, which means the time required for the wave to travel across the spatial displacement between the reference subarray and the s-th subarray [\ref{cite3}], can be expressed as 
\begin{equation}
\tau _s\left(  \varphi,\theta \right) =\frac{d}{c}M_S\left( s-1 \right) \cos \theta \cos \varphi 
\label{eq.10}
\end{equation}
Thereby the transmit signal synthesized by $S$ subarrays toward the target direction for FDA-MIMO radar can be expressed as
\begin{align}
z^{\left( \mathrm{F} \right)}\left( t \right) =\sum_{s=1}^S{\tilde{z}_s\left( t \right) }=\sqrt{M_S}\cdot \sum_{s=1}^S{u_s\left( t \right) e^{j2\pi f_0[t+\tau _s\left( \varphi_t ,\theta_t \right)]}}
\label{eq.11}
\end{align}

After reflection by the target, the monopulse signal received by the n-th element can be expressed as 
\begin{align}
y_n^{\left( \mathrm{F} \right)}\left( t \right) \approx &\xi _t\cdot \sqrt{M_S}\cdot e^{j2\pi \frac{d}{\lambda _0}\left( n-1 \right) \cos \varphi _t\cos \theta _t}
\nonumber\\
\times \sum_{s=1}^S&{A\left( t \right) e^{j2\pi f_0\tau _s\left( \varphi_t ,\theta_t \right)}}e^{j2\pi \left[ f_0+\left( s-1 \right) \varDelta f \right] \left( t-\tau_t \right)}
\label{eq.12}
\end{align}
Note that \eqref{eq.12} ignores $e^{j2\pi \frac{d}{c}\varDelta f\left( s-1 \right) \left( n-1 \right) \cos \varphi_t \cos \theta_t}$ due to $(S-1)\varDelta f\ll f_0$ [\ref{cite9}, \ref{cite10}, \ref{cite12}]. For FDA-MIMO radar receiver, each receive array samples discrete data sequence from $S$ channels, which correspond to $S$ transmit subarrays. After down-conversion $e^{-j2\pi f_0(t-\tau_t)}$ for the distance of $R_t$, each channel implements the MF process by using $\left\{ u_s\left( t \right) \right\} _{s=1}^{S}$ (orthogonality in \eqref{eq.2}). Then the fast-time snapshot of the target of interest can be expressed as a $SN\times 1$ virtual data vector [\ref{cite4}, \ref{cite12}].
\begin{align}
\boldsymbol{x}_{ \mathrm{F}}=&\xi _t\cdot \sqrt{M_S}\cdot \left[ \boldsymbol{b}\left( \varphi _t,\theta _t \right) \odot \boldsymbol{d}\left( R_t \right) \right] \otimes \boldsymbol{a}_r\left( \varphi _t,\theta _t \right)
\nonumber\\
 =&\xi _t\cdot \sqrt{M_S}\cdot \boldsymbol{c}\left( \varphi _t,\theta _t, R_t \right) \otimes \boldsymbol{a}_r\left( \varphi _t,\theta _t \right)
\label{eq.13}
\end{align}
where $\boldsymbol{b}\left( \varphi ,\theta \right)$ and $\boldsymbol{d}\left( R \right)$ are the subarray transmit spatial steering vector and range-dependent vector, respectively.
\begin{subequations}
\begin{align}
\boldsymbol{c}\left( \varphi ,\theta, R \right)=&\boldsymbol{b}\left( \varphi ,\theta  \right) \odot \boldsymbol{d}\left( R \right)\label{eq.14a}
\\
\boldsymbol{b}\left( \varphi ,\theta \right) =&\left[ \begin{matrix}
  1&    \cdots&   e^{j2\pi \frac{d}{\lambda _0}M_S\left( S-1 \right) \cos \varphi \cos \theta}\\
\end{matrix} \right] ^{\mathrm{T}}\label{eq.14b}
\\
\boldsymbol{d}\left( R \right) =&\left[ \begin{matrix}
  1&    \cdots&   e^{-j2\pi \frac{2R}{c}\left( S-1 \right) \varDelta f}\\
\end{matrix} \right] ^{\mathrm{T}}
\label{eq.14c}
\end{align}
\end{subequations}

From \eqref{eq.6} and \eqref{eq.13}, PA radar has a higher coherent gain but without range-dependency. FDA-MIMO radar with $S$ non-overlapping subarrays is a tradeoff radar between range-dependency, waveform diversity, and coherency. Moreover, FDA-MIMO radar with $S=M$ and $M_S=1$, which has no coherent processing gain, has been extensively researched due to its benefits on mainlobe interference suppression [\ref{cite9}, \ref{cite10}] and range-ambiguous clutter suppression [\ref{cite11}, \ref{cite12}]. 

\section{FDA JAMMING SIGNAL MODEL}

\begin{figure*}[t]
  \centering
  \vspace{-0.15in}
  \begin{minipage}{1\linewidth }
    \subfigure[FDA jammer structure]{
      \label{FIG.4(a)}
\includegraphics[width=18.8pc]{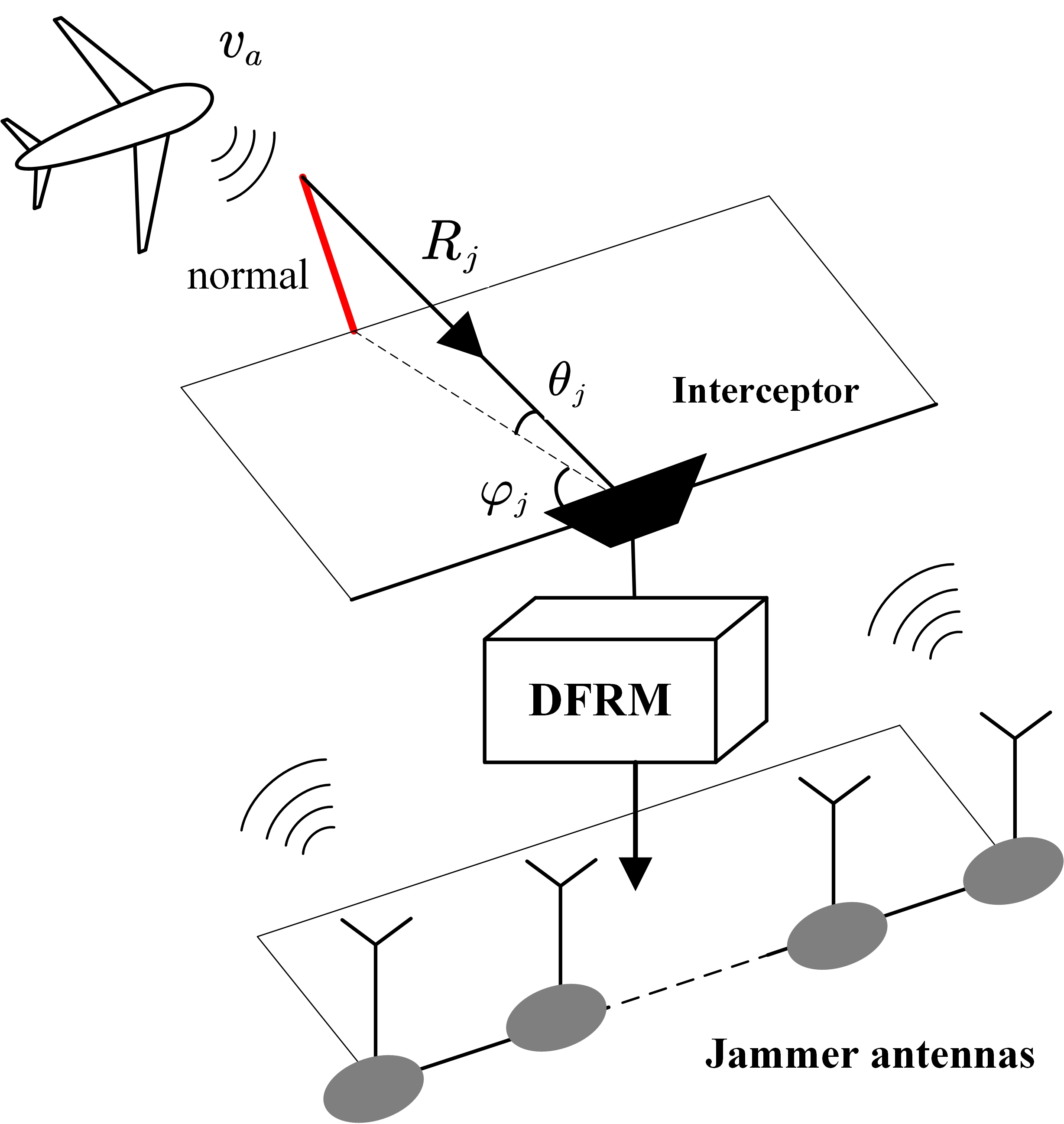}
      }
    \subfigure[Target parameters and radar parameters estimation]{\label{FIG.4(b)}  
    \includegraphics[width=18.8pc]{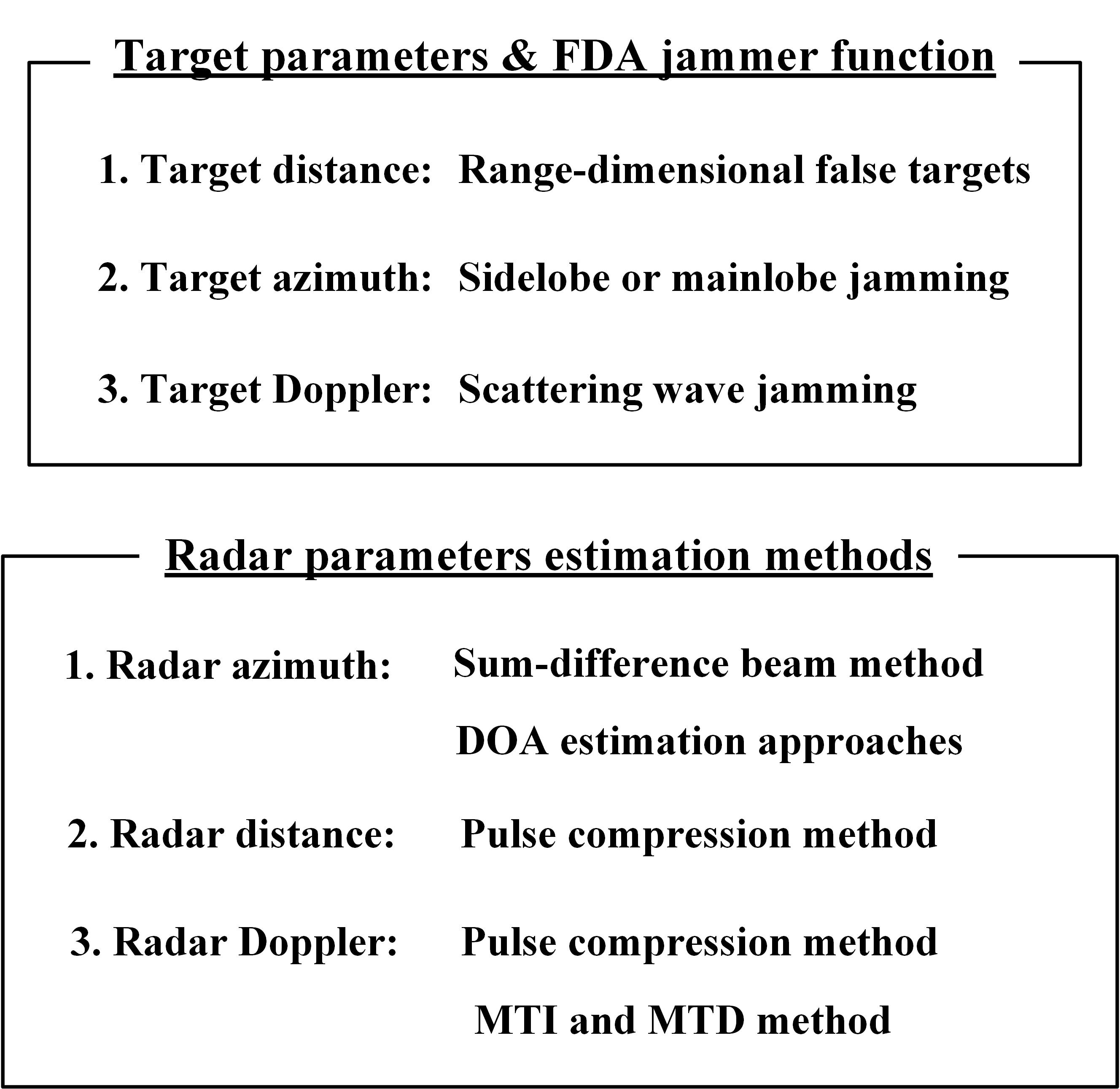}
   }
   \subfigure[SF and AF jammer transmitter array]{\label{FIG.4(c)}  
    \includegraphics[width=40.6pc]{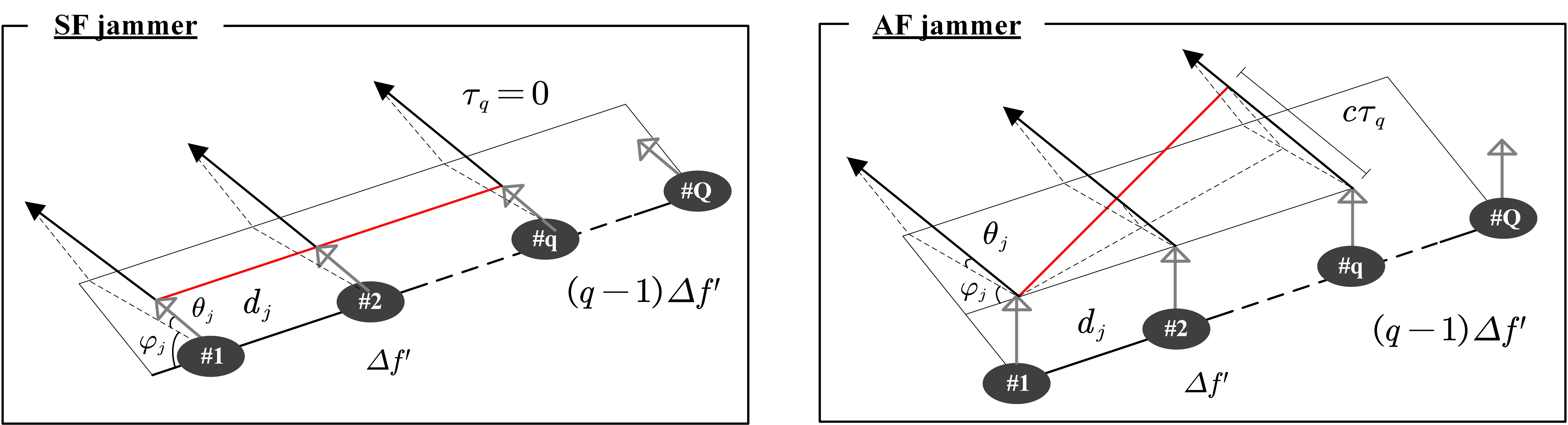}
   }
  \end{minipage}
  \vspace{-0.05in}  
  \caption{The principle of FDA jammer. (a) FDA jammer structure. (b) Target and radar parameters. (c) SF and AF jammer transmitter array. }
  \label{FIG.4}
\end{figure*}


This section introduces the FDA jamming signal model, including the principle of the FDA jammer and the derivation of the FDA jamming signal model. The FDA jammer is required to implement four steps, intercepting the radar signal, estimating the radar parameters, loading the jamming frequency offset, and transmitting the FDA jamming signal, as shown in Fig.\ref{FIG.4(a)}.
\begin{enumerate}
\def\labelenumi{\arabic{enumi})}
\item
  The radar signal is received by the jammer interceptor, and the intercepted signals are saved, copied, and sent to the digital radio frequency memory (DFRM) [\ref{cite37}, \ref{cite38}]. 
\item
  DFRM down-converts the signal and retains the envelope information to estimate the radar parameters. For the azimuth information of radar, it can use the sum-difference beam methods [\ref{cite20}] or the classical direction of arrival (DOA) estimation approaches [\ref{cite23}]. DRFM can estimate the pulse repetition frequency (PRF) and pulse width from multi-pulse radar signals, and use the envelope time delay [\ref{cite24}]. By using pulse compression, conventional moving target indicator (MTI), and moving target detection (MTD) method, the jammer can estimate the range and Doppler frequency of radar [\ref{cite26}]. Furthermore, prior information about the target can improve the efficiency of FDA jamming. When the FDA jammer knows the target range, it can design an appropriate jamming frequency offset to generate false targets adjacent to the target, increasing radar false alarms, as discussed in Section IV. When the Doppler information of the target is known to the FDA jammer, it can design an appropriate frequency offset and transmit the scattering wave jamming to the ground to worse the performance of STAP for radar clutter suppression, allowing the improvement factor (IF) notches to appear near the target Doppler frequency, which is described in Part II. The target prior information and radar parameters estimation methods are shown in Fig.\ref{FIG.4(b)}.
\item
  Before loading the jamming frequency offset, the FDA jammer can adjust the main beam orientation of the transmit antennas in advance by using the estimated azimuth of radar [\ref{cite13}]. However, the radar parameters estimation is not always completed in time or inaccurate [\ref{cite25}] [\ref{cite37}], thus there are two types of FDA jamming signals. One is that the main beam of transmit antennas is oriented toward the radar by using the estimated azimuth information, which is called the stepped frequency (SF) jammers as shown in Fig.\ref{FIG.4(c)}. The other is that the main beam of transmit antennas is not oriented toward the radar since the estimation of the radar parameters is not completed or is incorrect, which is called the arrayed frequency (AF) jammer as shown in Fig.\ref{FIG.4(c)}. The AF jammer antennas should consider the transmit array steering while the SF jammer does not. Hence, AF jammers require more power than SF jammers. In this paper, we consider loading the linear jamming frequency offset on each jammer antenna, e.g., the jamming frequency offset loaded on the q-th antenna is $(q-1)\varDelta f'$, where $\varDelta f'$ is the jamming frequency offset.
\item
  After loading the jamming frequency offset, the jammer antennas transmit the FDA jamming to phased-MIMO radar. We assume that the FDA jammer antennas are linearly arrayed as shown in Fig.\ref{FIG.4}.  
\end{enumerate}

Assume that the FDA jammer locates at ($\varphi_j$,$\theta_j$) as presented in Fig.\ref{FIG.2}. The number and spacing of jammer antennas are considered as $Q$ and $d_j=\lambda_0/2$. Let us discuss the FDA jamming signal model against two cases of the phased-MIMO radar, PA and FDA-MIMO radar. 

\subsection{Against PA radar}

The PA radar signal $\check{P}\left( t \right)$ intercepted by the FDA jammer can be represented as
\begin{equation}
\check{P}\left( t \right)=\xi _j\boldsymbol{w}_{t}^{\mathrm{H}}(\varphi_t, \theta_t)\boldsymbol{a}_t(\varphi_j, \theta_j) A\left( t \right) e^{j2\pi f_0(t-\tau_j')}
\label{eq.16}
\end{equation}
where $\xi_j$ and $\tau_j'=R_j/c$ are the propagation loss and time delay of the jamming signal. Note that $R_j$ is the one-way distance of the jammer relative to the radar. Then the jamming signal transmitted by the q-th antenna of the SF and AF jammer can be expressed as
\begin{subequations}
\begin{align}
\hat{P}_{q}^{\left( \mathrm{SF} \right)}\left( t \right)=&\rho _q\boldsymbol{w}_{{t}}^{\mathrm{H}}(\varphi _t,\theta _t)\boldsymbol{a}_t(\varphi _j,\theta _j)\nonumber\\
&~ \times A\left( t \right) e^{j2\pi \left[ f_0+\left( q-1 \right) \varDelta f' \right] \left( t-\tau_j' \right)}\label{eq.17a}
\\
\hat{P}_{q}^{\left( \mathrm{AF} \right)}\left( t \right)\approx&\rho _q\boldsymbol{w}_{{t}}^{\mathrm{H}}(\varphi _t,\theta _t)\boldsymbol{a}_t(\varphi _j,\theta _j)
\nonumber\\
&~\times  A\left( t \right) e^{j2\pi \left[ f_0+\left( q-1 \right) \varDelta f' \right] \left( t-\tau_j'+\tau_q \right)}
\label{eq.17b}
\end{align}
\end{subequations}
where $\rho _q$ denotes the amplitude coefficient of the jamming signal transmitted by the q-th jammer antenna, associating with the propagation loss, the antenna gain, and the transmitter power. In this paper, we assume that $\rho ^2_q$ is the jamming signal power related to the q-th jammer antenna. $\tau _q=(q-1)d_j\cos{\varphi_j}\cos{\theta_j}/c$ denotes the q-th jammer antenna delay related to the array displacement. \eqref{eq.17b} uses the narrow-band assumption, $A\left( t \right)\approx A\left( t+\tau_q \right)$. From \eqref{eq.17a} and \eqref{eq.17b}, the main difference between SF and AF jammer is $e^{j2\pi \left[ f_0+\left( q-1 \right) \varDelta f' \right] \tau _q}$ for the q-th jammer antenna, which denotes the transmit array steering related to the jamming frequency offset. For the SF jammer, the antennas have been oriented their main beam to the radar, thus there is no time delay associated with the array displacement for each jammer antenna, $\tau _q=0$ as shown in Fig.\ref{FIG.4(c)}. For the AF jammer, the main beam of the jammer antennas is not oriented to the radar, so there is a time delay related to the array displacement between each antenna and the reference antenna (The `1\#' antenna in Fig.\ref{FIG.4(c)} is the reference antenna). Therefore, after the radar receives the jamming signal, the AF jamming may have a lower power when the SF and AF jammer have the same jamming signal power.

After $\tau_j'=R_j/c$ of propagation, PA radar receives a synthetic jamming signal transmitted by $Q$ jammer antennas. For the SF and AF jammers, the jamming signals received by the n-th element of PA radar can be expressed as 
\begin{subequations}
\begin{align}
\bar{P}_n^{\left( \mathrm{SF} \right)}\left( t \right)=&\sum_{q=1}^Q{\xi _q\hat{P}_{q}^{\left(\mathrm{SF} \right)}\left( t-\tau_j'-\tau _{n}^{\left( R \right)} \right)}
\nonumber\\
\approx &E_tA\left( t \right) \sum_{q=1}^Q{\rho _qe^{j2\pi \left[ f_0+\left( q-1 \right) \varDelta f' \right] \left( t-\tau _j-\tau_n^{(R)} \right)}}
\label{eq.18a}
\\
\bar{P}_n^{\left( \mathrm{AF} \right)}\left( t \right)=&\sum_{q=1}^Q{\xi _q\hat{P}_{q}^{\left( \mathrm{AF} \right)}\left( t-\tau_j'-\tau _{n}^{\left( R \right)} \right)}
\nonumber\\
\approx &E_tA\left( t \right)  \sum_{q=1}^Q{\rho _qe^{j2\pi \left[ f_0+\left( q-1 \right) \varDelta f' \right] \left( t-\tau _j+\tau_q-\tau_n^{(R)} \right)}}
\label{eq.18b}
\end{align}
\end{subequations}
where $E_t=\boldsymbol{w}_{t}^{\mathrm{H}}(\varphi _t,\theta _t)\boldsymbol{a}_t(\varphi _j,\theta _j)$ and $\tau _j=2R_j/c$ is the two-way time delay and $\tau_n^{(R)}=-\frac{d}{c}\left( n-1 \right) \cos \varphi_j \cos \theta_j$ denotes the receive array delay for n-th receive element. The propagation loss $\xi _q$ for the q-th jamming signal is included in $\rho _q$, which can be considered as the q-th jammer antenna gain. Note that \eqref{eq.18a} and \eqref{eq.18b} use the narrow-band assumption, $A\left( t \right) \approx A\left( t-\tau_j'-\tau _{n}^{\left( R \right)} \right) $.

\setcounter{TempEqCnt}{\value{equation}} 
\setcounter{equation}{19}
\begin{figure*}[t]
\begin{subequations}
\begin{align}
\bar{F}_n^{\left( \mathrm{SF} \right)}\left( t \right)=&\sum_{q=1}^Q{\xi _q\hat{F}_{q}^{\left( \mathrm{SF} \right)}\left( t-\tau_j'-\tau _{n}^{\left( R \right)} \right)}
\approx\tilde{E}_t \sum_{s=1}^S{u_{\mathrm{s}}\left( t-\tau _j \right) e^{j2\pi f_0\tau _{\mathrm{s}}\left( \varphi _j,\theta _j \right)}}\sum_{q=1}^Q{\rho _qe^{j2\pi \left[ f_0+\left( q-1 \right) \varDelta f' \right] (t-\tau _j-\tau_n^{(R)})}}
\label{eq.21a}
\\
\bar{F}_n^{\left( \mathrm{AF} \right)}\left( t \right)=&\sum_{q=1}^Q{\xi _q\hat{F}_{q}^{\left( \mathrm{AF} \right)}\left( t-\tau_j'-\tau _{n}^{\left( R \right)} \right)}
\approx\tilde{E}_t \sum_{s=1}^S{u_{\mathrm{s}}\left( t-\tau _j \right) e^{j2\pi f_0\tau _{\mathrm{s}}\left( \varphi _j,\theta _j \right)}}\sum_{q=1}^Q{\rho _qe^{j2\pi \left[ f_0+\left( q-1 \right) \varDelta f' \right] (t-\tau _j+\tau_q-\tau_n^{(R)})}}
\label{eq.21b}
\end{align}
\end{subequations}
\hrulefill
\end{figure*}

\setcounter{TempEqCnt}{\value{equation}} 
\setcounter{equation}{17}

\subsection{Against FDA-MIMO radar}

The FDA-MIMO radar signal $\check{F}\left( t \right)$ intercepted by the FDA jammer can be represented as
\begin{align}
\check{F}\left( t \right) =&\xi _j\tilde{\boldsymbol{w}}_{{t}}^{\mathrm{H}}(\varphi _t,\theta _t)\tilde{\boldsymbol{a}}_t(\varphi _j,\theta _j)
\nonumber\\
&\times\sum_{s=1}^S{u_s\left( t-\tau_j' \right)e^{j2\pi f_0[t-\tau_j'+\tau _s\left( \varphi _j,\theta _j \right)]}}
\label{eq.19}
\end{align}
Each jammer antenna receives the synthetic of different baseband waveforms transmitted by the corresponding subarrays. Then the jamming signal transmitted by the q-th antenna of the SF and AF jammer can be expressed as 
\begin{subequations}
\begin{align}
\hat{F}_{q}^{\left( \mathrm{SF} \right)} (t)=&\tilde{\boldsymbol{w}}_{{t}}^{\mathrm{H}}(\varphi _t,\theta _t)\tilde{\boldsymbol{a}}_t(\varphi _j,\theta _j)e^{j2\pi \left[ f_0+\left( q-1 \right) \varDelta f' \right](t-\tau_j')}
\nonumber\\
&\times\rho _q\sum_{s=1}^S{u_s\left( t-\tau_j' \right) e^{j2\pi f_0\tau _s\left( \varphi _j,\theta _j \right)}}
\label{eq.20a}
\\
\hat{F}_{q}^{\left( \mathrm{AF} \right)} (t)\approx&\tilde{\boldsymbol{w}}_{{t}}^{\mathrm{H}}(\varphi _t,\theta _t)\tilde{\boldsymbol{a}}_t(\varphi _j,\theta _j)e^{j2\pi \left[ f_0+\left( q-1 \right) \varDelta f' \right]\left(t-\tau_j'+\tau_q \right)}
\nonumber\\
&\times \rho _q\sum_{s=1}^S{u_s\left( t-\tau_j' \right) e^{j2\pi f_0\tau _s\left( \varphi _j,\theta _j \right)}}
\label{eq.20b}
\end{align}
\end{subequations}
where $u_s\left( t-\tau_j' \right) \approx A\left( t \right) e^{j2\pi \left( s-1 \right) \varDelta f\left( t-\tau _j' \right)}$ by using the narrow-band assumption and $u_s\left( t-\tau_j' \right)\approx u_s\left( t-\tau_j'+\tau_q \right)$ by ignoring $e^{j2\pi \left( s-1 \right) \varDelta f\tau_q}$ due to $\varDelta f\ll f_0$. \eqref{eq.20a} and \eqref{eq.20b} indicate that the jamming frequency offset can be loaded on the each transmit waveform signal even if the baseband waveforms are different.

After $\tau_j'=R_j/c$ of propagation, the SF and AF jamming signals received by the n-th element of FDA-MIMO radar can be expressed as \eqref{eq.21a} and \eqref{eq.21b}, respectively, where $\tilde{E}_t=\tilde{\boldsymbol{w}}_{t}^{\mathrm{H}}(\varphi _t,\theta _t)\tilde{\boldsymbol{a}}_t(\varphi _j,\theta _j)$, and $\xi _q$ denotes the propagation loss for the q-th jammer antenna and it has been included in $\rho_q$. Note that $e^{-j2\pi \left( s-1 \right) \varDelta f\tau _{n}^{\left( R \right)}}$ is ignored in \eqref{eq.21a} and \eqref{eq.21b} due to $\varDelta f\ll f_0$.

\setcounter{TempEqCnt}{\value{equation}} 
\setcounter{equation}{20}

\section{MATCHED FILTERING FOR FDA JAMMING}

This section derives the FDA jamming signals after the MF processing based on the PA radar and the FDA-MIMO radar, discussing the effectiveness of the FDA jamming on the MF process.


\subsection{Case 1: PA radar}

The MF process of PA radar requires two steps, one is the down-conversion by using $e^{-j2\pi f_0 (t-\tau_j)}$ for the fast-time sampling at the distance of $R_j$, and the other is the MF by using the waveform $A(t)$ on each receive element. After the down-conversion in \eqref{eq.18a} and \eqref{eq.18b}, the SF and AF jamming signals for the n-th receive element can be expressed as 
\begin{subequations}
\begin{align}
\tilde{P}_{n}^{\left( \mathrm{SF} \right)}\left( t \right) \approx &E_tA\left( t \right)  e^{-j2\pi f_0 \tau _{n}^{(R)} }\sum_{q=1}^Q{\rho _qe^{j2\pi \left( q-1 \right) \varDelta f'\left( t-\tau _j \right)}}
\label{eq.22a}
\\
\tilde{P}_{n}^{\left( \mathrm{AF} \right)}\left( t \right) \approx &E_tA\left( t \right) e^{-j2\pi f_0\tau _{n}^{(R)}}
\nonumber\\
&\times \sum_{q=1}^Q{\rho _qe^{j2\pi f_0\tau _q}e^{j2\pi \left( q-1 \right) \varDelta f'\left( t-\tau _j \right)}}
\label{eq.22b}
\end{align}
\end{subequations}
where \eqref{eq.22a} ignores $e^{-j2\pi \left( q-1 \right) \varDelta f'\tau _{n}^{(R)}}$ and \eqref{eq.22b} ignores $e^{j2\pi \left( q-1 \right) \varDelta f'(\tau_q-\tau _{n}^{(R)})}$ due to $\varDelta f'\ll f_0$. Defining the following auxiliary vectors,
\begin{subequations}
\begin{align}
\boldsymbol{\rho }&=\left[ \begin{matrix}
  \rho _1&    \rho _2&    \cdots&   \rho _Q\\
\end{matrix} \right] \label{eq.23a}
\\
\boldsymbol{\vartheta}^{\left( \rm{SF} \right)}\left( t \right) &=\left[ \begin{matrix}
  1&      \cdots&   e^{j2\pi \left( Q-1 \right) \varDelta f'  \left( t-\tau _j \right)}\\
\end{matrix} \right] ^{\mathrm{T}}\label{eq.23b}
\\
\boldsymbol{\vartheta}^{\left( \rm{AF} \right)}\left( t \right)&=\left[ \begin{matrix}
  1&      \cdots&   e^{j2\pi \left[ f_0\tau _Q+\left( Q-1 \right) \varDelta f'\left( t-\tau _j \right) \right]}\\
\end{matrix} \right] ^{\mathrm{T}}\label{eq.23c}
\end{align}
\end{subequations}
where $\tau _Q=(Q-1)d_j\cos{\varphi_j}\cos{\theta_j}/c$, then the SF and AF jamming signals after MF on the n-th receive element of PA radar can be expressed as 
\begin{subequations}
\begin{align}
P_{n}^{\left( \mathrm{SF} \right)}=&e^{-j2\pi f_0\tau _{n}^{(R)}}\int_{T_p}{E_t\left[ \boldsymbol{\rho}\cdot \boldsymbol{\vartheta }^{\left( \mathrm{SF} \right)}\left( t \right) \right] \left| A\left( t \right) \right|^2\mathrm{d}t}\label{eq.24a}
\\
P_{n}^{\left( \mathrm{AF} \right)}=&e^{-j2\pi f_0\tau _{n}^{(R)}}\int_{T_p}{E_t\left[ \boldsymbol{\rho}\cdot \boldsymbol{\vartheta }^{\left( \mathrm{AF} \right)}\left( t \right) \right] \left| A\left( t \right) \right|^2\mathrm{d}t}
\label{eq.24b}
\end{align}
\end{subequations}
For PA radar, the $N\times 1$ receive jamming vector can be expressed as
\begin{equation}
\boldsymbol{j}_{\mathrm{P}}=\varUpsilon _{\mathrm{P}}^{\left( \mathrm{FDA} \right)}\boldsymbol{a}_r\left( \varphi _j,\theta _j \right) 
\label{eq.25}
\end{equation}
where $\boldsymbol{a}_r\left( \varphi _j,\theta _j \right)$ can be referred to \eqref{eq.7}. $\varUpsilon _{\mathrm{P}}^{\left( \mathrm{FDA} \right)}$ denotes a scalar factor against PA radar, impacting on the amplitude and phase of FDA jamming signal. The superscript `$^{(\rm{FDA})}$' is used to replace the superscript `$^{(\rm{SF})}$' and `$^{(\rm{AF})}$' for simplicity.
\begin{equation}
\varUpsilon _{\mathrm{P}}^{\left( \mathrm{FDA} \right)}=\int_{T_p}{E_t\left[ \boldsymbol{\rho }\cdot\boldsymbol{\vartheta }^{\left( \mathrm{FDA} \right)}\left( t \right) \right] \left| A\left( t \right) \right|^2\mathrm{d}t}
\label{eq.26}
\end{equation}
where $\boldsymbol{\vartheta }^{\left( \mathrm{FDA} \right)}\left( t \right)$ can be replaced by $\boldsymbol{\vartheta }^{\left( \mathrm{SF} \right)}\left( t \right)$ and $\boldsymbol{\vartheta }^{\left( \mathrm{AF} \right)}\left( t \right)$ in \eqref{eq.23b} and \eqref{eq.23c}.  

\subsection{Case 2: FDA-MIMO radar}

The MF process of FDA-MIMO radar requires two steps, one is the down-conversion similar to the PA radar, and the other is the multi-channel MF by using the waveforms $\left\{ u_s\left( t \right) \right\} _{s=1}^{S}$ for $S$ channels on each receive element. After the down-conversion in \eqref{eq.21a} and \eqref{eq.21b}, the SF and AF jamming signals on the n-th receive element can be expressed as
\begin{subequations}
\begin{align}
\tilde{F}_{n}^{\left( \mathrm{SF} \right)}\left( t \right) \approx &\tilde{E}_t\sum_{s=1}^S{u_{\mathrm{s}}\left( t-\tau _j \right) e^{j2\pi f_0\tau _{\mathrm{s}}\left( \varphi _j,\theta _j \right)}}
\nonumber\\
&\times\sum_{q=1}^Q{\rho _qe^{j2\pi \left( q-1 \right) \varDelta f'\left( t-\tau _j \right)}}
\label{eq.27a}
\\
\tilde{F}_{n}^{\left( \mathrm{AF} \right)}\left( t \right) \approx &\tilde{E}_t\sum_{s=1}^S{u_{\mathrm{s}}\left( t-\tau _j\right) e^{j2\pi f_0\tau _{\mathrm{s}}\left( \varphi _j,\theta _j \right)}}
\nonumber\\
&\times\sum_{q=1}^Q{\rho _qe^{j2\pi f_0\tau _q}e^{j2\pi \left( q-1 \right) \varDelta f'\left( t-\tau _j \right)}}
\label{eq.27b}
\end{align}
\end{subequations}
where \eqref{eq.27a} ignores $e^{j2\pi \left( q-1 \right) \varDelta f'\tau _{n}^{(R)}}$ and \eqref{eq.27b} ignores $e^{j2\pi \left( q-1 \right) \varDelta f'(\tau_q-\tau _{n}^{(R)})}$ due to $\varDelta f'\ll f_0$.  By matched-filtering with $u_s(t)$ on the s-th channel of the n-th receive element, the FDA jamming signals can be expressed as
\begin{align}
F_{n,s}^{\left( \mathrm{FDA} \right)}=\int_{T_p}{\tilde{F}_{n}^{\left( \mathrm{FDA} \right)}\left( t \right) u_{s}^{*}\left( t \right) \mathrm{d}t}
\label{eq.28}
\end{align}
Defining the following auxiliary matrix, 
\begin{align}
&\boldsymbol{X}\left( t \right) 
\nonumber\\
=&\left[ \begin{matrix}
  \left|A\left( t \right)\right|^2&     \cdots&   \left|A\left( t \right)\right|^2e^{j2\mathrm{\pi}\left( S-1 \right) \varDelta ft}\\
  \vdots&     \ddots&   \vdots\\
  \left|A\left( t \right)\right|^2e^{-j2\mathrm{\pi}\left( S-1 \right) \varDelta ft}&     \cdots&   \left|A\left( t \right)\right|^2\\
\end{matrix} \right]
\label{eq.29}
\end{align}
and using the auxiliary vectors in \eqref{eq.14a}, \eqref{eq.14b}, \eqref{eq.14c}, \eqref{eq.23a}, \eqref{eq.23b}, and \eqref{eq.23c}, the jamming signals received by the n-th element for FDA-MIMO radar can be modeled as a $S\times 1$ vector.
\begin{align}
\boldsymbol{j}_{\mathrm{F,n}}^{\left( \mathrm{FDA} \right)}&=\tilde{E}_t\cdot e^{-j2\pi f_0\tau_n^{(R)}}\cdot
\nonumber\\
&  \int_{T_p}{\left[ \boldsymbol{\rho  }\cdot\boldsymbol{\vartheta}^{\left( \rm{FDA} \right)}\left( t \right) \right]}\boldsymbol{X}\left( t \right)\mathrm{d}t \cdot\boldsymbol{c}\left( \varphi_j ,\theta_j, R_j \right)
\label{eq.30}
\end{align}
Thereby the $SN\times 1$ receive jamming vector for FDA-MIMO radar can be expressed as 
\begin{equation}
\boldsymbol{j}_{\mathrm{F}}=\left[ \boldsymbol{\varUpsilon }_{\mathrm{F}}^{\left( \mathrm{FDA} \right)}\boldsymbol{c}\left( \varphi_j ,\theta_j, R_j \right) \right] \otimes \boldsymbol{a}_r\left( \varphi _j,\theta _j \right)
\label{eq.31}
\end{equation}
where $\boldsymbol{\varUpsilon }_{\mathrm{F}}^{\left( \mathrm{FDA} \right)}\in \mathbb{C}^{S\times S}$ is a spectral leakage matrix for the FDA jamming against the FDA-MIMO radar.
\begin{equation}
\boldsymbol{\varUpsilon }_{\mathrm{F}}^{\left( \mathrm{FDA} \right)}=\int_{T_p}{  \tilde{E}_t\left[ \boldsymbol{\rho  }\cdot\boldsymbol{\vartheta}^{\left( \rm{FDA} \right)}\left( t \right) \right]}\boldsymbol{X}\left( t \right)\mathrm{d}t
\label{eq.32}
\end{equation}

\subsection{Multiple false targets}

From \eqref{eq.25} and \eqref{eq.31}, ${\varUpsilon }_{\mathrm{P}}^{\left( \mathrm{FDA} \right)}$ and $\boldsymbol{\varUpsilon }_{\mathrm{F}}^{\left( \mathrm{FDA} \right)}$ affect the MF process for PA and FDA-MIMO radar, respectively, indicating the difference between the FDA jamming signal and the target signal. For the fast-time snapshot data, the target and FDA jamming can be observed in the range dimension by using the Fast Fourier Transform (FFT) to calculate the MF outputs. To clarify the effectiveness of FDA jamming on the MF process, the following Proposition 1 gives the distribution of range-dimensional outputs for SF and AF jamming after the MF processing and determines the corresponding conditions of the jamming frequency offset $\varDelta f'$ and the jamming power $\rho^2_q$ for the q-th jammer antenna.


\vspace{0.1in}\textbf{Proposition~1}\label{Pro.1}: Consider an FDA jammer with $Q$ jamming antennas against a phased-MIMO radar, which has two different cases, PA radar with $M$ transmit elements and FDA-MIMO radar with $S$ non-overlapping subarrays and $M_S=M/S$ elements within each subarray. Assume that the azimuths and elevations of the target and the FDA jammer relative to the radar are $\varphi _t$, $\theta _t$, $\varphi _j$, and $\theta _j$, respectively. When the jamming frequency offset $\varDelta f'$ and the q-th jamming power satisfy \eqref{eq.33a} and \eqref{eq.33b} against PA radar, or satisfy \eqref{eq.33a} and \eqref{eq.33c} against FDA-MIMO radar, 
\begin{subequations}
\begin{align}
\frac{1}{T_p(Q-1)}\leqslant &\varDelta f'\leqslant \frac{B}{Q-1}\leqslant \frac{\varDelta f}{Q-1} \label{eq.33a}
\\
\left( \rho _{q}^{\left( \mathrm{P} \right)} \right) ^2&\geqslant M\frac{\sigma _{t}^{2}}{\left( E_t \right) ^2}
 \label{eq.33b}
\\
\left( \rho _{q}^{\left( \mathrm{F} \right)} \right) ^2&\geqslant M_S\frac{\sigma _{t}^{2}}{\left( \tilde{E}_t \right) ^2}
\label{eq.33c}
\end{align}
\end{subequations}
where $\sigma _{t}^{2}=\mathrm{E}\left\{ \left| \xi _t \right|^2 \right\} $, and $T_p$, $B$, and $\xi_t$ denote the pulse width, baseband bandwidth, target scattering coefficient, respectively, $E_t=\boldsymbol{w}_{t}^{\mathrm{H}}(\varphi _t,\theta _t)\boldsymbol{a}_t(\varphi _j,\theta _j)$ as mentioned in \eqref{eq.3a} and \eqref{eq.3b} and $\tilde{E}_t=\tilde{\boldsymbol{w}}_{t}^{\mathrm{H}}(\varphi _t,\theta _t)\tilde{\boldsymbol{a}}_t(\varphi _j,\theta _j)$ as mentioned in \eqref{eq.8a} and \eqref{eq.8b}, then the SF and AF jammer can generate $Q$ false targets in the range dimension after MF process for both two cases of phased-MIMO radar, where the q-th false target will appear at 
\begin{subequations}
\begin{align}
R _{q}^{\left( \mathrm{SF} \right)}=&R _j-T_p\varDelta f'\varDelta R\left( q-1 \right) \label{eq.34a}
\\
R _{q}^{\left( \mathrm{AF} \right)}=&R _j-\frac{(q-1)T_p}{4B}\cos{\theta_j}\cos{\varphi_j}
\nonumber\\
&-T_p\varDelta f'\varDelta R\left( q-1 \right)\label{eq.34b}
\end{align}
\end{subequations}
where $\varDelta R=c/2B$ is the range resolution of radar and $R_j$ denotes the range of the FDA jammer relative to the radar.


\vspace{0.05in}\textbf{Proof}: See Appendix A.\vspace{0.05in}

Proposition 1 indicates the effect of FDA jamming on the MF of phased-MIMO radar and lists the constraints of the jamming frequency offset and the jamming power against the phased-MIMO radar. Moreover, \eqref{eq.34a} and \eqref{eq.34b} indicate that the false target peaks can be adjusted by the jamming frequency offset and the number of jammer antennas. Given a known target distance, the FDA jamming can generate multiple false targets near the target, increasing the radar false alarms. Even though FDA jamming is more expensive than the conventional delayed copy jamming for generating the range deceptive interference [\ref{cite29}], reporting on the properties of FDA jamming after the MF process is necessary. Apart from the MF processing, it can also deteriorate the spatial filtering and the STAP of the phased-MIMO radar under different conditions of the jamming frequency offset. Next section we will discuss the effectiveness of FDA jamming against the radar spatial filtering on the condition of $\varDelta f'< {1}/[(Q-1){T_p}]$.

\section{SPATIAL FILTERING FOR FDA JAMMING}

This section introduces the spatial filtering process of the phased-MIMO radar under the presence of FDA jamming, discussing the effects of FDA jamming on the beampattern and the output SINR based on the PA radar and FDA-MIMO radar.

The existing anti-jamming algorithms are based on the fact that the target signal and the jamming signal can be distinguished by the radar from the range dimension, the azimuth dimension, or the Doppler dimension [\ref{cite15}]. From the perspective of ECM, the jamming signal can cover the target signal in a specific dimension, which causes the radar to suppress the jamming in this dimension while also suppressing the target energy [\ref{cite31}]. Furthermore, even if the jamming does not cover the target in a particular dimension, the jamming signal can also worsen the performance of radar anti-jamming algorithms through its property [\ref{cite25}]. Apart from the range-dimensional false targets as described in Section IV, FDA jamming can be considered as sidelobe or mainlobe interference when the spatial positions of the target and jammer maintain a certain relationship. Specifically, sidelobe interference requires that the jammer and target are at a close distance but different azimuths while mainlobe interference requires that the jammer and target are at the same azimuth but different ranges. Radar adaptive spatial filtering can use the training data to estimate the interference covariance matrix and suppress the jamming within sidelobe [\ref{cite21}] or mainlobe [\ref{cite9}]. However, the conventional mainlobe or sidelobe deceptive jamming signals are generated by forwarding the original radar signals, while the FDA jamming signals load a small jamming frequency offset on the intercepted radar signals, which has a different interference covariance matrix and a significant influence on the performance of radar anti-interference. 

In this section, we assume that the interference covariance matrix is known for the radar, which means the spatial filtering vector is optimal [\ref{cite15}], avoiding the performance degradation due to the covariance matrix estimation from the training data. As a sidelobe or mainlobe jamming, the phased-MIMO radar can receive the echoes reflected by the jammer itself. For the PA radar and the FDA-MIMO radar, the reflected jamming signal after the MF process can be expressed as 
\begin{subequations}
\begin{align}
\boldsymbol{v}_{\mathrm{P}}=&\xi _v\cdot E_t\cdot\boldsymbol{a}_r\left( \varphi _j,\theta _j \right) 
\label{eq.35a}
\\
\boldsymbol{v}_{\mathrm{F}}=&\xi _v\cdot\tilde{E}_t \cdot\boldsymbol{c}\left( \varphi _j,\theta _j,R_j \right) \otimes \boldsymbol{a}_r\left( \varphi _j,\theta _j \right) 
\label{eq.35b}
\end{align}
\end{subequations}
where $\xi _v$ denotes the jamming reflection coefficient. The disturbance (interference plus noise) covariance matrix is
\begin{align}
\boldsymbol{R}_w=&\mathrm{E}\left\{ \left( \boldsymbol{j}+\boldsymbol{v}+\boldsymbol{n} \right) \left( \boldsymbol{j}+\boldsymbol{v}+\boldsymbol{n} \right) ^{\mathrm{H}} \right\} 
\nonumber\\
=&\boldsymbol{R}_j+\boldsymbol{R}_v+ \boldsymbol{R}_n
\label{eq.36}
\end{align}
where $\boldsymbol{j}$ can be replaced by $\boldsymbol{j}_\mathrm{P}$ in \eqref{eq.25} and $\boldsymbol{j}_\mathrm{F}$ in \eqref{eq.31} corresponding to PA and FDA-MIMO radar, respectively. $\boldsymbol{v}$ can be replaced by $\boldsymbol{v}_\mathrm{P}$ in \eqref{eq.35a} and $\boldsymbol{v}_\mathrm{F}$ in \eqref{eq.35b}, respectively. Similarly, $\boldsymbol{R}_j=\mathrm{E}\{ \boldsymbol{j} \boldsymbol{j} ^{\mathrm{H}} \}$ and $\boldsymbol{R}_v=\mathrm{E}\{ \boldsymbol{v} \boldsymbol{v} ^{\mathrm{H}} \}$ can be replaced as $\boldsymbol{R}_j^{(\mathrm{P})}=\mathrm{E}\{ \boldsymbol{j}_{\mathrm{P}} (\boldsymbol{j}_{\mathrm{P}}) ^{\mathrm{H}} \}$ and $\boldsymbol{R}_v^{(\mathrm{P})}=\mathrm{E}\{ \boldsymbol{v}_{\mathrm{P}} (\boldsymbol{v}_{\mathrm{P}}) ^{\mathrm{H}} \}$ for PA radar, and $\boldsymbol{R}_j^{(\mathrm{F})}=\mathrm{E}\{ \boldsymbol{j}_{\mathrm{F}} (\boldsymbol{j}_{\mathrm{F}}) ^{\mathrm{H}} \}$ and $\boldsymbol{R}_v^{(\mathrm{F})}=\mathrm{E}\{ \boldsymbol{v}_{\mathrm{F}} (\boldsymbol{v}_{\mathrm{F}}) ^{\mathrm{H}} \}$ for FDA-MIMO radar. $\boldsymbol{R}_n=\mathrm{E}\{  \boldsymbol{n}  \boldsymbol{n} ^{\mathrm{H}} \} $ and $\boldsymbol{n}$ denotes the noise satisfies the white Gaussian noise with zero mean and variance $\sigma ^2_n$. Suppose that the interference covariance matrix for radar spatial filtering is dominated by FDA jamming. 
\begin{align}
\boldsymbol{R}_u=\boldsymbol{R}_j+ \boldsymbol{R}_n
\label{eq.37}
\end{align}

Adaptive spatial filtering aims at maximizing the output SINR for target detection. Hence, we adapt the minimum variance distortionless response (MVDR) beamformer to minimize the disturbance power while maintaining a distortionless power toward the direction of the target of interest. The optimal $N\times 1$ and $SN\times 1$ weight vectors for the MVDR filter of PA and FDA-MIMO radar can be expressed as
\begin{subequations}
\begin{align}
\boldsymbol{w}^{(\mathrm{P})}_o=& \frac{1}{\xi _t \sqrt{M}}\cdot\boldsymbol{R}_{u}^{-1}\boldsymbol{x}_{\mathrm{P}}
\label{eq.38a}
\\
\boldsymbol{w}^{(\mathrm{F})}_o=& \frac{1}{\xi _t \sqrt{M_S}}\cdot\boldsymbol{R}_{u}^{-1}\boldsymbol{x}_{\mathrm{F}}
\label{eq.38b}
\end{align}
\end{subequations}
where $\boldsymbol{x}_{\mathrm{P}}$ and $\boldsymbol{x}_{\mathrm{F}}$ are referred to \eqref{eq.6} and \eqref{eq.13}. $\frac{1}{\xi _t \sqrt{M}}$ and $\frac{1}{\xi _t \sqrt{M_S}}$ guarantee $[ \boldsymbol{w}_{o}^{\left( \mathrm{P} \right)} ] ^{\mathrm{H}}\boldsymbol{x}_{\mathrm{P}}=1$ and $[ \boldsymbol{w}_{o}^{\left( \mathrm{F} \right)} ] ^{\mathrm{H}}\boldsymbol{x}_{\mathrm{F}}=1$, respectively. 

We investigate the effects of the FDA jamming on the beampattern and output SINR for radar spatial filtering from the following three measurements.
\begin{subequations}
\begin{align}
\left|Y_{\varphi _j}\left( \varDelta f' \right)\right| 
&\triangleq \left| \boldsymbol{w}_{o}^{\mathrm{H}}\boldsymbol{v} \right|\left| _{R =R _t,\varphi =\varphi _j} \right. 
\label{eq.39a}
\\
\left|Y_{R _j}\left( \varDelta f' \right)\right| 
&\triangleq \left| \boldsymbol{w}_{o}^{\mathrm{H}}\boldsymbol{v} \right|\left| _{R =R _j,\varphi =\varphi _t} \right. 
\label{eq.39b}
\\
\mathrm{SINR}_o\left( \varDelta f' \right) &\triangleq \frac{\left| \boldsymbol{w}_{o}^{\mathrm{H}}\boldsymbol{x} \right|^2}{\boldsymbol{w}_{o}^{\mathrm{H}}(\boldsymbol{R}_v+\boldsymbol{R}_n)\boldsymbol{w}_o}\approx \frac{1}{\boldsymbol{w}_{o}^{\mathrm{H}}\boldsymbol{R}_v\boldsymbol{w}_o}
\label{eq.39c}
\end{align}
\end{subequations}
where $\left. \left( \cdot \right) \right|_{\theta =\theta _t,\varphi =\varphi _j}$ denotes the calculation under the conditions of $\theta =\theta _t$ and $\varphi =\varphi _j$. $\boldsymbol{w}_{o}$ and $\boldsymbol{x}$ are two integrated expressions, which can be replaced by $\boldsymbol{w}^{(\mathrm{P})}_o$ in \eqref{eq.38a} and $\boldsymbol{x}^{(\mathrm{P})}$ in \eqref{eq.6} for PA, and $\boldsymbol{w}^{(\mathrm{F})}_o$ in \eqref{eq.38b} and $\boldsymbol{x}^{(\mathrm{F})}$ in \eqref{eq.13} for FDA-MIMO radar. $\left|Y_{\varphi _j}\left( \varDelta f' \right)\right|$ denotes the jamming notch at $\varphi_j$ of radar azimuthal beampattern and $\left|Y_{\varphi _j}\left( \varDelta f' \right)\right|^2$ is known as the sidelobe jamming power after the spatial filtering under the conditions of $R_j = R _t$ and $\varphi_t \ne\varphi _j$ [\ref{cite21}]. $\left|Y_{R _j}\left( \varDelta f' \right)\right|$ denotes the jamming notch at $R_j$ of radar range-dimensional beampattern and $\left|Y_{R _j}\left( \varDelta f' \right)\right|^2$ is known as the mainlobe jamming power after the spatial filtering under the conditions of $R_j \ne R _t$ and $\varphi_t =\varphi _j$ [\ref{cite9}, \ref{cite10}]. $\mathrm{SINR}_o\left( \varDelta f' \right)$ represents the output SINR after the spatial filtering, where the noise can be ignored in the strong jamming environment [\ref{cite3}]. Moreover, it is worth noting that the effects of the FDA jamming on these measurements originate from $\varUpsilon _{\mathrm{P}}^{\left( \mathrm{FDA} \right)}$ in \eqref{eq.26} for PA radar and $\boldsymbol{\varUpsilon }_{\mathrm{F}}^{\left( \mathrm{FDA} \right)}$ in \eqref{eq.32} for FDA-MIMO radar. To simplify and analyze these measurements, Proposition 2 presents the properties of $\varUpsilon _{\mathrm{P}}^{\left( \mathrm{FDA} \right)}$ and $\boldsymbol{\varUpsilon }_{\mathrm{F}}^{\left( \mathrm{FDA} \right)}$ of two types of FDA jammer, SF and AF jammer. 

\vspace{0.1in}\textbf{Proposition~2}\label{Pro.2}: As $\varDelta f'< {1}/[(Q-1){T_p}]$ and $T_p \Delta f$ is a positive integer, the frequency spectrum leakage matrix $\boldsymbol{\varUpsilon }_{\mathrm{F}}^{\left( \mathrm{SF} \right)}$ and $\boldsymbol{\varUpsilon }_{\mathrm{F}}^{\left( \mathrm{AF} \right)}$ of AF and SF jammer against FDA-MIMO radar with $S$ subarrays can be expressed as 
\begin{subequations}
\begin{align}
&\boldsymbol{\varUpsilon }_{\mathrm{F}}^{\left( \mathrm{SF} \right)}\approx \frac{\tilde{E}_t}{E_t}\varUpsilon _{\mathrm{P}}^{\left( \mathrm{SF} \right)}\boldsymbol{I}_S
\label{eq.40a}
\\
&\boldsymbol{\varUpsilon }_{\mathrm{F}}^{\left( \mathrm{AF} \right)}\approx \frac{\tilde{E}_t}{E_t}\varUpsilon _{\mathrm{P}}^{\left( \mathrm{AF} \right)}\boldsymbol{I}_S
\label{eq.40b}
\end{align}
\end{subequations}
where $E_t=\boldsymbol{w}_{t}^{\mathrm{H}}(\varphi _t,\theta _t)\boldsymbol{a}_t(\varphi _j,\theta _j)$ as mentioned in \eqref{eq.3a} and \eqref{eq.3b} and $\tilde{E}_t=\tilde{\boldsymbol{w}}_{t}^{\mathrm{H}}(\varphi _t,\theta _t)\tilde{\boldsymbol{a}}_t(\varphi _j,\theta _j)$ as mentioned in \eqref{eq.8a} and \eqref{eq.8b}. $\varUpsilon _{\mathrm{P}}^{\left( \mathrm{SF} \right)}$ and $\varUpsilon _{\mathrm{P}}^{\left( \mathrm{AF} \right)}$ are the jamming factors of SF and AF jammer against PA radar, respectively, both monotonically decreasing with respect to $\Delta f'\in \left[ 0,\frac{1}{\left( Q-1 \right) T_p} \right) $.
\begin{subequations}
\begin{align}
\varUpsilon _{\mathrm{P}}^{\left( \mathrm{SF} \right)}\approx&E_t\cdot\boldsymbol{\rho }\cdot \boldsymbol{\varPhi }\left( \varDelta f' \right) 
\label{eq.41a}
\\
\varUpsilon _{\mathrm{P}}^{\left( \mathrm{AF} \right)}\approx&E_t\cdot\left[ \boldsymbol{\rho }\odot \boldsymbol{g}^{\mathrm{T}}\left( \varphi _j,\theta _j \right) \right] \cdot \boldsymbol{\varPhi }\left( \varDelta f' \right) 
\label{eq.41b}
\end{align}
\end{subequations}
where $A(t)$ is the rectangular baseband signal with pulse width of $T_p$ and the unit energy. $\boldsymbol{\rho }$ can be referred to \eqref{eq.23a}.
\begin{subequations}
\begin{align}
\boldsymbol{g}\left( \varphi ,\theta \right) =\left[ \begin{matrix}
  1&    \cdots&   e^{j2\pi \frac{d_j}{\lambda _0}\left( Q-1 \right) \cos \varphi \cos \theta}\\
\end{matrix} \right] ^{\mathrm{T}}
\label{eq.42a}
\\
\boldsymbol{\varPhi }\left( \varDelta f' \right) =\left[ \begin{matrix}
  1&    \cdots&   \frac{\sin \left[ \pi T_p\left( Q-1 \right) \varDelta f' \right]}{\pi T_p\left( Q-1 \right) \varDelta f'}\\
\end{matrix} \right] ^{\mathrm{T}}
\label{eq.42b}
\end{align}
\end{subequations}

\vspace{0.05in}\textbf{Proof}: See Appendix B.\vspace{0.05in}

By using three measurements and Proposition 2, let us discuss the spatial filtering of the PA radar and the FDA-MIMO radar under the presence of FDA jamming, respectively. To simplify the expressions, we use $\boldsymbol{a}_{r}^{\left( \varphi _j \right)}$, $\boldsymbol{a}_{r}^{\left( R_j \right)}$, $\boldsymbol{a}_{r}^{\left( T \right)}$, and $\boldsymbol{a}_{r}^{\left( J \right)}$ to  represent $\boldsymbol{a}_r\left( \varphi _j,\theta _t \right)$, $\boldsymbol{a}_r\left( \varphi _t,\theta _j \right)$, $\boldsymbol{a}_r\left( \varphi _t,\theta _t \right)$, and $\boldsymbol{a}_r\left( \varphi _j,\theta _j \right)$, respectively. Similarly, we use $\boldsymbol{c}^{\left( \varphi _j \right)}$, $\boldsymbol{c}^{\left( R_j \right)}$, $\boldsymbol{c}^{\left( T \right)}$, and $\boldsymbol{c}^{\left( J \right)}$ to represent $\boldsymbol{c}\left( \varphi _j,\theta _t,R_t \right)$, $\boldsymbol{c}\left( \varphi _t,\theta _j,R_j \right)$, $\boldsymbol{c}\left( \varphi _t,\theta _t,R_t \right)$, and $\boldsymbol{c}\left( \varphi _j,\theta _j,R_j \right)$, respectively.

\vspace{-0.05in}\subsection{Case 1: PA radar}

Since PA radar has no range-dependency, we consider $\left|Y_{\varphi _j}\left( \varDelta f' \right)\right|$ and $\mathrm{SINR}_o\left( \varDelta f' \right)$ by substituting \eqref{eq.35a} and \eqref{eq.38a} into \eqref{eq.39a} and \eqref{eq.39c}, respectively.
\begin{subequations}
\begin{align}
\left| Y_{\varphi _j}\left( \varDelta f' \right) \right|=&\left| \frac{\mathcal{X} _{r}^{\left( \varphi _j \right)}}{\sigma _{n}^{2}}-\frac{\varPsi _{\mathrm{P}}\left( \varDelta f' \right) \mathcal{X} _{r}^{\left( \varphi _j \right)}\left( \sum_{q=1}^Q{\rho _{q}^{2}} \right)}{\sigma _{n}^{2}\left( \sigma _{n}^{2}+N\sum_{q=1}^Q{\rho _{q}^{2}} \right)} \right|
\label{eq.43a}
\\
\mathrm{SINR}_o\left( \varDelta f' \right) &=M\sigma _{t}^{2}\times
\nonumber\\
&{\left| \frac{\mathcal{X} _{r}^{\left( J \right)}}{\sigma _{n}^{2}}-\frac{\varPsi _{\mathrm{P}}^{\left( J \right)}\left( \varDelta f' \right) \mathcal{X} _{r}^{\left( J \right)}\left( \sum_{q=1}^Q{\rho _{q}^{2}} \right)}{\sigma _{n}^{2}\left( \sigma _{n}^{2}+N\sum_{q=1}^Q{\rho _{q}^{2}} \right)} \right|^{-2}}
\label{eq.43b}
\end{align}
\end{subequations}
where 
\begin{subequations}
\begin{align}
\mathcal{X} _{r}^{\left( \varphi _j \right)}=&\left[ \boldsymbol{a}_{r}^{\left( \varphi _j \right)} \right] ^{\mathrm{H}}\boldsymbol{a}_{r}^{\left( T \right)}
\label{eq.44a}
\\
\varPsi^{( \varphi _j )} _{\mathrm{P}}\left( \varDelta f' \right) =&\left| \varUpsilon _{\mathrm{P}}^{\left( \mathrm{FDA} \right)}\boldsymbol{a}_{r}^{\left( \varphi _j \right)} \right|^{2}
\label{eq.44b}
\\
\mathcal{X} _{r}^{\left( J \right)}=&\left[ \boldsymbol{a}_{r}^{\left( J \right)} \right] ^{\mathrm{H}}\boldsymbol{a}_{r}^{\left( T \right)}
\label{eq.44c}
\\
\varPsi _{\mathrm{P}}^{\left( J \right)}\left( \varDelta f' \right) =&\left| \varUpsilon _{\mathrm{P}}^{\left( \mathrm{FDA} \right)}\boldsymbol{a}_{r}^{\left( J \right)} \right|^{2}
\label{eq.44d}
\end{align}
\end{subequations}
See Appendix C for the derivations of \eqref{eq.43a} and \eqref{eq.43b}. Using the monotonicity of $\varUpsilon _{\mathrm{P}}^{\left( \mathrm{SF} \right)}$ and $\varUpsilon _{\mathrm{P}}^{\left( \mathrm{AF} \right)}$ in Proposition 2, we can summarize the relationship between two measurements and the jamming frequency offset $\varDelta f'$ for the sidelobe jamming suppression of PA radar.

\vspace{0.05in}\noindent\textbf{Conclusion:}\vspace{0.05in}

Under the condition of $\varDelta f'\in \left[ 0,\frac{1}{\left( Q-1 \right) T_p} \right) $, as $\Delta f'$ increases, $\varPsi^{( \varphi _j )} _{\mathrm{P}}\left( \varDelta f' \right)$ and $\varPsi^{( J )} _{\mathrm{P}}\left( \varDelta f' \right)$ will decrease with the decreasing of $\varUpsilon _{\mathrm{P}}^{\left( \mathrm{FDA} \right)}$ as shown in \eqref{eq.44b} and \eqref{eq.44d}, thereby $Y_{\varphi _j}\left( \varDelta f' \right)$ in (\ref{eq.43a}) becomes large, which means that the jamming notch of azimuthal beampattern at $\varphi_j$ will rise and the jamming power after the spatial filtering will increase. Accordingly, $\mathrm{SINR}_o\left( \varDelta f' \right)$ in (\ref{eq.43b}) will decrease, which means the performance deterioration of the sidelobe jamming suppression for PA radar.

\subsection{Case 2: FDA-MIMO radar}

FDA-MIMO radar has a range-dependent beampattern, thus we consider $\left|Y_{\varphi _j}\left( \varDelta f' \right)\right|$, $\left|Y_{R_j}\left( \varDelta f' \right)\right|$, and $\mathrm{SINR}_o\left( \varDelta f' \right)$, respectively.
\begin{subequations}
\begin{align}
&\left| Y_{\varphi _j}\left( \varDelta f' \right) \right|
\nonumber\\
=&\left|\frac{\mathcal{X} _{t}^{\left( \varphi _j \right)}\mathcal{X} _{r}^{\left( \varphi _j \right)}}{\sigma _{\mathrm{n}}^{2}}-\frac{\varPsi _{\mathrm{F}}^{\left( \varphi _j \right)}\left( \varDelta f' \right) \mathcal{X} _{r}^{\left( \varphi _j \right)}(\sum_{q=1}^Q{\rho _{q}^{2}})}{\sigma _{n}^{2}\left( \sigma _{n}^{2}+S\cdot N\cdot \sum_{q=1}^Q{\rho _{q}^{2}} \right)}\right|
\label{eq.45a}
\\
&\left| Y_{R _j}\left( \varDelta f' \right) \right|
\nonumber\\
=&\left|\frac{\mathcal{X} _{t}^{\left( R _j \right)}\mathcal{X} _{r}^{\left( R _j \right)}}{\sigma _{\mathrm{n}}^{2}}-
\frac{\varPsi _{\mathrm{F}}^{\left( R _j \right)}\left( \varDelta f' \right) \mathcal{X} _{r}^{\left( R _j \right)}(\sum_{q=1}^Q{\rho _{q}^{2}})}{\sigma _{n}^{2}\left( \sigma _{n}^{2}+S\cdot N\cdot \sum_{q=1}^Q{\rho _{q}^{2}} \right)}\right|\label{eq.45b}
\\
&\mathrm{SINR}_o\left( \varDelta f' \right) 
\nonumber\\
=&{M_S\sigma _{t}^{2}}
{\left| \frac{\mathcal{X} _{t}^{\left( J \right)}\mathcal{X} _{r}^{\left( J \right)}}{\sigma _{\mathrm{n}}^{2}}-\frac{\varPsi _{\mathrm{F}}^{\left( J \right)}\left( \varDelta f' \right) \mathcal{X} _{r}^{\left( J \right)}(\sum_{q=1}^Q{\rho _{q}^{2}})}{\sigma _{n}^{2}\left( \sigma _{n}^{2}+S\cdot N\cdot \sum_{q=1}^Q{\rho _{q}^{2}} \right)} \right|^{-2}}
\label{eq.45c}
\end{align}
\end{subequations}
where $\mathcal{X} _{r}^{\left( \varphi _j \right)}$ and $\mathcal{X} _{r}^{\left( J \right)}$ can be referred to \eqref{eq.44a} and \eqref{eq.44c}, respectively.
\begin{subequations}
\begin{align}
\mathcal{X} _{t}^{\left( \varphi _j \right)}=&\left[ \boldsymbol{c}^{\left( \varphi _j \right)} \right] ^{\mathrm{H}}\boldsymbol{c}^{\left( T \right)}
\label{eq.46a}
\\
\mathcal{X} _{t}^{\left( R_j \right)}=&\left[ \boldsymbol{c}^{\left( R_j \right)} \right] ^{\mathrm{H}}\boldsymbol{c}^{\left( T \right)}
\label{eq.46b}
\\
\mathcal{X}_{r}^{\left( R_j \right)}=&\left[ \boldsymbol{a}_{r}^{\left( R_j \right)} \right] ^{\mathrm{H}}\boldsymbol{a}_{r}^{\left( T \right)}
\label{eq.46c}
\\
\mathcal{X} _{t}^{\left( J \right)}=&\left[ \boldsymbol{c}^{\left( J \right)} \right] ^{\mathrm{H}}\boldsymbol{c}^{\left( T \right)}\label{eq.46d}
\\
\varPsi _{\mathrm{F}}^{\left( \varphi_j \right)}\left( \varDelta f' \right) =&\left[ \boldsymbol{c}^{\left( \varphi_j \right)} \right] ^{\mathrm{H}} \boldsymbol{\varUpsilon }_{\mathrm{F}}^{\left( \mathrm{FDA} \right)}\boldsymbol{c}^{\left( \varphi_j \right)}\left[ \boldsymbol{c}^{\left( \varphi_j \right)} \right] ^{\mathrm{H}}\left[ \boldsymbol{\varUpsilon }_{\mathrm{F}}^{\left( \mathrm{FDA} \right)} \right] ^{\mathrm{H}}  
\nonumber\\
&\cdot\boldsymbol{c}^{\left(T \right)}\left| \boldsymbol{a}_{r}^{\left( \varphi_j \right)} \right|^2
\label{eq.46e}
\\
\varPsi _{\mathrm{F}}^{\left( R_j \right)}\left( \varDelta f' \right) =&\left[ \boldsymbol{c}^{\left( R_j \right)} \right] ^{\mathrm{H}} \boldsymbol{\varUpsilon }_{\mathrm{F}}^{\left( \mathrm{FDA} \right)}\boldsymbol{c}^{\left( R_j \right)}\left[ \boldsymbol{c}^{\left( R_j \right)} \right] ^{\mathrm{H}}\left[ \boldsymbol{\varUpsilon }_{\mathrm{F}}^{\left( \mathrm{FDA} \right)} \right] ^{\mathrm{H}} \nonumber\\
&\cdot \boldsymbol{c}^{\left( T \right)}\left| \boldsymbol{a}_{r}^{\left( R_j \right)} \right|^2
\label{eq.46f}
\\
\varPsi ^{\left( J \right)}_{\mathrm{F}}\left( \varDelta f'\right)=&\left[ \boldsymbol{c}^{\left( J \right)} \right] ^{\mathrm{H}} \boldsymbol{\varUpsilon }_{\mathrm{F}}^{\left( \mathrm{FDA} \right)}\boldsymbol{c}^{\left( J \right)}\left[ \boldsymbol{c}^{\left( J \right)} \right] ^{\mathrm{H}}\left[ \boldsymbol{\varUpsilon }_{\mathrm{F}}^{\left( \mathrm{FDA} \right)} \right] ^{\mathrm{H}} \nonumber\\
&\cdot \boldsymbol{c}^{\left( T \right)}\left| \boldsymbol{a}_{r}^{\left( J \right)} \right|^2\label{eq.46g}
\end{align}
\end{subequations}
See Appendix C for derivations of \eqref{eq.45a}, \eqref{eq.45b}, and \eqref{eq.45c}. By using \eqref{eq.40a} and \eqref{eq.40b} in Proposition 2, $\varPsi _{\mathrm{F}}^{\left( R_j \right)}\left( \varDelta f' \right)$ can be modified as
\begin{equation}
\varPsi _{\mathrm{F}}^{\left( R_j \right)}\left( \varDelta f' \right) =\varPsi _{\mathrm{P}}^{\left( R_j \right)}\left( \varDelta f' \right) \mathcal{X} _{t}^{\left( R_j \right)}\left| \boldsymbol{a}_{r}^{\left( J \right)} \right|^2
\label{eq.47}
\end{equation}
which are similar to $\varPsi _{\mathrm{F}}^{\left( \varphi_j \right)}\left( \varDelta f' \right)$ and $\varPsi _{\mathrm{F}}^{\left( J \right)}\left( \varDelta f' \right)$. According to the monotonicity of $\varUpsilon _{\mathrm{P}}^{\left( \mathrm{SF} \right)}$ and $\varUpsilon _{\mathrm{P}}^{\left( \mathrm{AF} \right)}$ as described in Proposition 2, we can summarize the relationship between three measurements and the jamming frequency offset $\Delta f'$ for the jamming suppression of FDA-MIMO radar.

\vspace{0.05in}\noindent\textbf{Conclusion:}\vspace{0.05in}

Under the condition of $\varDelta f'\in \left[ 0,\frac{1}{\left( Q-1 \right) T_p} \right) $, as $\varDelta f'$ increases, $\varPsi^{( \varphi _j )} _{\mathrm{F}}\left( \varDelta f' \right)$ in \eqref{eq.46e}, $\varPsi^{( R _j )} _{\mathrm{F}}\left( \varDelta f' \right)$ in \eqref{eq.46f}, and $\varPsi^{( J )} _{\mathrm{F}}\left( \varDelta f' \right)$ in \eqref{eq.46g} will decrease with the decreasing of $\varUpsilon _{\mathrm{P}}^{\left( \mathrm{FDA} \right)}$ as described in \eqref{eq.47}, thereby $Y_{\varphi _j}\left( \varDelta f' \right)$ in (\ref{eq.45a}) and $Y_{R _j}\left( \varDelta f' \right)$ in (\ref{eq.45b}) become large, which means the jamming notch of the azimuthal beampattern at $\varphi_j$ and the range-dimensional beampattern at $R_j$ will rise. Accordingly, $\mathrm{SINR}_o\left( \varDelta f' \right)$ in (\ref{eq.45c}) will become small, which means the performance deterioration of jamming suppression for FDA-MIMO radar. 



\begin{table}[b]
 \centering
 \caption{Simulation Parameters}
 \label{tab.1}
 \begin{tabular}{lll}
  \toprule
 Parameter & Symbol & Value  \\
  \midrule
  Carrier frequency & ~~$f_0$ & 10~GHz \\
  Platform height & ~~$H$ & 2000~m\\
  Platform velocity & ~~$v_a$ & 75~m/s\\
  Baseband signal bandwidth & ~~$B$ & 10~MHz\\
  Frequency offset for FDA-MIMO radar  & ~~$\varDelta f$ & 10~MHz\\
  Radar antenna spacing & ~~$d$ & 15~mm  \\
  FDA jammer antenna spacing & ~~$d_j$ & 15~mm  \\
  Pulse duration & ~~$T_p$ & 10~us\\
  Number of transmitting antennas & ~~$M$ & 16\\
  Number of receiving antennas & ~~$N$ & 16\\
  Azimuth of target & ~~$\varphi_t $ & 0$^\circ$\\
  Range of target & ~~$R_t$ & 6000~m\\
  Velocity of target & ~~$v_t$ & 25~m/s\\
  \bottomrule
 \end{tabular}
\end{table}

\section{NUMERICAL RESULTS}

In this section, numerical results are presented to verify the effectiveness of two types of FDA jammers against the phased-MIMO radar, which is divided into two cases, PA radar and FDA-MIMO radar. The simulation parameters of the phased-MIMO radar and the target are listed in Table \ref{tab.1}. 

The numerical results consist of three parts, matched filtering, spatial filtering beampattern, and output SINR. In matched filtering, the jamming frequency offset is constrained from ${100}/{(Q-1)}$ kHz to ${10}/{(Q-1)}$ MHz. The case of $\varDelta f'=0~\mathrm{kHz}$ denotes the conventional time delay jammer that copies the envelope to generate a range false target as mentioned in [\ref{cite27}]. In the simulation of spatial filtering beampattern and output SINR, $\varDelta f'$ is constrained from $0$ kHz to ${100}/{(Q-1)}$ kHz. The case of $\varDelta f'=0~\mathrm{kHz}$ denotes the classical sidelobe or mainlobe deceptive jamming as mentioned in [\ref{cite14}, \ref{cite15}, \ref{cite28}, \ref{cite33}, \ref{cite34}]. As a sidelobe interference, the FDA jammer is located at the same distance as the target but at a different azimuth, $\varphi_j=15\degree$. As a mainlobe interference, the FDA jammer is located at the same azimuth as the target but at a different range, $R_j=6010$ m. Fig.\ref{FIG.8}, Fig.\ref{FIG.9}, Fig.\ref{FIG.10}, Fig.\ref{FIG.11}, and Fig.\ref{FIG.13} correspond to the sidelobe interference suppression. Fig.\ref{FIG.12} and Fig.\ref{FIG.14} correspond to the mainlobe interference suppression. The signal-to-noise ratio (SNR) is 10 dB and the jamming-to-noise ratio (JNR), which is defined as a ratio of the sum of the FDA jamming signal power to noise, is 30 dB. The interference-to-noise ratio (INR), which is defined as the ratio of the reflected power of the jammer to noise, is 15 dB.
\begin{subequations}
\begin{align}
\mathrm{SNR}\triangleq &10\lg \frac{\sigma _{t}^{2}}{\sigma _{n}^{2}}
\label{eq.48a}
\\
\mathrm{JNR}\triangleq &10\lg \frac{\sum_{q=1}^Q{\rho _{q}^{2}}}{\sigma _{n}^{2}}\label{eq.48b}
\\
\mathrm{INR}\triangleq &10\lg \frac{\sigma _{j}^{2}}{\sigma _{n}^{2}}\label{eq.48c}
\end{align}
\end{subequations}
where $\sigma _{t}^{2}=\mathrm{E}\left\{ \left| \xi _t \right|^2 \right\} $ and $\sigma _{j}^{2}=\mathrm{E}\left\{ \left| \xi _j \right|^2 \right\} $.

\begin{figure*}[t]
  \centering
  \vspace{-0.15in}
  \begin{minipage}{1\linewidth}  
    \subfigure[$\varDelta f'=0~\mathrm{kHz}$]{
      \label{FIG.6(a)}
\includegraphics[width=0.231\linewidth,height=1.2in]{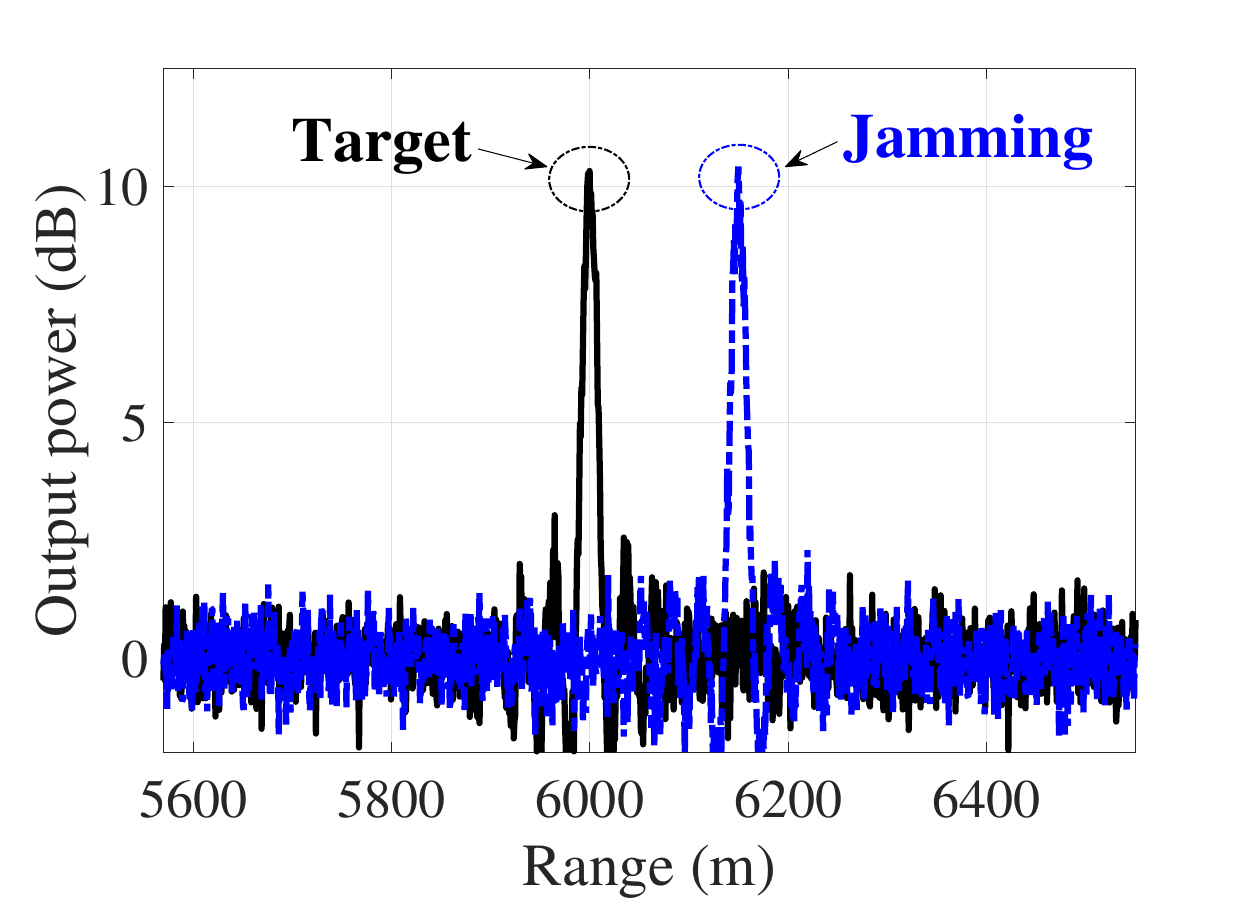}    
 }
    \subfigure[$\varDelta f'=500~\mathrm{kHz}$, $Q=4$]{
      \label{FIG.6(b)}
\includegraphics[width=0.231\linewidth,height=1.2in]{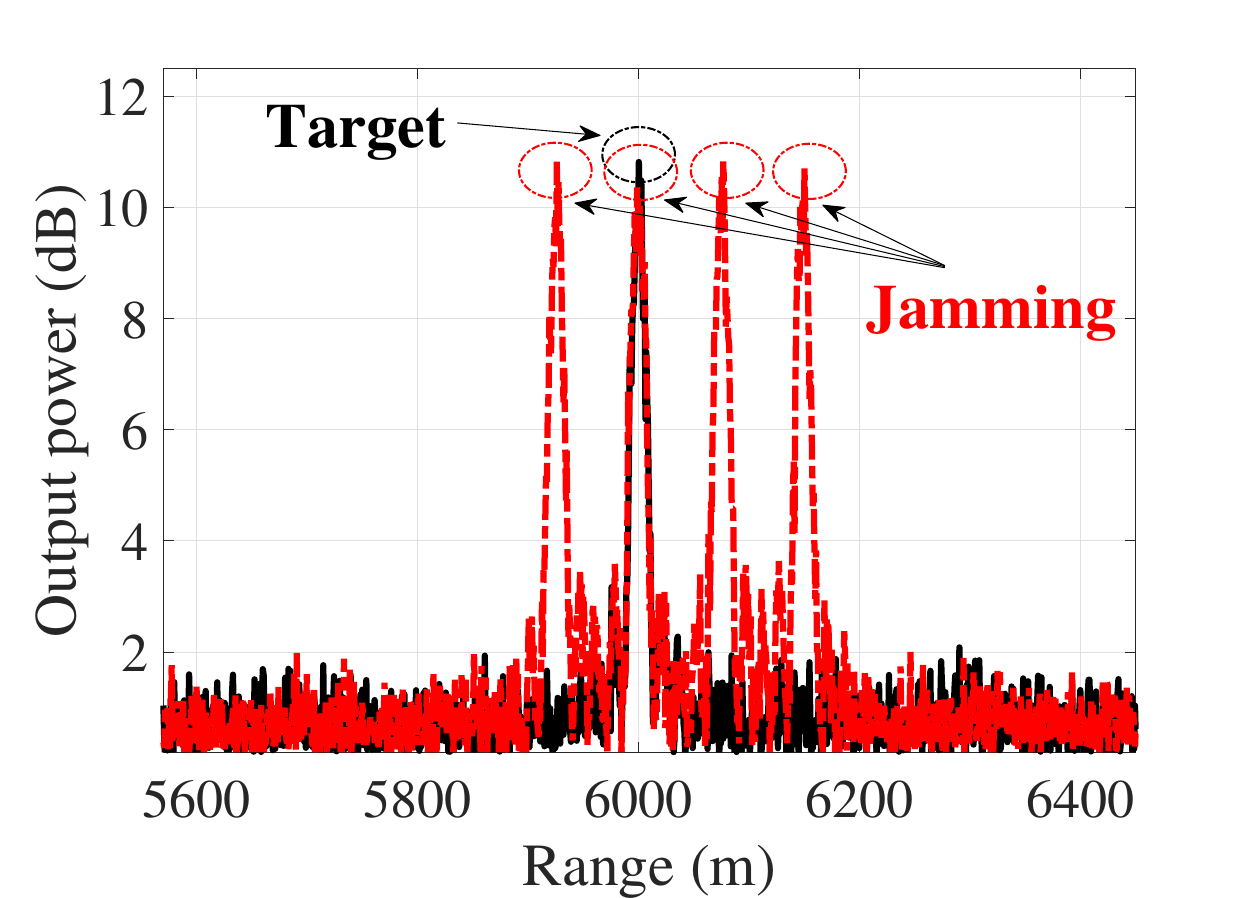}
    }
    \subfigure[$\varDelta f'=400~\mathrm{kHz}$, $Q=8$]{
      \label{FIG.6(c)}
\includegraphics[width=0.231\linewidth,height=1.2in]{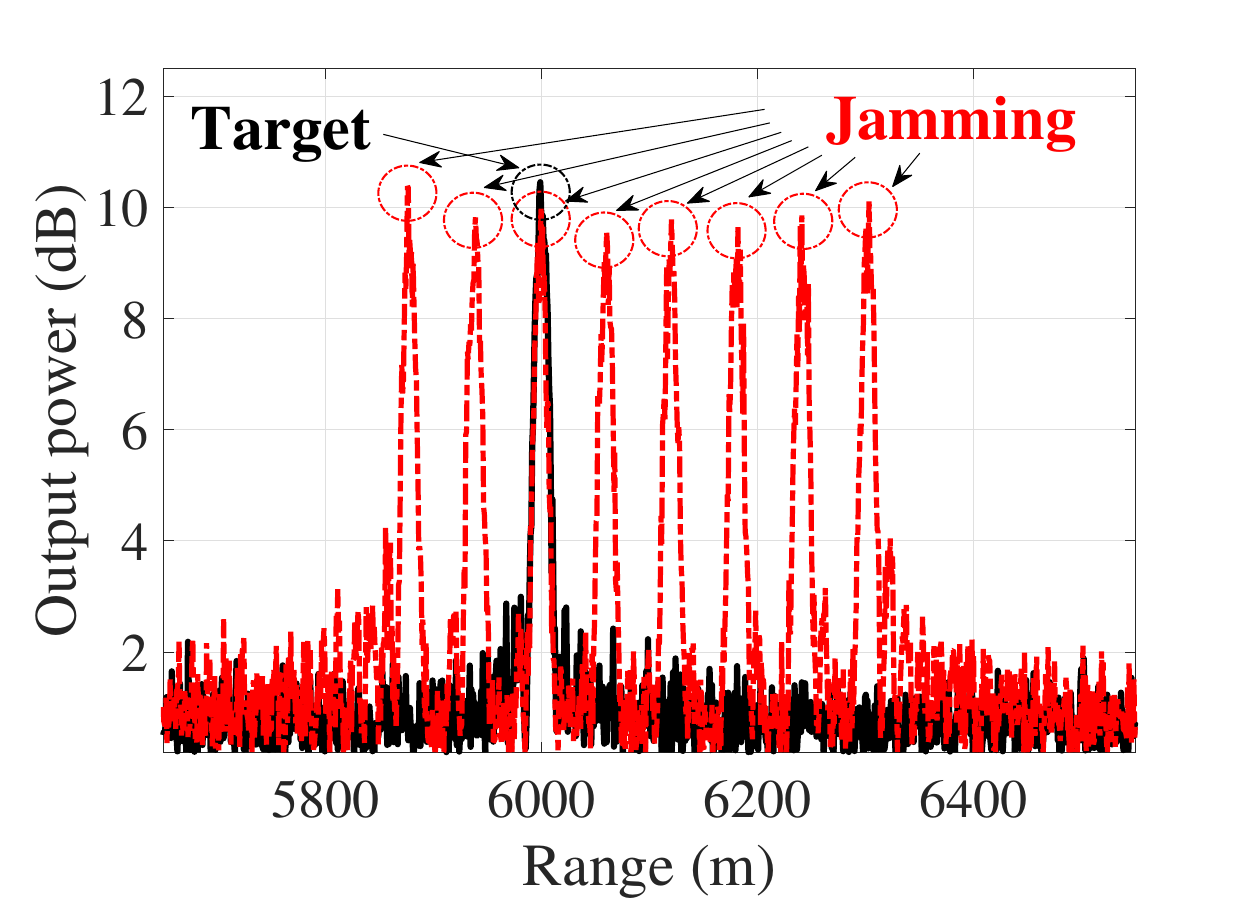}    
 }
 \subfigure[$\varDelta f'=500~\mathrm{kHz}$, $Q=4$]{
      \label{FIG.6(d)}
\includegraphics[width=0.231\linewidth,height=1.2in]{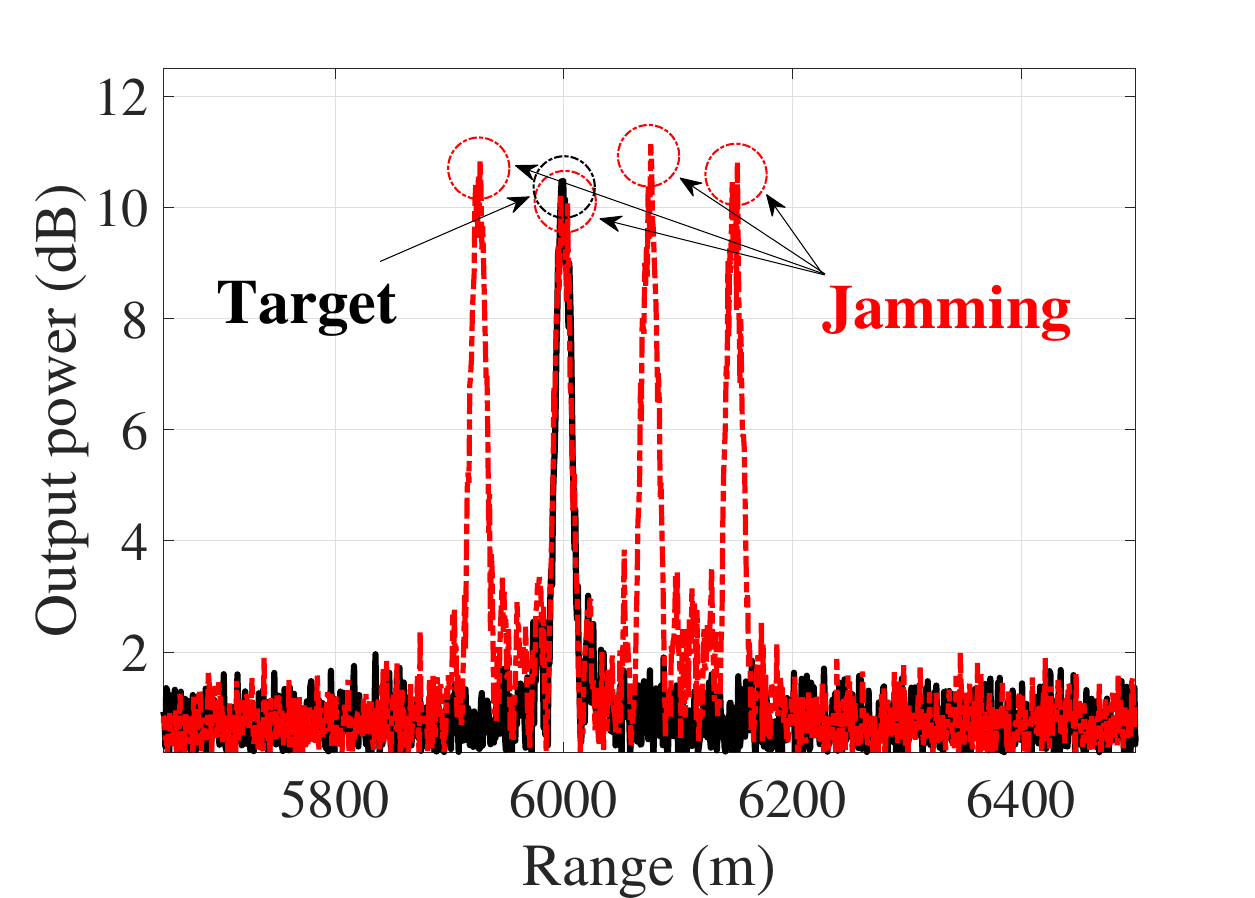}    
 }
  \end{minipage}
  \vspace{-0.05in}  
  \caption{Matched filtering range-dimensional output of the jamming and target for PA radar with $M=16$. (a) Conventional jammer. (b) SF jammer. (c) SF jammer. (d) AF jammer.}
  \label{FIG.6}
\end{figure*}

\begin{figure*}[t]
  \centering
  \begin{minipage}{1\linewidth }
    \subfigure[$\varDelta f'=0~\mathrm{kHz}$]{
      \label{FIG.7(a)}
\includegraphics[width=0.231\linewidth,height=1.2in]{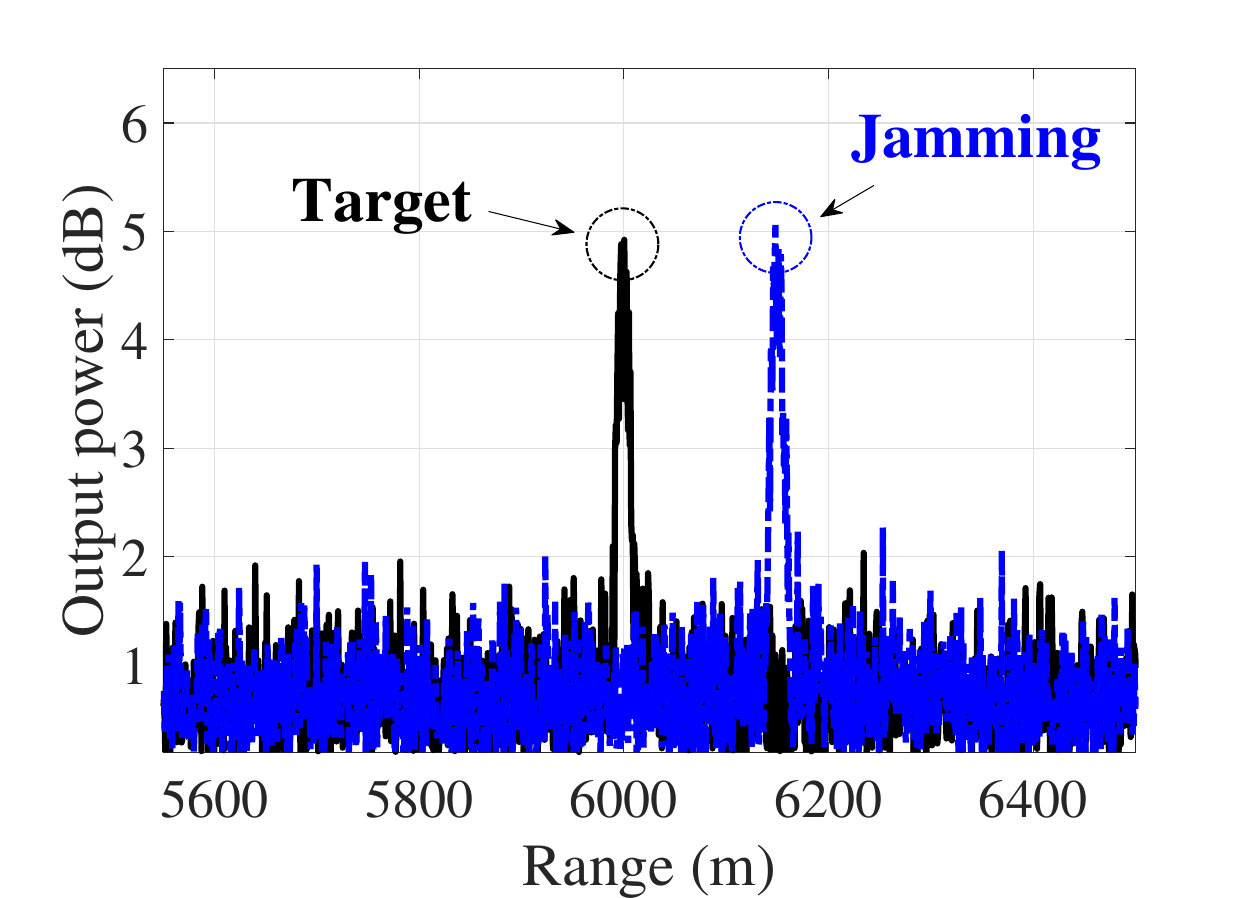}
      }
    \subfigure[$\varDelta f'=500~\mathrm{kHz}$, $Q=4$]{\label{FIG.7(b)}  
    \includegraphics[width=0.231\linewidth,height=1.2in]{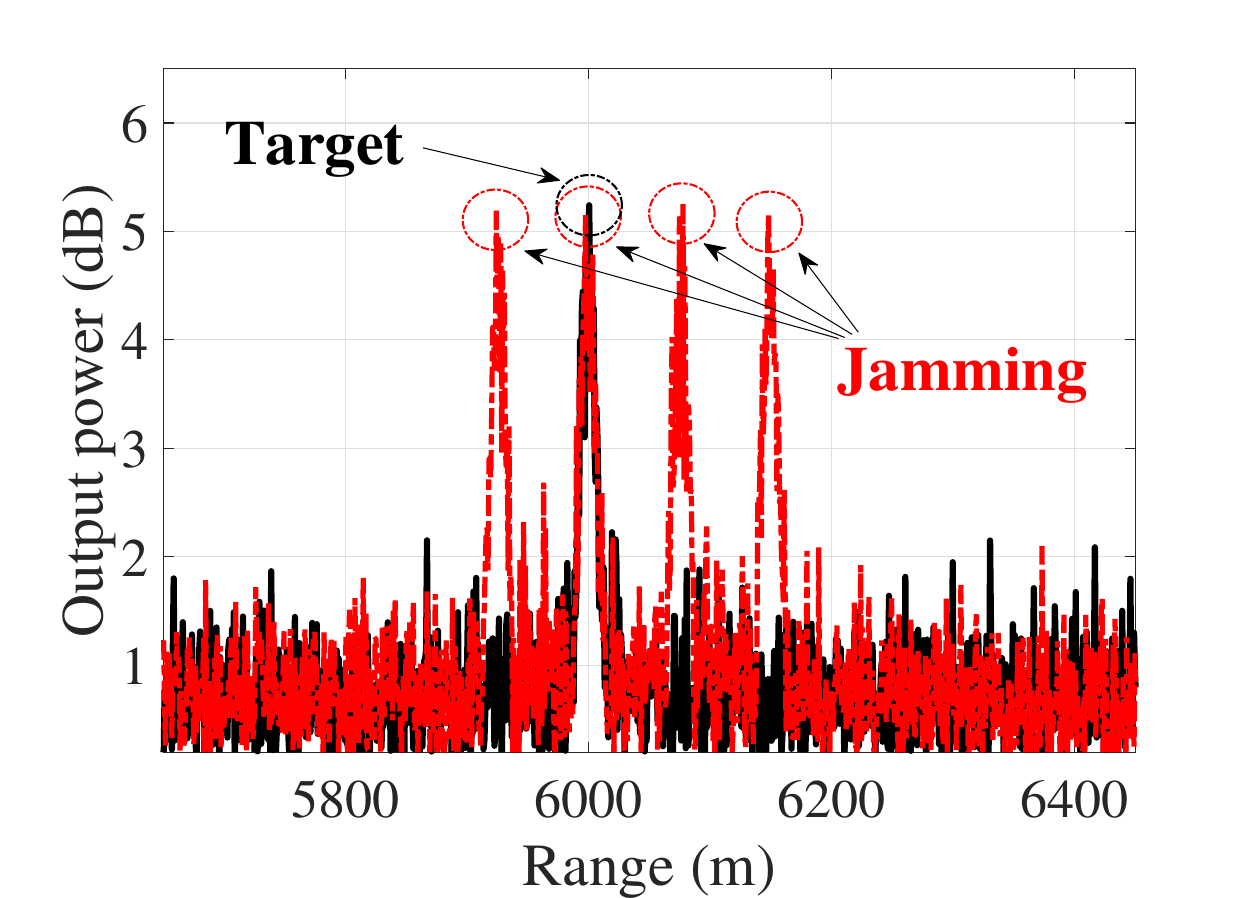}
    }
    \subfigure[$\varDelta f'=400~\mathrm{kHz}$, $Q=8$]{
      \label{FIG.7(c)}
\includegraphics[width=0.231\linewidth,height=1.2in]{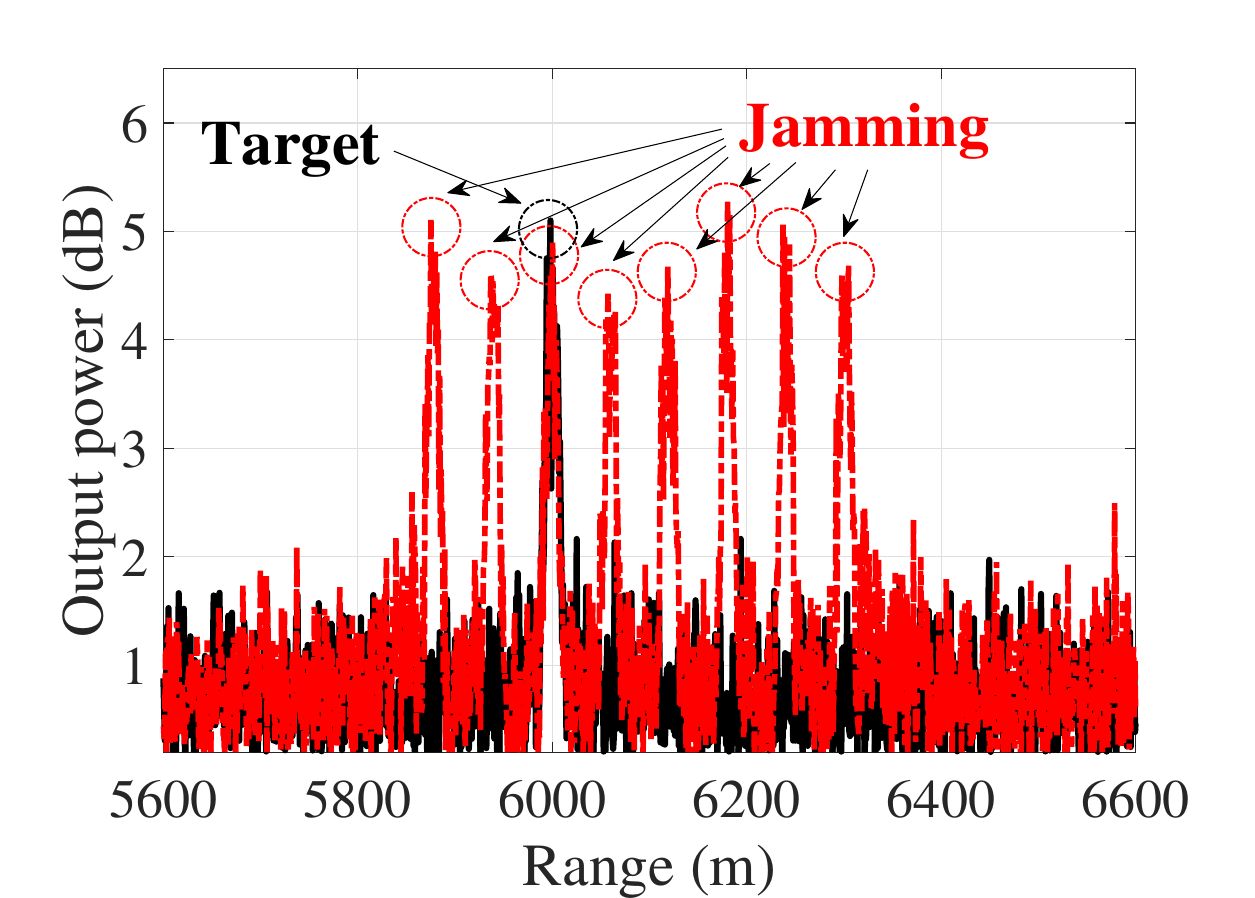}    
 }
 \subfigure[$\varDelta f'=500~\mathrm{kHz}$, $Q=4$]{
      \label{FIG.7(d)}
\includegraphics[width=0.231\linewidth,height=1.2in]{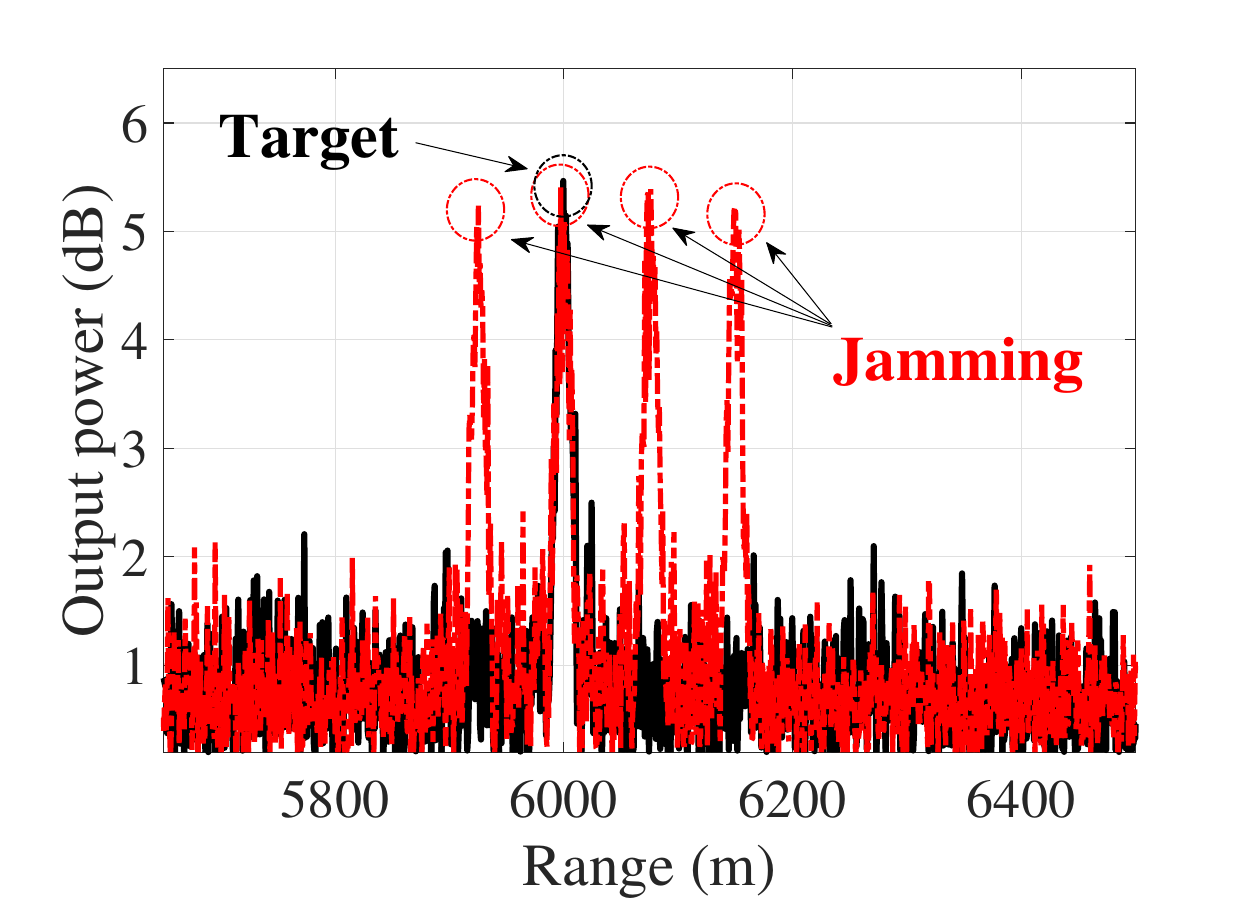} }
  \end{minipage}
  \vspace{-0.05in}  
  \caption{Matched filtering range-dimensional output of the jamming and target for FDA-MIMO radar with $S=2$ and $M_S=8$. (a) Conventional jammer. (b) SF jammer. (c) SF jammer. (d) AF jammer.}
  \label{FIG.7}
\end{figure*}

\subsection{Matched filtering}

In this example, we show the MF outputs of the phased-MIMO radar after receiving the signals reflected by a target and transmitted by the SF or the AF jammer. We use the black solid line, blue dashed line, and red dashed line to represent the target signal, the conventional jamming signal, and FDA jamming signal, respectively. The Doppler shift due to the movement of platform and target causes a frequency mismatch with the match filter. Since the Doppler frequency is much smaller than the signal bandwidth, i.e., $2\left| v_a-v_t \right|/{\lambda_0} \approx 33~\mathrm{kHz}\ll B=10~\mathrm{MHz}$, the Doppler spreading caused by this mismatch is not obvious and can be neglected [\ref{cite15}].

In Fig.\ref{FIG.6}, we present the MF range-dimensional output for PA radar with $M=16$ under different jammers, the conventional jammer in Fig.\ref{FIG.6(a)}, SF jammer with different $\varDelta f'$ and $Q$ in Fig.\ref{FIG.6(b)} and Fig.\ref{FIG.6(c)}, and AF jammer in Fig.\ref{FIG.6(d)}. In Fig.\ref{FIG.6(a)}, the target and the conventional jammer are located at $6000$ m and $6150$ m, respectively. The PA radar receiver can obtain two peaks with the output powers about 10 dB corresponding to the target and jamming at different range bins, indicating PA radar can separate them in range dimension. Fig.\ref{FIG.6(b)} shows the MF output of the same target signal and the jamming signal transmitted by the SF jammer with $\varDelta f'=500$ kHz and $Q=4$, where the ranges of target and jammer are consistent with Fig.\ref{FIG.6(a)}. PA radar receiver obtains a target peak and four jamming peaks in $5925$ m, $6000$ m, $6075$ m, and $6150$ m, which verify \eqref{eq.34a} in Proposition 1. Moreover, the target peak and one of the jamming peaks both appear at $6000$ m, indicating that PA radar cannot distinguish them and suppress the jamming from the range dimension. Fig.\ref{FIG.6(c)} shows the MF output when the target and the jammer are located at $6000$ m and $6300$ m, respectively. By adjusting the jamming frequency offset to $\varDelta f'=400$ kHz and the number of jammer antennas to $Q=8$, PA radar receiver obtains a target peak and eight jamming peaks in $5880$ m, $5940$ m, $6000$ m, $6060$ m, $6120$ m, $6180$ m, $6240$ m, and $6300$ m, which also verify \eqref{eq.34a} in Proposition 1. Meanwhile, the target peak is also covered by one of the jamming peaks based on these parameters of the SF jammer, resulting in the inability of PA radar to distinguish them and suppress the jamming. Fig.\ref{FIG.6(d)} shows the MF output of the AF jammer with the same parameters as Fig.\ref{FIG.6(b)}. Since $\frac{(q-1)T_p}{4B}\cos{\theta_j}\cos{\varphi_j}$ is much smaller than $T_p\varDelta f'\varDelta R\left( q-1 \right)$ in \eqref{eq.34b}, the locations where the jamming peaks appear are consistent with Fig.\ref{FIG.6(b)}, which verifies \eqref{eq.34b} in Proposition 1 for PA radar.

In Fig.\ref{FIG.7}, we present the MF range-dimensional output for FDA-MIMO radar with $S=2$ and $M_S=8$ under different jammers, where the parameters in these four subplots are consistent with the corresponding subplots in Fig.\ref{FIG.6}. Fig.\ref{FIG.7(a)} shows that the FDA-MIMO radar receiver obtains two peaks with the output powers about 5 dB corresponding to the target and jamming at different range bins, indicating that FDA-MIMO radar has a lower coherent gain than PA radar, which can also be observed in Fig.\ref{FIG.7(b)}, Fig.\ref{FIG.7(c)}, and Fig.\ref{FIG.7(d)}. Similarly, Fig.\ref{FIG.7(b)} and Fig.\ref{FIG.7(c)} verify \eqref{eq.34a} in Proposition 1 and indicate that the jamming peak can cover the target by adjusting $\varDelta f'$ and $Q$ to avoid FDA-MIMO radar distinguishing them and suppressing the jamming in the range dimension. Fig.\ref{FIG.7(d)} shows that the jamming peaks appear at about $5925$ m, $6000$ m, $6075$ m, and $6150$ m, which verifies \eqref{eq.34b} in Proposition 1 for FDA-MIMO radar. 

\begin{figure*}[t]
  \centering
  \begin{minipage}{1\linewidth }
    \subfigure[SF jammer]{
      \label{FIG.8(a)}
\includegraphics[width=20pc]{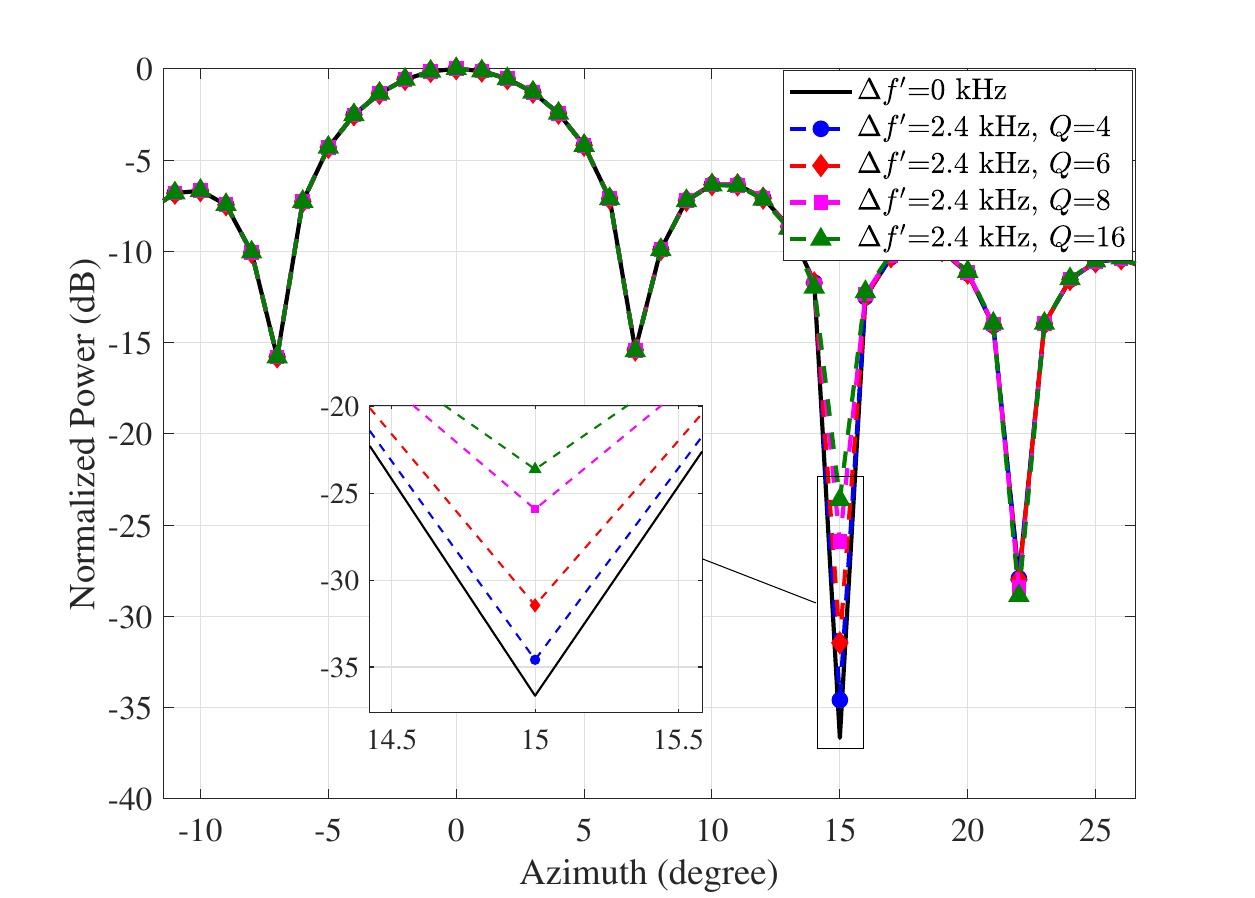}
      }
    \subfigure[AF jammer]{\label{FIG.8(b)}  
    \includegraphics[width=20pc]{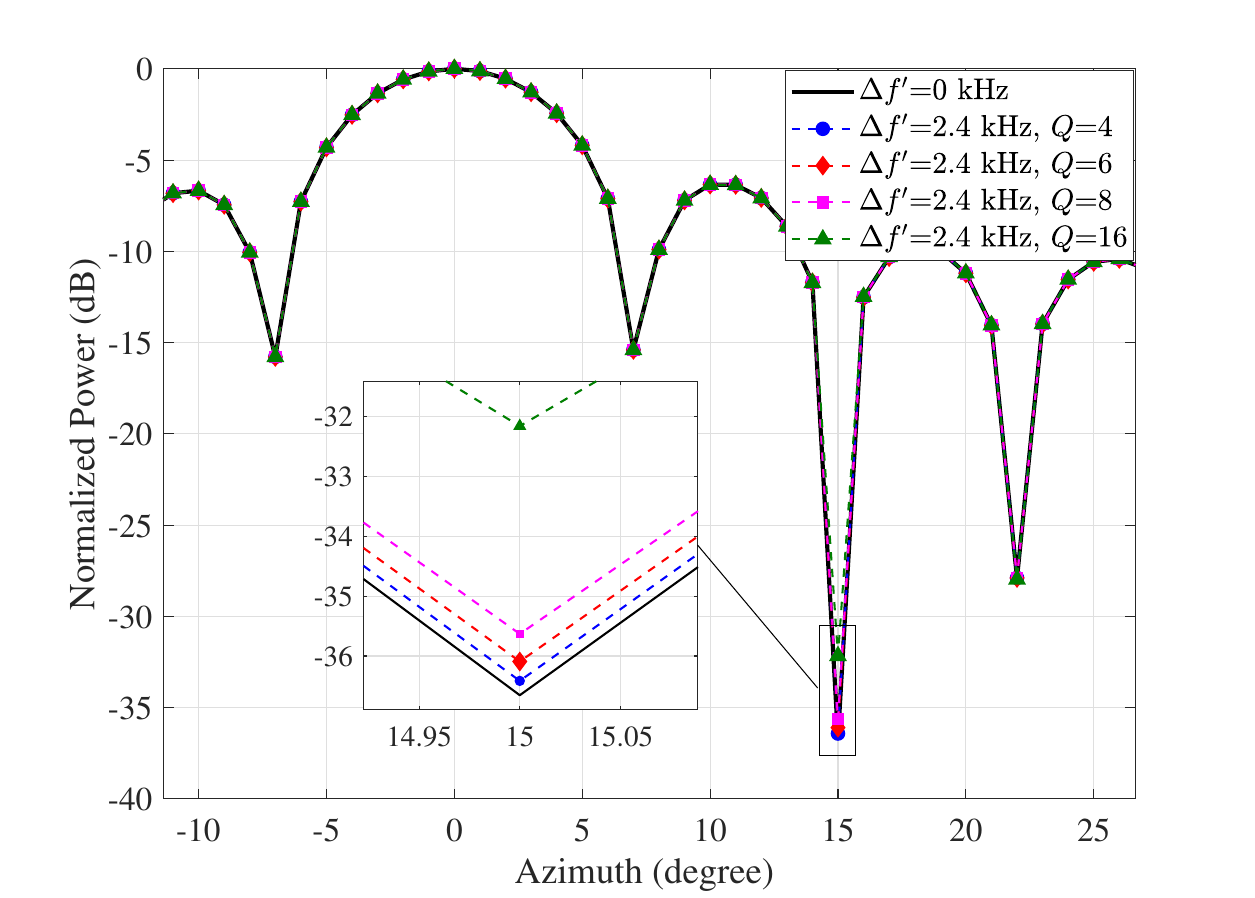}
   }
  \end{minipage}
  \vspace{-0.05in}  
  \caption{Effect of $Q$ for FDA jamming on PA radar. (a) SF jammer. (b) AF jammer.}
  \label{FIG.8}
\end{figure*}

\begin{figure*}[t]
  \centering
  \vspace{-0.15in}
  \begin{minipage}{1\linewidth }
    \subfigure[SF jammer]{
      \label{FIG.9(a)}
\includegraphics[width=20pc]{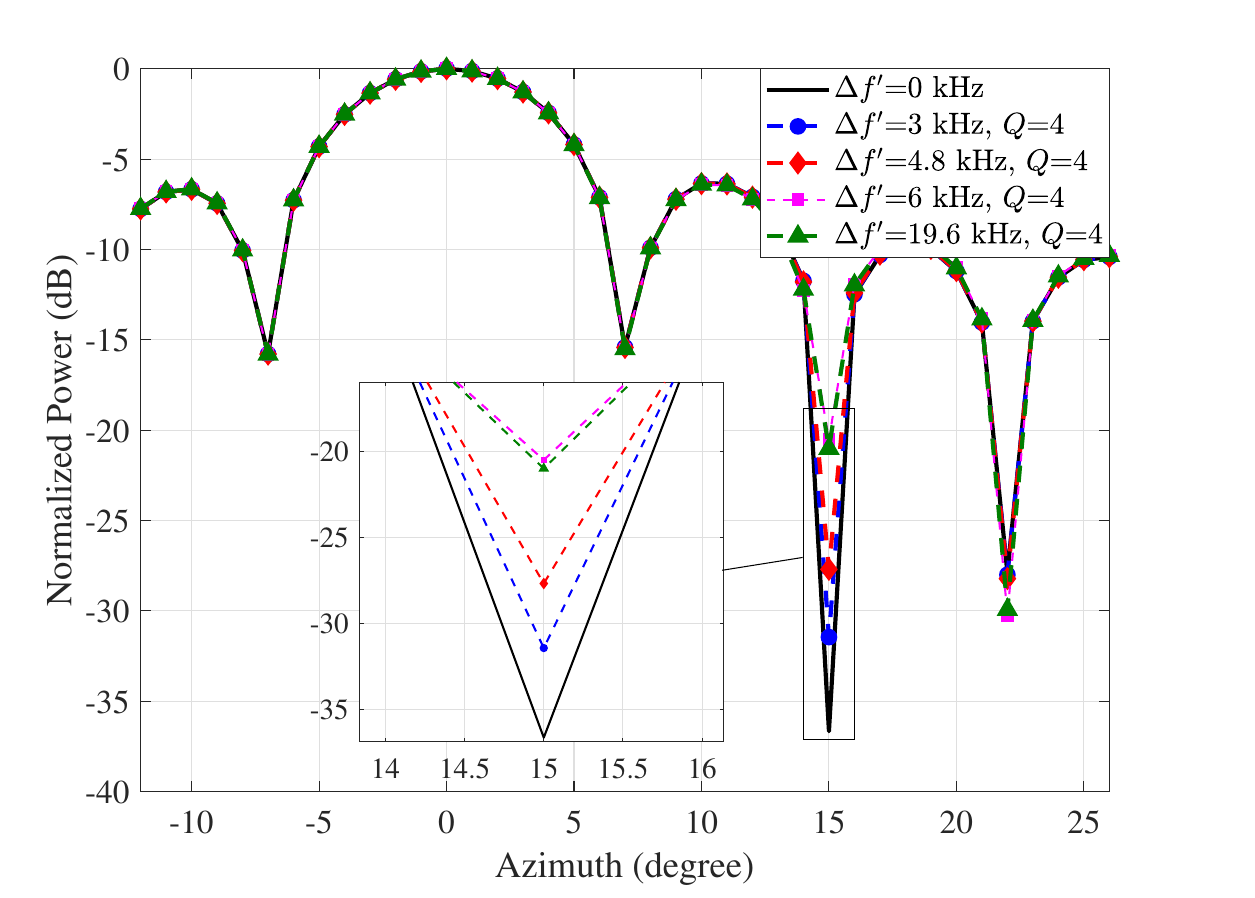}
      }
    \subfigure[AF jammer]{\label{FIG.9(b)}  
    \includegraphics[width=20pc]{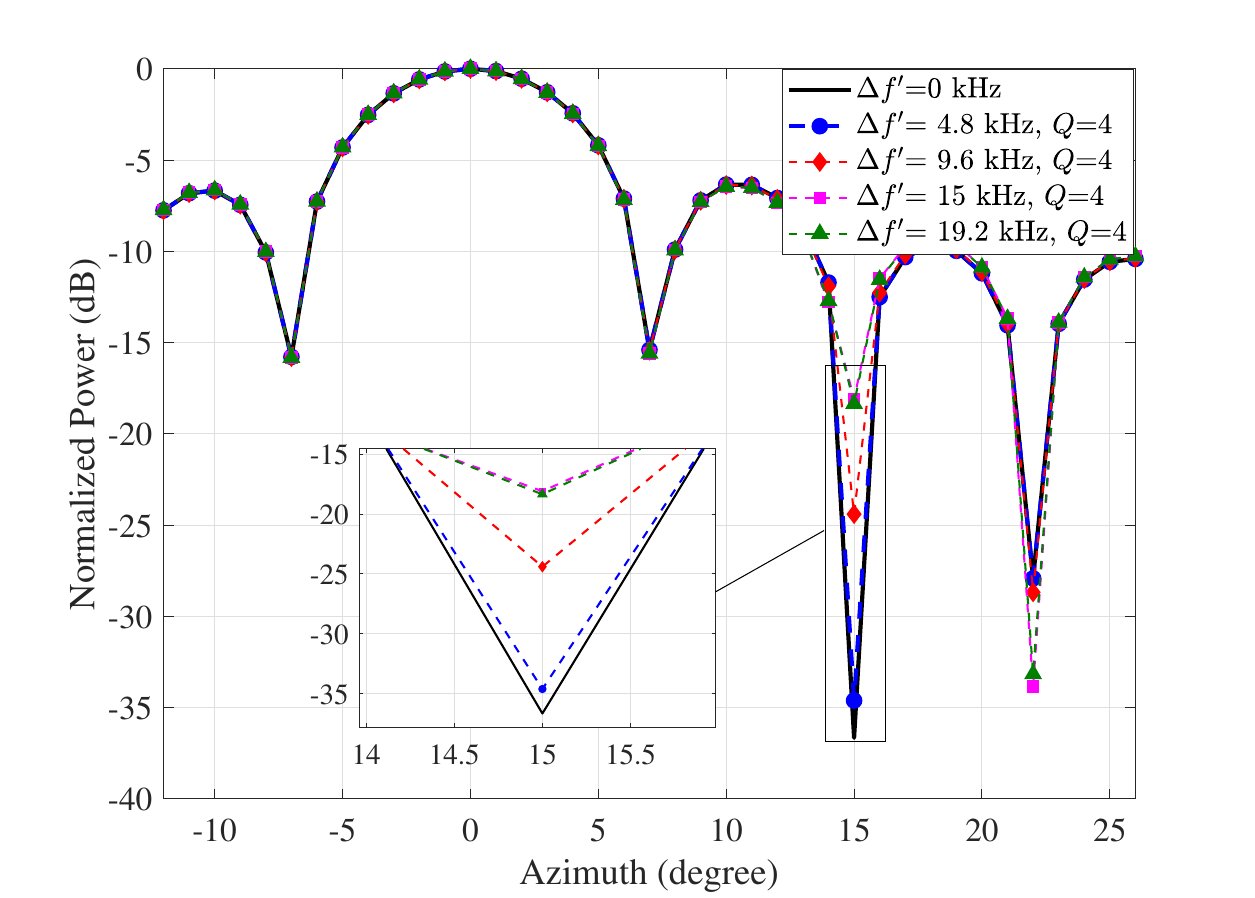}
   }
  \end{minipage}
  \vspace{-0.05in}  
  \caption{Effect of $\varDelta f'$ for FDA jamming on PA radar. (a) SF jammer. (b) AF jammer.}
  \label{FIG.9}
\end{figure*}

\vspace{-0.05in}
\subsection{Spatial filtering}

In this example, we show the spatial filtering process of phased-MIMO radar to suppress the FDA jamming from sidelobe or mainlobe, where the phased-MIMO radar is divided into PA radar and FDA-MIMO radar. We assume that the radar has a known covariance matrix of FDA jamming to calculate the optimal vector for spatial filtering. Typically, the interference covariance matrix estimation by using the training data has a significant influence on the jamming suppression performance. In this paper, we focus on the effect of the FDA jamming on spatial filtering with a known interference covariance matrix, excluding the performance loss due to the covariance matrix estimation.

In Fig.\ref{FIG.8}, we present the azimuth beampattern of PA radar to illustrate the effect of the number of jammer antennas $Q$ for FDA jamming on spatial filtering, where Fig.\ref{FIG.8(a)} and Fig.\ref{FIG.8(b)} corresponds to SF and AF jammer, respectively. We set the jamming frequency offset $\varDelta f'$ as $2.4~{\mathrm{kHz}}$. It can be seen that the spatial filtering can generate a notch at $15\degree$ in the azimuth beampattern, which can suppress the jamming. The jamming notch depth will decrease with the increasing number of antennas for both two FDA jammers. Compared to conventional sidelobe jamming, the proposed FDA jamming can raise the jamming null, indicating that the jamming power increases after spatial filtering. 

In Fig.\ref{FIG.9}, we show the azimuth beampattern of PA radar to illustrate the effect of the jamming frequency offset $\varDelta f'$ for FDA jamming on spatial filtering, where Fig.\ref{FIG.9(a)} and Fig.\ref{FIG.9(b)} corresponds to SF and AF jammer, respectively. We use four jammer antennas to transmit FDA jamming signals. In Fig.\ref{FIG.9(a)}, the jamming notch rises with the increasing jamming frequency offset, indicating that $\left| Y_{\varphi _j}\left( \varDelta f' \right) \right|$ in \eqref{eq.43a} will increase as the increasing of $\varDelta f'$, which verifies the conclusion in the case 1 of Section V. In Fig.\ref{FIG.9(b)}, the jamming notch rises with the increasing of jamming frequency offset for AF jammer, which is similar to Fig.\ref{FIG.9(a)}, indicating that the increasing of jamming frequency offset can worsen the performance of spatial filtering and increase the jamming power after spatial filtering.

\begin{figure*}[t]
  \centering
  \begin{minipage}{1\linewidth }
    \subfigure[SF jammer]{
      \label{FIG.10(a)}
\includegraphics[width=20pc]{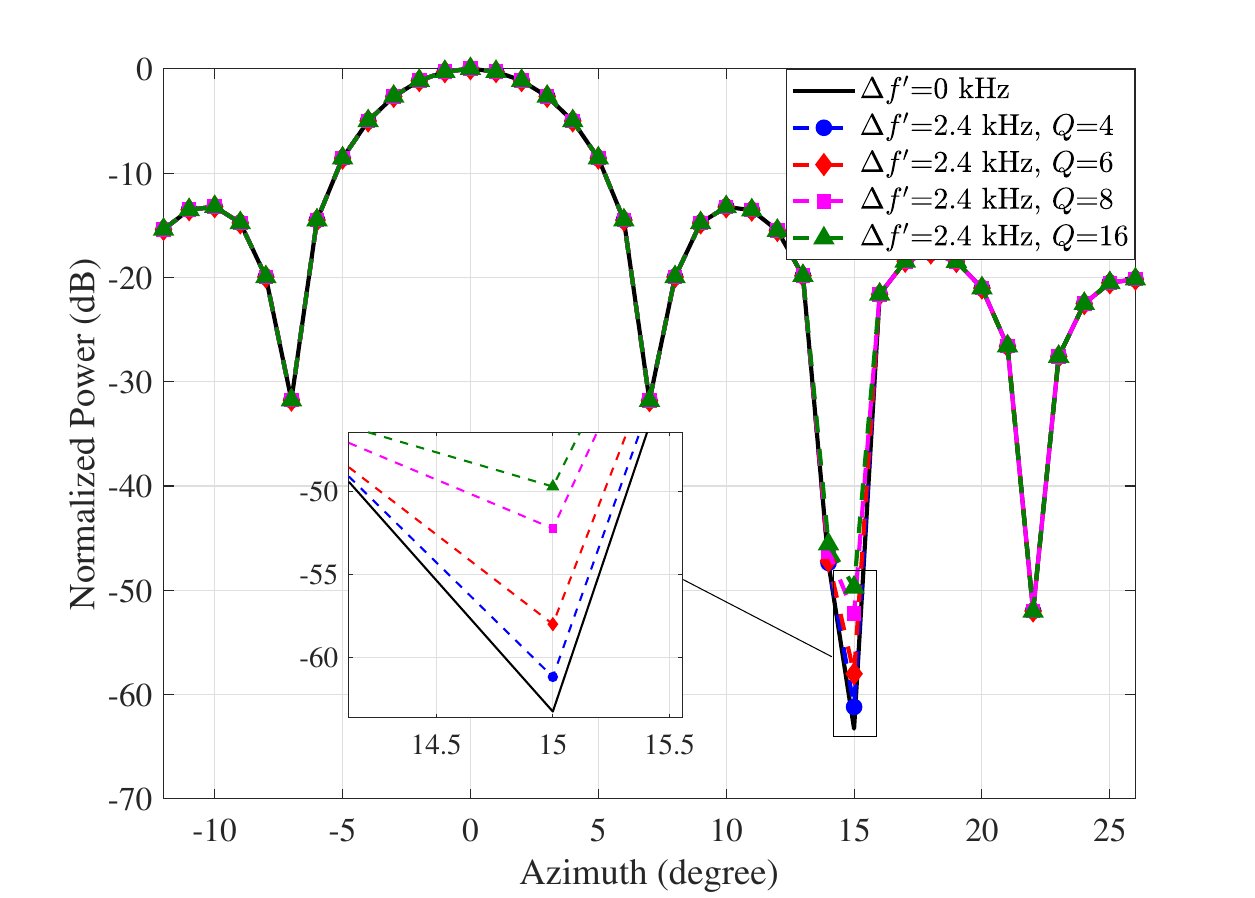}
      }
    \subfigure[AF jammer]{\label{FIG.10(b)}  
    \includegraphics[width=20pc]{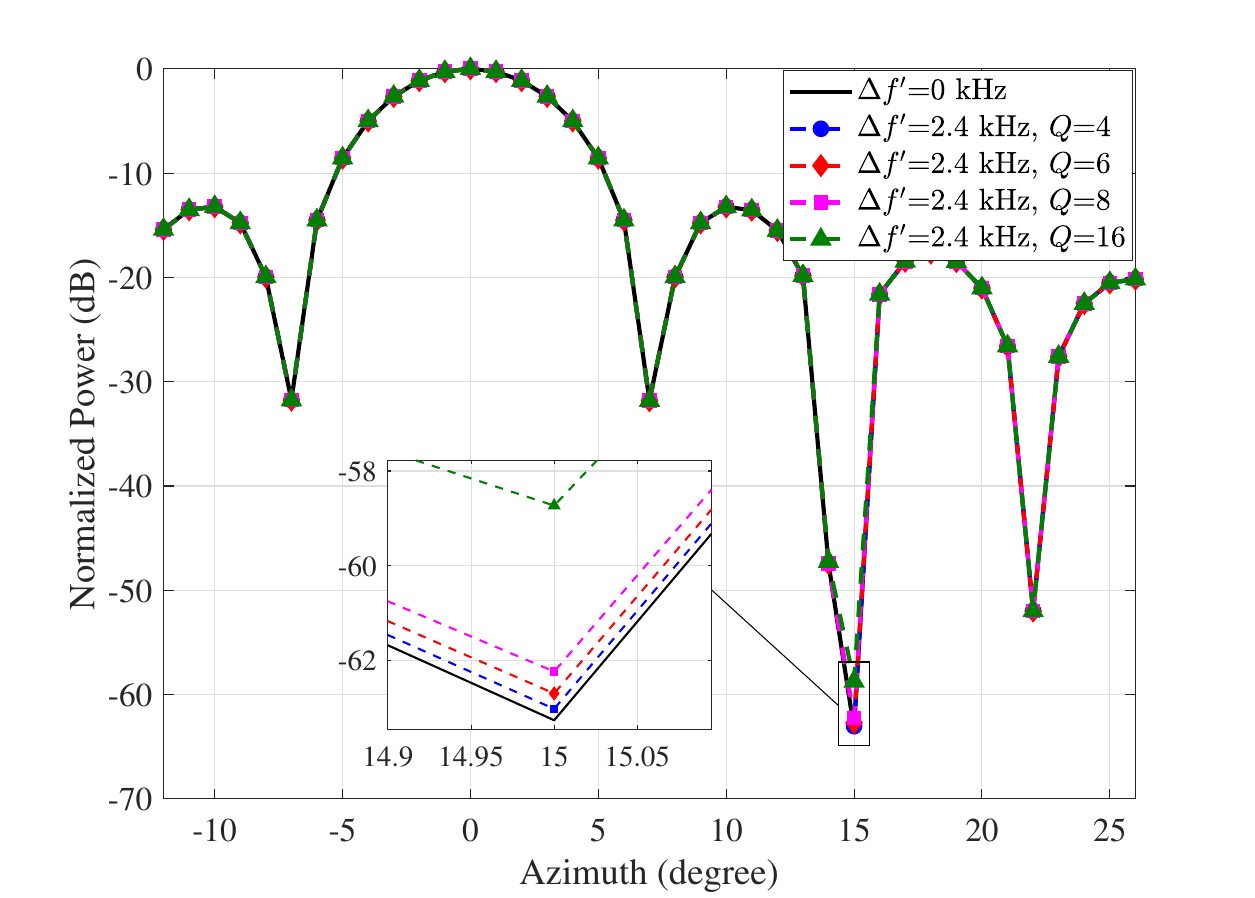}
   }
  \end{minipage}
  \vspace{-0.05in}  
  \caption{Effect of $Q$ for FDA jamming on FDA-MIMO radar with $S=4$. (a) SF jammer. (b) AF jammer.}
  \label{FIG.10}
\end{figure*}

\begin{figure*}[t]
  \centering
  \vspace{-0.15in}
  \begin{minipage}{1\linewidth }
    \subfigure[SF jammer]{
      \label{FIG.11(a)}
\includegraphics[width=20pc]{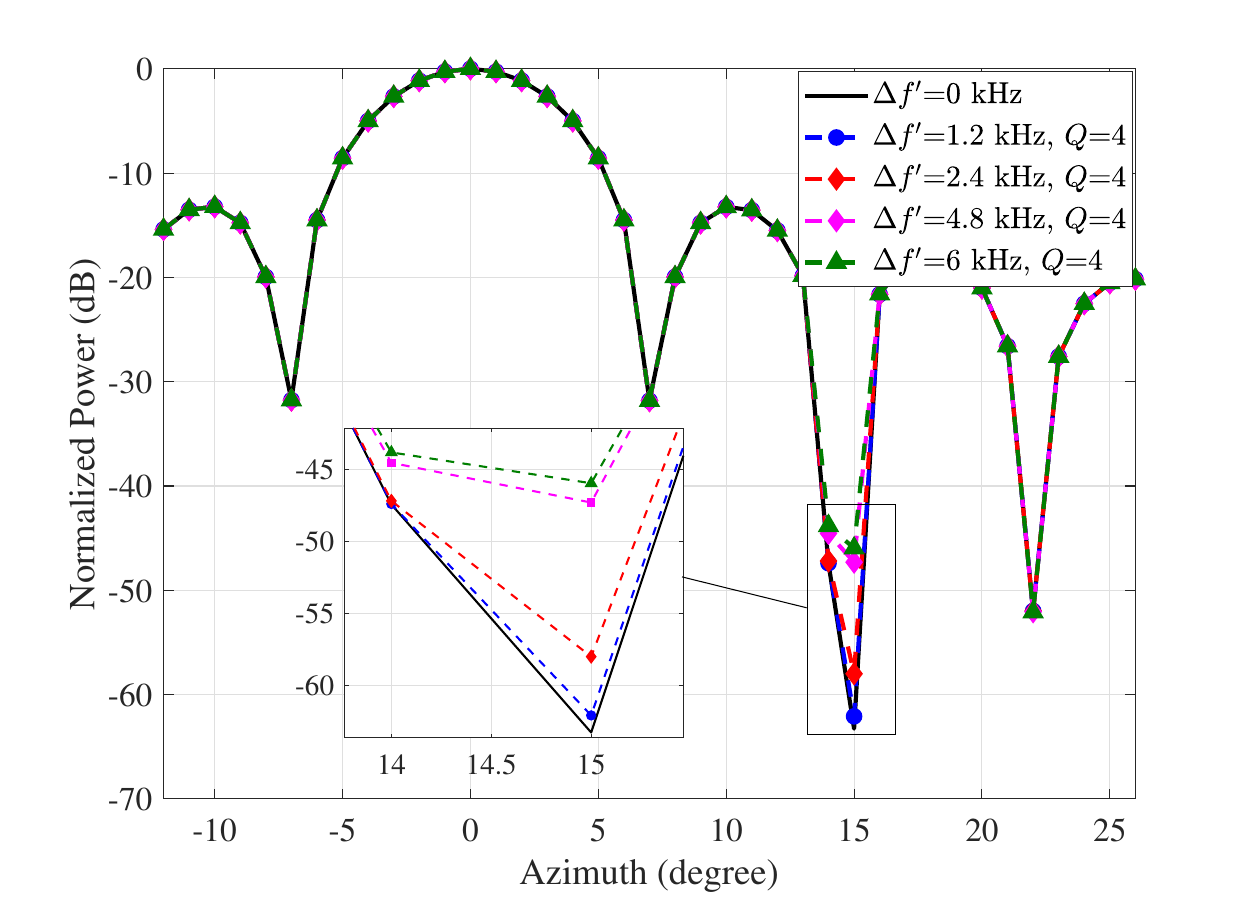}
      }
    \subfigure[AF jammer]{\label{FIG.11(b)}  
    \includegraphics[width=20pc]{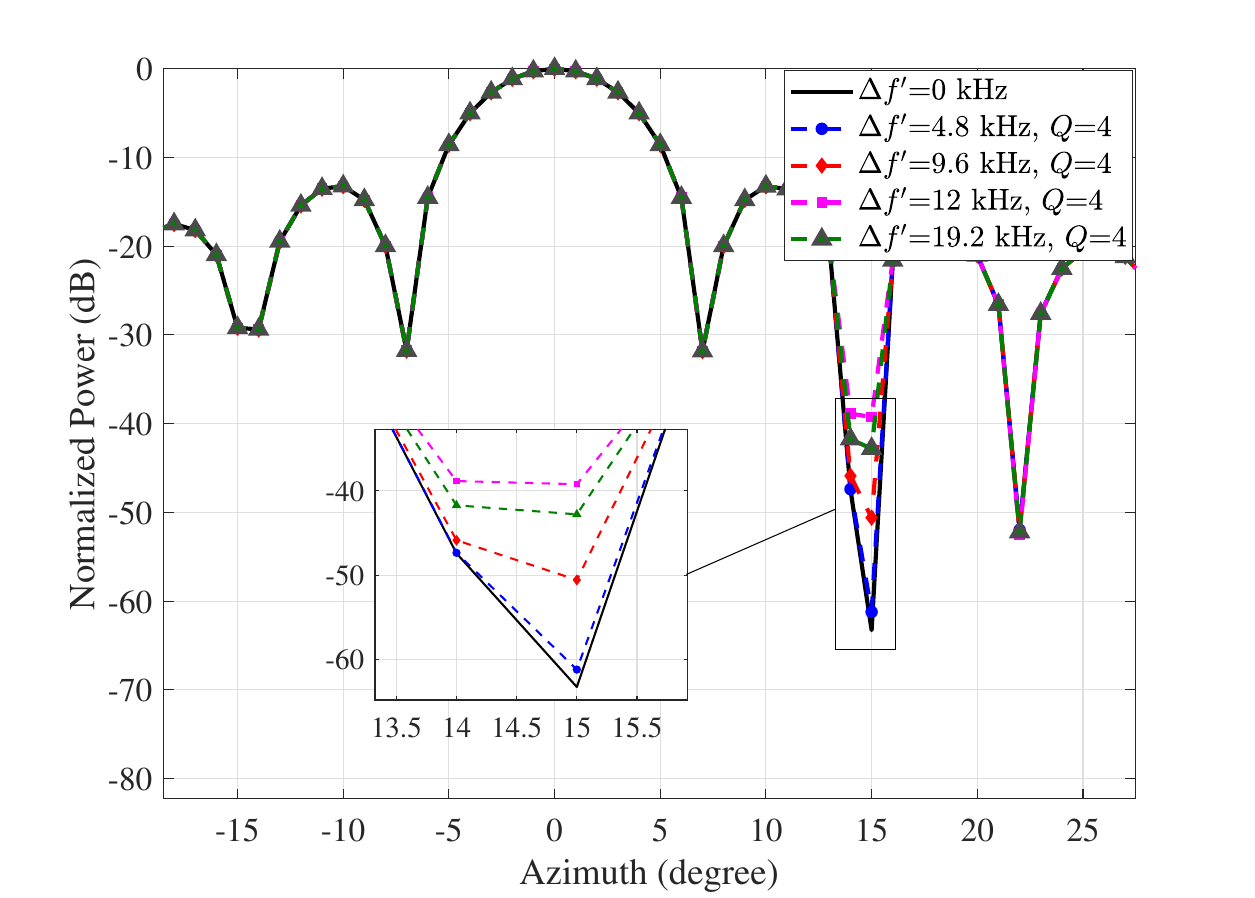}
   }
  \end{minipage}
  \vspace{-0.05in}  
  \caption{Effect of $\varDelta f'$ for FDA jamming on FDA-MIMO radar with $S=4$. (a) SF jammer. (b) AF jammer.}
  \label{FIG.11}
\end{figure*}

\begin{figure*}[t]
  \centering
  \vspace{-0.15in}
  \begin{minipage}{1\linewidth }
    \subfigure[SF jammer]{
      \label{FIG.12(a)}
\includegraphics[width=20pc]{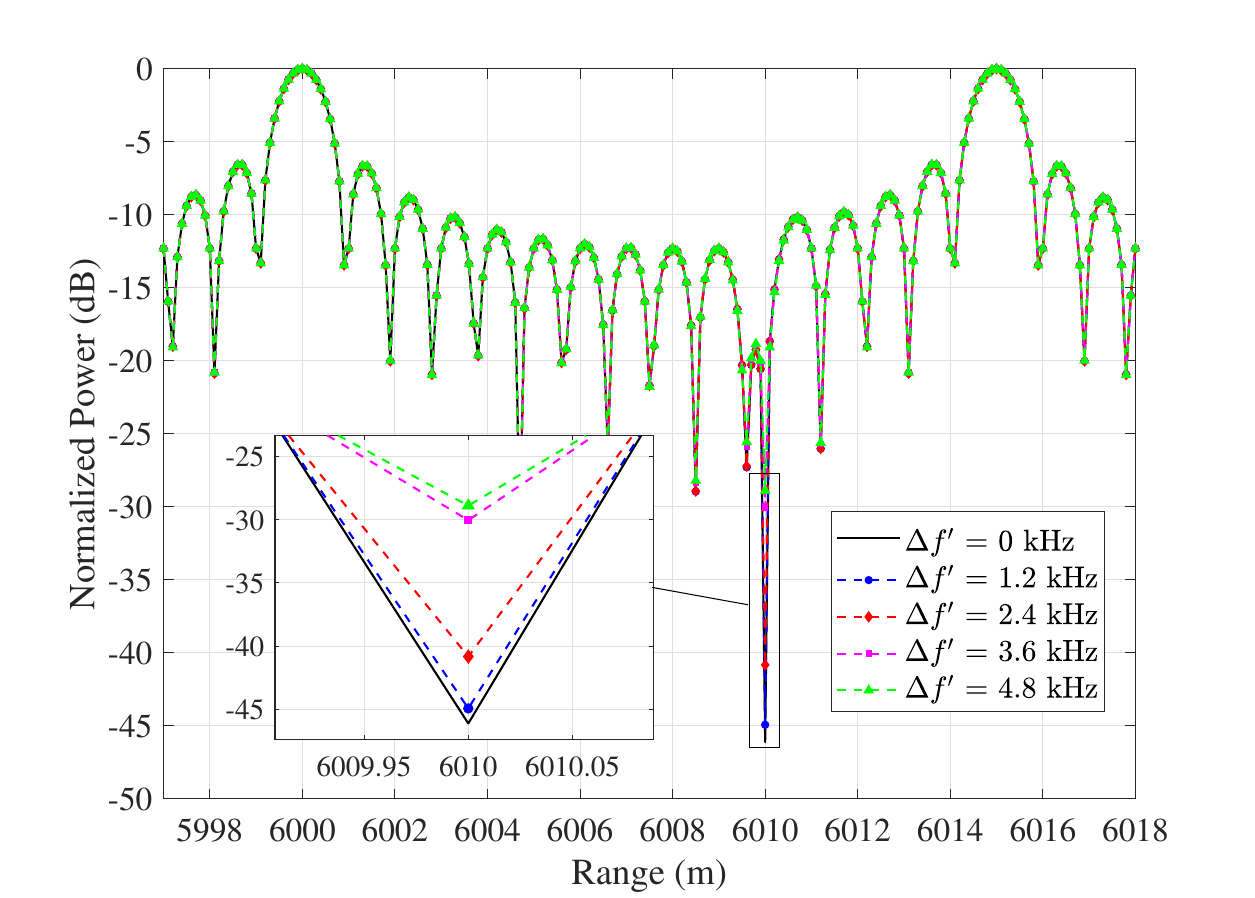}
      }
    \subfigure[AF jammer]{\label{FIG.12(b)}  
    \includegraphics[width=20pc]{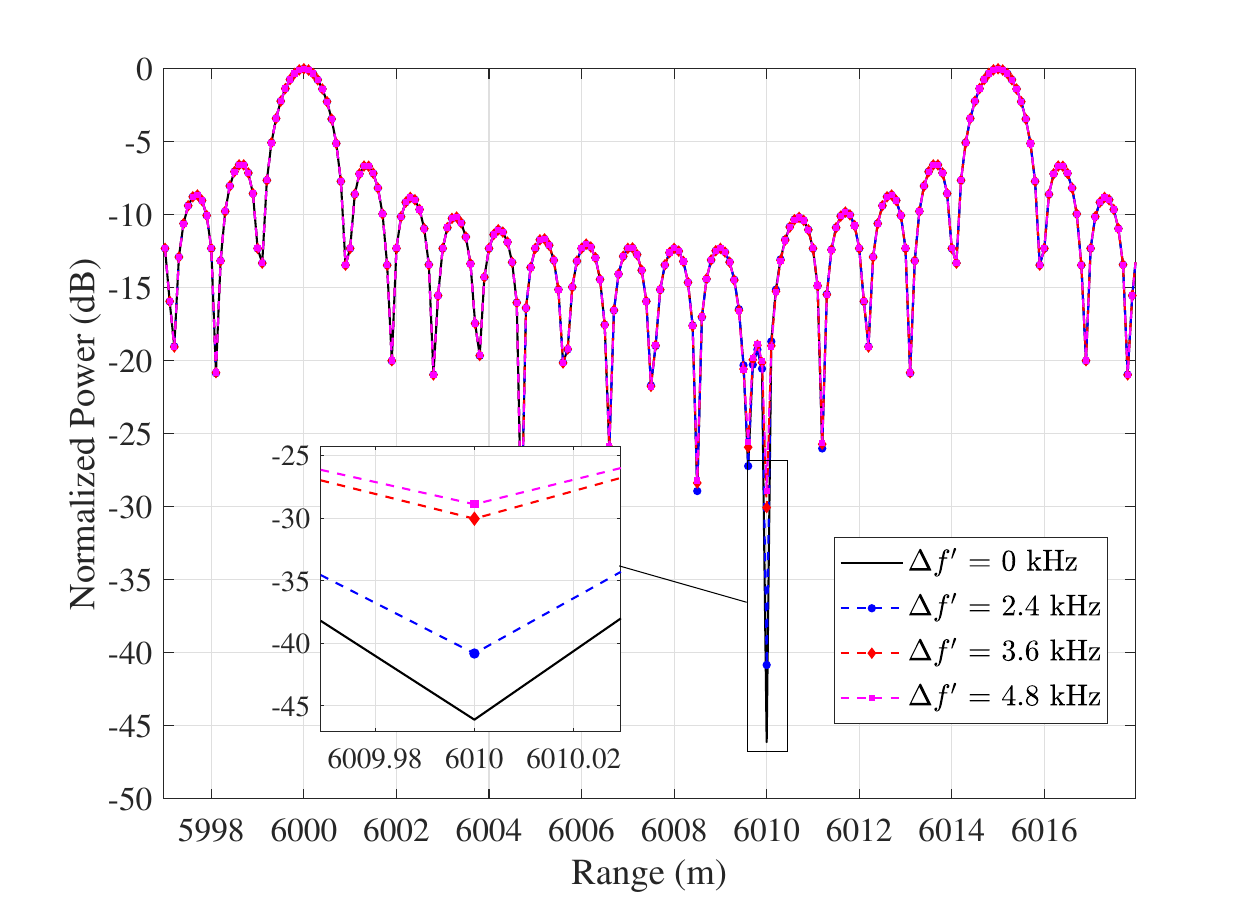}
   }
  \end{minipage}
  \vspace{-0.05in}  
  \caption{Effect of $\varDelta f'$ for FDA jamming on range-dimensional filtering of FDA-MIMO radar with $S=16$. (a) SF jammer. (b) AF jammer.}
  \label{FIG.12}
\end{figure*}

In Fig.\ref{FIG.10}, we show the azimuth beampattern of FDA-MIMO radar to illustrate the effect of $Q$ for FDA jamming on spatial filtering, where Fig.\ref{FIG.10(a)} and Fig.\ref{FIG.10(b)} corresponds to SF and AF jammer, respectively. We set the jamming frequency offset of FDA jammers as $2.4~{\mathrm{kHz}}$. In Fig.\ref{FIG.10(a)} and Fig.\ref{FIG.10(b)}, the jamming notch rises with the increasing of $Q$ for both SF and AF jamming, which is consistent with Fig.\ref{FIG.8(a)} and Fig.\ref{FIG.8(b)}. Morerover, the sidelobes of FDA-MIMO radar in Fig.\ref{FIG.10} are lower than PA radar in Fig.\ref{FIG.8}, which indicates that FDA-MIMO radar enjoys a higher spatial resolution from the benefits of waveform diversity.

In Fig.\ref{FIG.11}, we show the azimuth beampattern of FDA-MIMO radar to illustrate the effect of $Q$ for FDA jamming on spatial filtering, where Fig.\ref{FIG.11(a)} and Fig.\ref{FIG.11(b)} corresponds to SF and AF jammer, respectively. We use four jammer antennas to transmit FDA jamming signals. The jamming notch rises as the jamming frequency offset increases, indicating that $\left| Y_{\varphi _j}\left( \varDelta f' \right) \right|$ in \eqref{eq.45a} will increase as the increasing of $\varDelta f'$, which verifies the conclusion in the case 2 of Section V.

Fig.\ref{FIG.12} shows the range-dimensional beampattern of FDA-MIMO radar to illustrate the effect of $\varDelta f'$ for FDA jamming on spatial filtering, where Fig.\ref{FIG.12(a)} and Fig.\ref{FIG.12(b)} corresponds to SF and AF jammer, respectively. Note that the FDA jammer is considered as a mainlobe interference for FDA-MIMO radar since it has the same azimuth as the target but a different range, $R_j=6010$ m. Accordingly, we use $\left| Y_{R_j}\left( \varDelta f' \right) \right|$ in \eqref{eq.45b} to measure the range-dimensional jamming notch for FDA-MIMO radar. From Fig.\ref{FIG.12(a)} and Fig.\ref{FIG.12(b)}, the range-dimensional spatial filtering of FDA-MIMO radar can generate a jamming notch at $6010~\mathrm{m}$. As the increasing jamming frequency offset, the jamming notch rises for both SF and AF jamming, indicating that $\left| Y_{R_j}\left( \varDelta f' \right) \right|$ in \eqref{eq.45b} and the jamming power after spatial filtering increases with the increasing of $\varDelta f'$, which is consistent with the  conclusion of case 2 in Section V.


\begin{figure*}[t]
  \centering
  \vspace{-0.15in}
  \begin{minipage}{1\linewidth }
    \subfigure[SF jammer]{
      \label{FIG.13(a)}
\includegraphics[width=20pc]{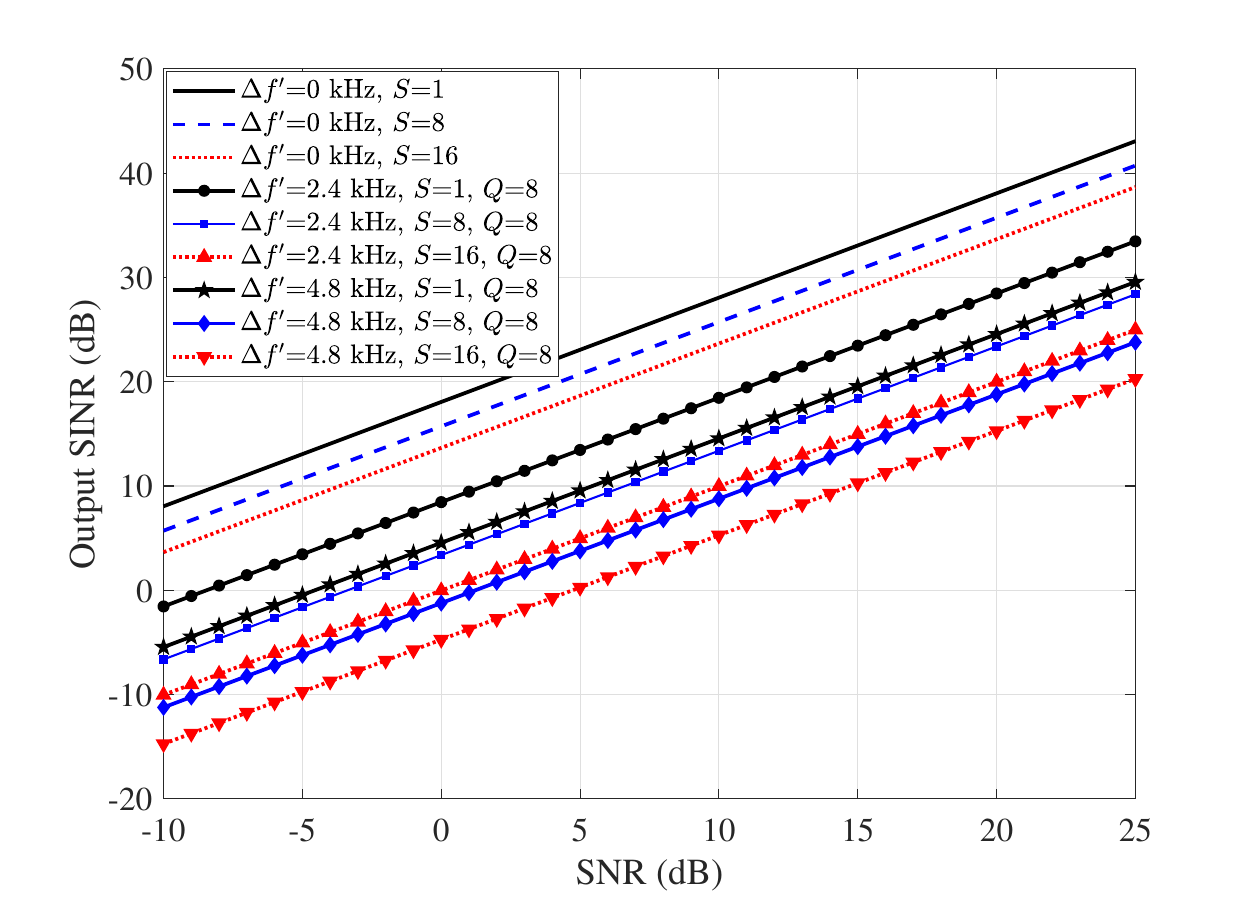}
      }
    \subfigure[AF jammer]{\label{FIG.13(b)}  
    \includegraphics[width=20pc]{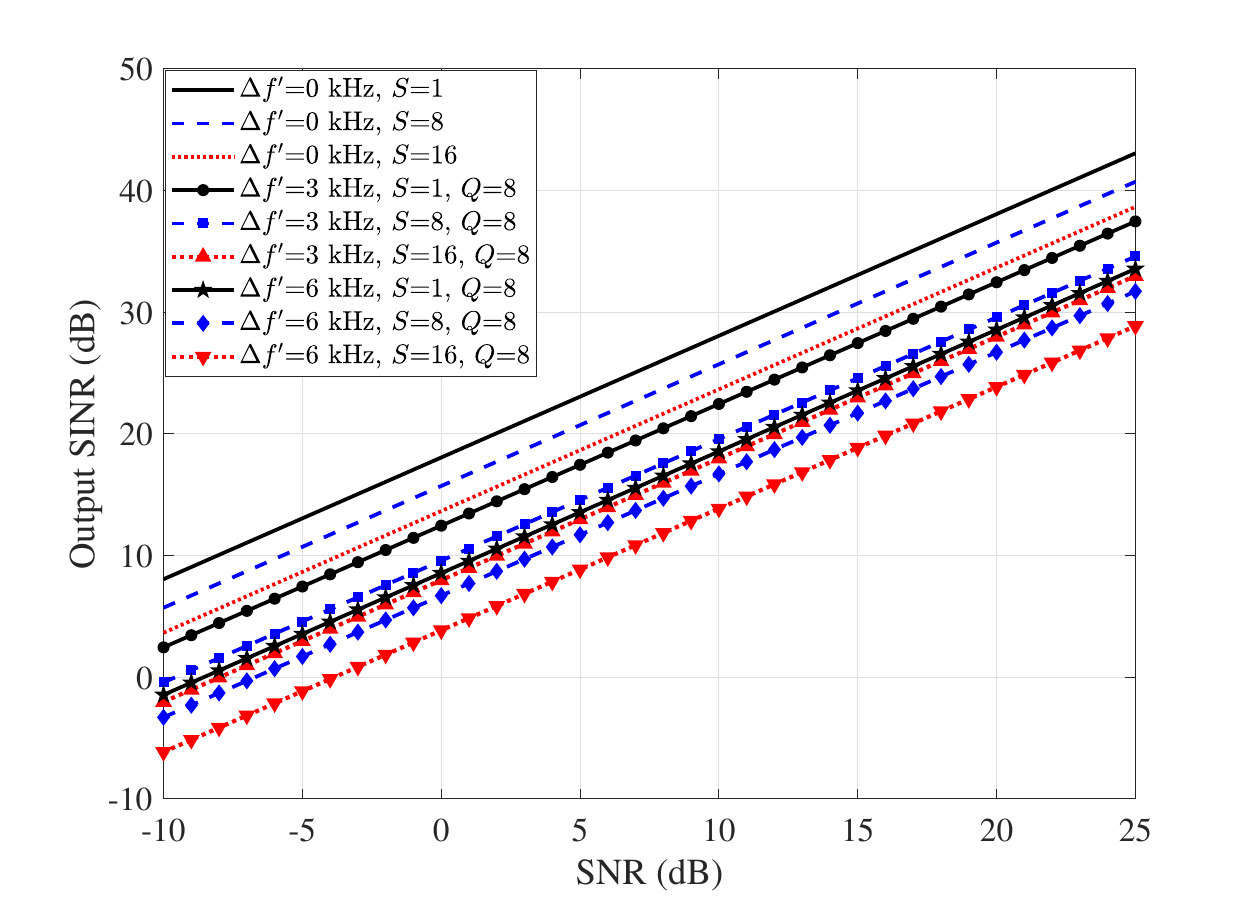}
   }
  \end{minipage}
  \vspace{-0.05in}  
  \caption{The output SINR of phased-MIMO radar with different transmit partitionings against sidelobe interference. (a) SF jammer. (b) AF jammer.}
  \label{FIG.13}
\end{figure*}

\vspace{-0.1in}
\subsection{Output SINR}

In this example, we use 200 Monte Carlo simulations to calculate the output SINR for phased-MIMO radar with different transmit partitionings against two types of FDA jammer, where $S=1$ represents the PA radar, $S=8$ and $S=16$ represents FDA-MIMO radar with different subarray partitionings. Meanwhile, we use this example to verify the  conclusions on $\mathrm{SINR}_o\left( \varDelta f' \right)$ in \eqref{eq.43b} and \eqref{eq.45c} corresponding to PA and FDA-MIMO radar, respectively.

In Fig.\ref{FIG.13}, we show the output SINR of phased-MIMO radar against sidelobe interference to illustrate the effect of FDA jamming on spatial filtering, where Fig.\ref{FIG.13(a)} and Fig.\ref{FIG.13(b)} corresponds to the SF and the AF jammer, respectively. In Fig.\ref{FIG.13(a)}, PA radar has a higher coherent gain, thus it has a higher output SINR than FDA-MIMO radar when $\varDelta f'=0$ kHz. With the increasing number of subarrays for FDA-MIMO radar, the output SINR curve drops as the decreasing of the coherent gain. As the increasing of the jamming frequency offset, output SINR for both PA and FDA-MIMO radar decreases since the jamming power increases after spatial filtering, which is consistent with Fig.\ref{FIG.9(a)} and Fig.\ref{FIG.11(a)}. In Fig.\ref{FIG.13(b)}, AF jammers have the same performance against phased-MIMO radar. Moreover, since the AF jammer is less powerful than the SF jammer when $\sum_{q=1}^Q{\rho _{q}^{2}}$ is fixed, AF jammers require a larger jamming frequency offset to achieve the same effect on output SINR as SF jammers.

\begin{figure*}[t]
  \centering
  \begin{minipage}{1\linewidth }
    \subfigure[SF jammer]{
      \label{FIG.14(a)}
\includegraphics[width=20pc]{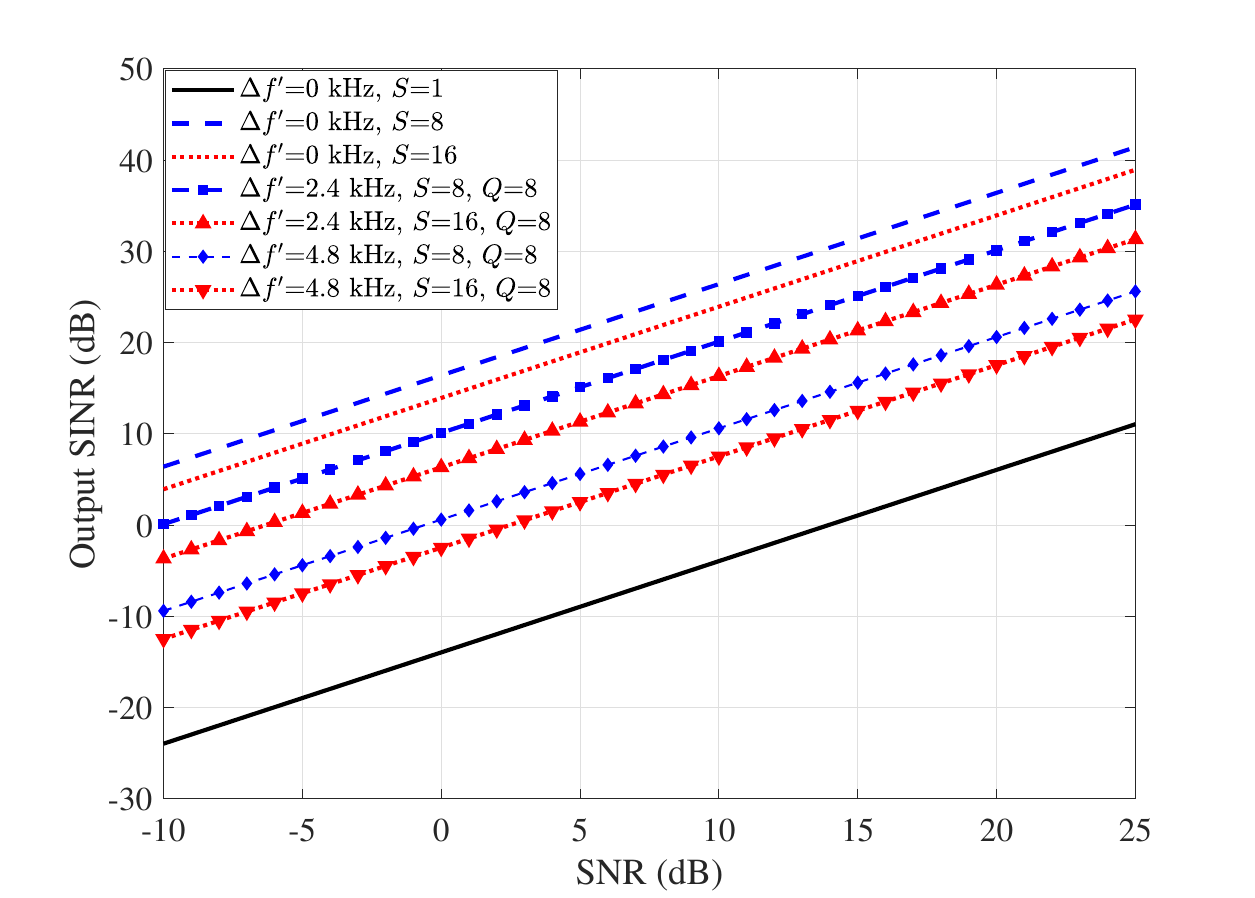}
      }
    \subfigure[AF jammer]{\label{FIG.14(b)}  
    \includegraphics[width=20pc]{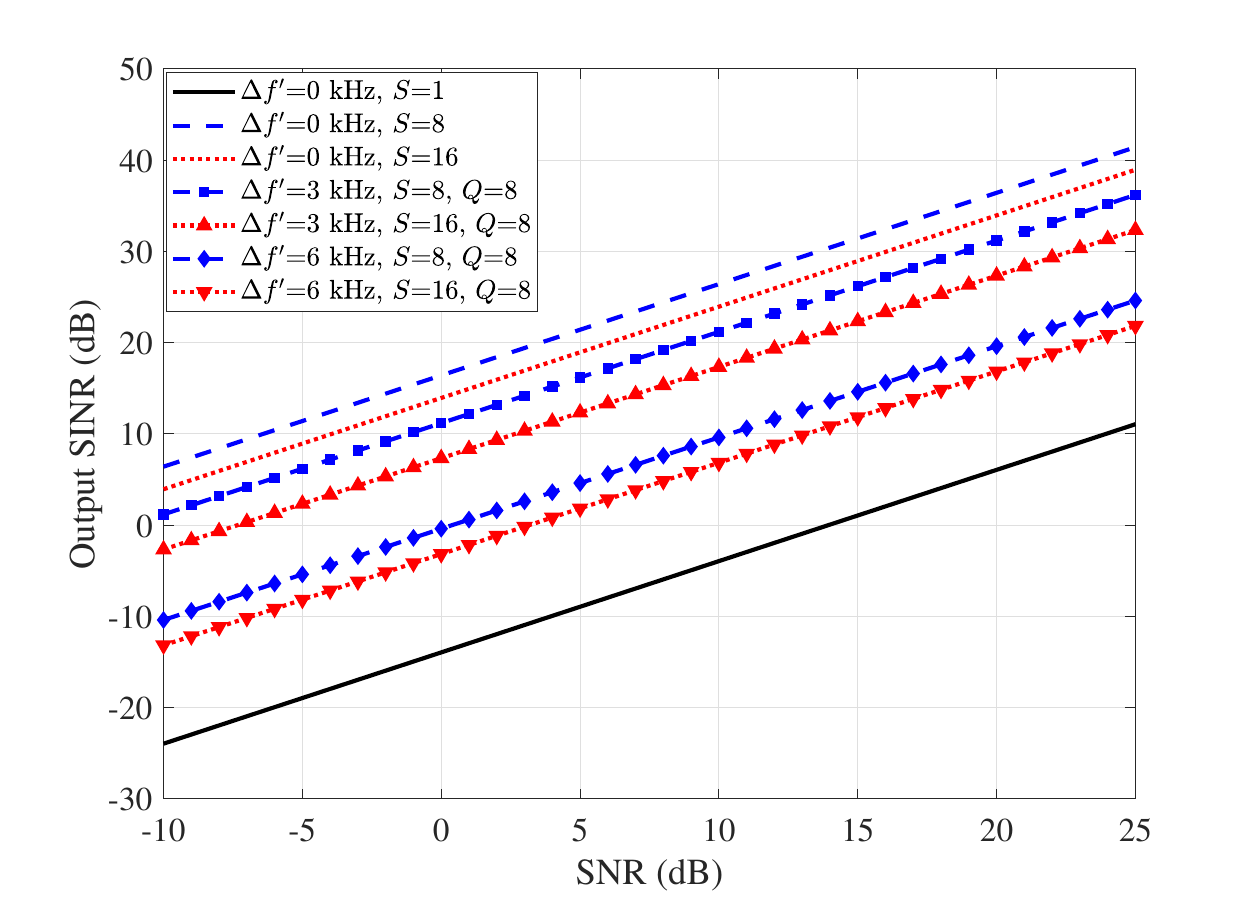}
   }
  \end{minipage}
  \caption{The output SINR of phased-MIMO radar with different transmit partitionings against mainlobe interference. (a) SF jammer. (b) AF jammer.}
  \label{FIG.14}
\end{figure*}

In Fig.\ref{FIG.14}, we show the output SINR of the phased-MIMO radar against mainlobe interference to illustrate the effect of FDA jamming on spatial filtering, where Fig.\ref{FIG.14(a)} and Fig.\ref{FIG.14(b)} corresponds to the SF and the AF jammer, respectively. In Fig.\ref{FIG.14(a)} and Fig.\ref{FIG.14(b)}, PA radar is unable to suppress the mainlobe jamming from the range-dimensional spatial filtering since it does not have a range-dependent transmit spatial frequency. FDA-MIMO radar has a better performance since it can suppress mainlobe interference due to its range-dependency [\ref{cite9}, \ref{cite10}]. Moreover, FDA-MIMO radar has a higher output SINR than PA radar even it is in the presence of FDA jamming, which indicates that FDA-MIMO radar can still suppress some mainlobe jamming energy, but the performance of mainlobe jamming suppression decreases as the jamming frequency offset increases. With the increasing number of subarrays for FDA-MIMO radar, the output SINR curves in Fig.\ref{FIG.14(a)} and Fig.\ref{FIG.14(b)} drop as the decreasing of the coherent gain. As the increasing of the jamming frequency offset, output SINR for FDA-MIMO radar decreases since the jamming power increases after spatial filtering, which is consistent with Fig.\ref{FIG.12(a)} and Fig.\ref{FIG.12(b)}. Fig.\ref{FIG.13} and Fig.\ref{FIG.14} verify the effectiveness of FDA jamming on spatial filtering against the phased-MIMO radar. 

\begin{table*}[!t]\centering
 \caption{Comparison of the FDA jamming and existing jamming techniques}
 \label{table.2}
 \renewcommand\arraystretch{1.8}
 \begin{minipage}{\textwidth}\centering
 \scalebox{0.88}{\begin{tabular}{cccccccc}
 \hline
 \textbf{Jamming Technique}  &  \textbf{Counter Objective}  &  \textbf{Function}   &  \textbf{Results} &  \textbf{Drawbacks} \\
 \hline
\multirow{2}{*}{\makecell{FDA jamming\\(No scene constraints)}} &\multirow{2}{*}{Radar detection} & Increase false alarms    & Intensive false targets & \multirow{2}{*}{Sophisticated jamming antennas}\\
      & & Reduce output SINR    & Raise interference null &  \\                               
\hline
\multirow{4}{*}{\makecell{Deceptive jamming\\ (No scene constraints)}} & Radar recognition\textsuperscript{[\ref{cite40}]}  & Reduce recognition probability   & Different time-frequency response  & \multirow{4}{*}{\makecell{Widely and systematically studied\\Easily suppressed and recognized}} \\
             &  Radar detection\textsuperscript{[\ref{cite34}]}  & Reduce detection probability    & False target information &    \\
             &  Radar imaging\textsuperscript{[\ref{cite41}]} &  Disrupte imaging   & False SAR images   &    \\
             &  Radar tracking\textsuperscript{[\ref{cite42}]} & Reduce tracking accuracy   & False trajectory &    \\
\hline
\multirow{4}{*}{\makecell{Towed jamming\\ (Missiles or vessels)}} & \multirow{2}{*}{Radar detection\textsuperscript{[\ref{cite43}]}} &  Fake range information   &  \multirow{2}{*}{Cover decoy}  & \multirow{2}{*}{Desirable hardware materials}  \\
             &    & Fake azimuth information    &  &    \\
             &  \multirow{2}{*}{Radar tracking\textsuperscript{[\ref{cite44}]}} & Fake wave gate    & \multirow{2}{*}{False target decoy}   &   \multirow{2}{*}{Complex trajectory deception}  \\
             &   & False trajectory  &   & &     \\
 \hline
 \multirow{2}{*}{\makecell{Modulation jamming\\ (Multipulse signal)}} & \multirow{2}{*}{Radar detection} &  Phase modulation jamming\textsuperscript{[\ref{cite27}]} &  High-energy suppression  &\multirow{2}{*}{\makecell{Pulse modulation is highly regular\\ and easily recognized}}   \\
             &    & Intermittent sampling repeater\textsuperscript{[\ref{cite33}]}   & False point target string &    \\
 \hline
 \end{tabular}}
 \end{minipage}
 \vspace{0mm}
 \vspace{0.12in}
 \end{table*}

\section{DISCUSSION}

The jamming techniques against airborne radar can be categorized into three major groups, deceptive jamming [\ref{cite29}, \ref{cite40}, \ref{cite41}, \ref{cite42}], towed jamming [\ref{cite43}, \ref{cite44}], and modulation jamming [\ref{cite33}, \ref{cite45}]. The proposed FDA jamming is a special kind of deceptive jamming. Other classical sidelobe or mainlobe deceptive jammings has been widely researched as an objective for the anti-interference approaches of radar [\ref{cite9}, \ref{cite21}]. The towed jamming is used to counter radar detection and radar tracking, especially on missiles and vessels. The modulation jamming is designed for the different pulse waveforms in multipulse signals, forming the strong suppressive power or false pulse information. In table \ref{table.2}, we make a detailed summary of the proposed FDA jamming techniques and the existing jamming techniques, describing the applicable scenarios, jammer functions, and drawbacks. Specifically, the proposed FDA jamming can be combined with other kinds of jamming techniques to extend the applicability and capability. For example, FDA jamming can be integrated with towed deceptive jamming, overcoming the drawbacks of deceptive jamming that are easy to suppress and recognize. FDA jamming can also be combined with decoy jamming, increasing the probability of surprise defense.

In this paper, we focus on the range-dimensional deceptive jamming and azimuth-dimensional deceptive jamming generated by FDA jamming. For the range-dimensional deceptive jamming, the existing jamming technique is to use a delayed copy of the envelope to generate range-dimensional false targets [\ref{cite33}], which is more affordable than the proposed FDA jamming technique. For the azimuth dimensional deceptive jamming, the existing jamming technique is to forward radar signal, depending on the spatial position of the jammer to confuse the radar, such as the towed jamming [\ref{cite43}] and the dropped jamming [\ref{cite44}], which has been extensively studied by proposing effective sidelobe or mainlobe jamming suppression methods. The proposed FDA jamming is more effective than these existing jamming approaches for azimuth dimensional deceptive jamming. Although the FDA jamming is indeed more expensive and complex than the existing jamming, it is multi-functional and can be employed in different ECM scenarios by adjusting the frequency offset. Furthermore, we believe that the combination of the concept of FDA jamming with the existing jamming techniques is instructive for ECM study.


\section{CONCLUSION}

As the first part of this series, this paper introduced the working principle of FDA jammer against airborne phased-MIMO radar, including SF jamming signal and AF jamming signal models. For the MF process of radar receiver, the FDA jamming can use the mismatch caused by jamming frequency offset to generate the dense false targets surrounding to the target in range dimension. From the perspective of spatial beamforming, the FDA jamming can decrease the interference null depth and reduce the output SINR of spatial filtering. Simulations verified the correctness of theoretical derivation and effectiveness of FDA jamming.  



\section*{APPENDIX A}
This appendix presents the proof of Proposition 1.

We use the monopulse linear frequency modulation (LFM) signal $A(t)=G( \frac{t}{T_p}) e^{ j\pi \kappa t^2}$ to prove Proposition 1, where $G( \frac{t}{T_p})$ is the rectangular envelope with time width $T_p$ and unit energy, $\int_{-\infty}^{+\infty}{\left| G\left( \frac{t}{T_p} \right) \right|}^2\mathrm{d}t=1$, and $\kappa=B/T_p$ is the frequency modulation radio. Let us derive the MF outputs of the FDA jamming for two different cases, PA and FDA-MIMO radar.

\subsection*{Case 1: PA radar}

According to \eqref{eq.18a} and \eqref{eq.18b}, for the n-th radar receive element, the SF and AF jamming signals transmitted by the q-th jammer antenna can be expressed as
\begin{subequations}
\begin{align}
\bar{P}_{n,q}^{\left( \mathrm{SF} \right)}\left( t \right)\approx &
E_t\cdot \rho _q\cdot G( \frac{t}{T_p}) e^{ j\pi \kappa (t-\tau _j)^2}
\nonumber\\
&\cdot e^{j2\pi \left[ f_0+\left( q-1 \right) \varDelta f' \right] \left( t-\tau _j-\tau _{n}^{(R)} \right)}
\label{eq.A1a}
\\
\bar{P}_{n,q}^{\left( \mathrm{AF} \right)}\left( t \right)\approx &
E_t\cdot \rho _q\cdot G( \frac{t}{T_p}) e^{ j\pi \kappa (t-\tau _j)^2}
\nonumber\\
&\cdot e^{j2\pi \left[ f_0+\left( q-1 \right) \varDelta f' \right] \left( t-\tau _j+\tau_q-\tau _{n}^{(R)} \right)}
\label{eq.A1b}
\end{align}
\end{subequations}
After using $e^{-j2\pi f_0 (t-\tau_j)}$ for the down-conversion as mentioned in \eqref{eq.22a} and \eqref{eq.22b}, we can get
\begin{subequations}
\begin{align}
\tilde{P}_{n,q}^{\left( \mathrm{SF} \right)}\left( t \right) \approx &E_t\cdot \rho _q\cdot G( \frac{t}{T_p}) e^{ j\pi \kappa (t-\tau _j)^2}
\nonumber\\ 
&\cdot e^{-j2\pi f_0 \tau _{n}^{(R)} }e^{j2\pi \left( q-1 \right) \varDelta f'\left( t-\tau _j \right)}
\label{eq.A2a}
\\
\tilde{P}_{n,q}^{\left( \mathrm{AF} \right)}\left( t \right) \approx &E_t\cdot \rho _q\cdot G( \frac{t}{T_p}) e^{ j\pi \kappa (t-\tau _j)^2}
\nonumber\\ 
&\cdot e^{-j2\pi f_0 \tau _{n}^{(R)}}e^{j2\pi \left[ f_0\tau _q+\left( q-1 \right) \varDelta f'\left( t-\tau _j \right) \right]}
\label{eq.A2b}
\end{align}
\end{subequations}
Through the convolution with $G( \frac{t}{T_p}) e^{ -j\pi \kappa t^2}$, we can get two integrals for SF and AF jamming.
\begin{subequations}
\begin{align}
{P}_{n,q}^{\left( \mathrm{SF} \right)}\left( t \right) =\int_{-\infty}^{+\infty}{\tilde{P}_{n,q}^{\left( \mathrm{SF} \right)}\left( \tau \right) \left[ G(\frac{t-\tau}{T_p})e^{-j\pi \kappa (t-\tau)^2} \right] \mathrm{d}\tau}
\label{eq.A3a}
\\
{P}_{n,q}^{\left( \mathrm{AF} \right)}\left( t \right) =\int_{-\infty}^{+\infty}{\tilde{P}_{n,q}^{\left( \mathrm{AF} \right)}\left( \tau \right) \left[ G(\frac{t-\tau}{T_p})e^{-j\pi \kappa (t-\tau)^2} \right] \mathrm{d}\tau}
\label{eq.A3b}
\end{align}
\end{subequations}
Substituting \eqref{eq.A2a} and \eqref{eq.A2b} into \eqref{eq.A3a} and \eqref{eq.A3b}, respectively, the integrals yield
\begin{subequations}
\begin{align}
{P}_{n,q}^{\left( \mathrm{SF} \right)}\left( t \right) =&E_t\rho _qT_pe^{-j2\pi f_0\tau _{n}^{(R)}}
\nonumber\\
&\times \frac{\sin \pi T_p\left[ \kappa \left( t-\tau _j \right) +\left( q-1 \right) \varDelta f' \right]}{\pi T_p\left[ \kappa \left( t-\tau _j \right) +\left( q-1 \right) \varDelta f' \right]}
\label{eq.A4a}
\\
{P}_{n,q}^{\left( \mathrm{AF} \right)}\left( t \right) =&E_t\rho _qT_pe^{-j2\pi f_0\tau _{n}^{(R)}}\nonumber\\
&\times\frac{\sin \pi T_p\left[ \kappa \left( t-\tau _j \right) +\left( q-1 \right) \varDelta f' +\tau_q f_0\right]}{\pi T_p\left[ \kappa \left( t-\tau _j \right) +\left( q-1 \right) \varDelta f'+\tau_q f_0\right]}
\label{eq.A4b}
\end{align}
\end{subequations}

For a target signal received by the n-th element for PA radar as mentioned in \eqref{eq.5}, the result of the convolution integral can be expressed as
\begin{align}
&\int_{T_p}{y_{n}^{\left( \mathrm{P} \right)}\left( \tau \right) \left[ G(\frac{t}{T_p})e^{-j\pi \kappa (t-\tau )^2} \right] \mathrm{d}\tau}
\nonumber\\
=&\xi _t\sqrt{M}T_pe^{-j2\pi f_0\tau _{n}^{(R)}}\times\frac{\sin \pi T_p\kappa \left( t-\tau _t \right)}{\pi T_p\kappa \left( t-\tau _t \right)}
\label{eq.A5}
\end{align}
 
\subsection*{Case 2: FDA-MIMO radar}

For FDA-MIMO radar, the MF process is implemented in $S$ channels of each receive element. Focusing on the s-th channel of the n-th receive element, the SF and AF jamming signals transmitted by the q-th jammer antenna can be expressed as 
\begin{subequations}
\begin{align}
\bar{F}_{n,s,q}^{\left( \mathrm{SF} \right)}\left( t \right)\approx &
\tilde{E}_t\rho _q{G(\frac{t}{T_p})e^{j\pi \kappa (t-\tau _j)^2}e^{j2\pi \left( s-1 \right) \varDelta f(t-\tau _j)}}
\nonumber\\
\times  e&^{j2\pi f_0\tau _{\mathrm{s}}\left( \varphi _j,\theta _j \right)}e^{j2\pi \left[ f_0+\left( q-1 \right) \varDelta f' \right] (t-\tau _j-\tau _{n}^{(R)})}
\label{eq.A6a}
\\
\bar{F}_{n,s,q}^{\left( \mathrm{AF} \right)}\left( t \right)\approx &
\tilde{E}_t\rho _q{G(\frac{t}{T_p})e^{j\pi \kappa (t-\tau _j)^2}e^{j2\pi \left( s-1 \right) \varDelta f(t-\tau _j)} }
\nonumber\\
\times e&^{j2\pi f_0\tau _{\mathrm{s}}\left( \varphi _j,\theta _j \right)}e^{j2\pi \left[ f_0+\left( q-1 \right) \varDelta f' \right] (t-\tau _j-\tau _q-\tau _{n}^{(R)})}
\label{eq.A6b}
\end{align}
\end{subequations}
After the down-conversion as mentioned in \eqref{eq.27a} and \eqref{eq.27b}, then 
\begin{subequations}
\begin{align}
\tilde{F}_{n,s,q}^{\left( \mathrm{SF} \right)}\left( t \right) \approx &\tilde{E}_t \cdot\rho _q\cdot{ e^{-j2\pi f_0 \tau _{n}^{(R)}}\cdot e^{j2\pi f_0\tau _{\mathrm{s}}\left( \varphi _j,\theta _j \right)}}
\nonumber\\
\times G&(\frac{t}{T_p})e^{j\pi \kappa (t-\tau _j)^2}\cdot e^{j2\pi [\left( q-1 \right) \varDelta f'+\left( s-1 \right) \varDelta f]\left( t-\tau _j \right)}
\label{eq.A7a}
\\
\tilde{F}_{n,s,q}^{\left( \mathrm{AF} \right)}\left( t \right) \approx &\tilde{E}_t \cdot\rho _q\cdot{ e^{-j2\pi f_0 \tau _{n}^{(R)}}\cdot e^{j2\pi f_0\tau _{\mathrm{s}}\left( \varphi _j,\theta _j \right)}}\cdot e^{j2\pi f_0\tau _q}
\nonumber\\
\times G&(\frac{t}{T_p})e^{j\pi \kappa (t-\tau _j)^2}\cdot e^{j2\pi [\left( q-1 \right) \varDelta f'+\left( s-1 \right) \varDelta f]\left( t-\tau _j \right)}
\label{eq.A7b}
\end{align}
\end{subequations}
By using $G( \frac{t}{T_p}) e^{ -j\pi \kappa t^2}e^{ -j2\pi (s-1)\Delta f t}$ to operate a convolution with $\tilde{F}_{n,s,q}^{\left( \mathrm{SF} \right)}\left( t \right)$ and $\tilde{F}_{n,s,q}^{\left( \mathrm{AF} \right)}\left( t \right)$ for the s-th channel, we can get the results of two integrals corresponding to the n-th receive element, the s-th channel, and the q-th jamming signal.
\begin{subequations}
\begin{align}
&{F}_{n,s,q}^{\left( \mathrm{SF} \right)}\left( t \right) 
\nonumber\\
=&\tilde{E}_t\cdot \rho _q\cdot T_p\cdot e^{-j2\pi f_0\tau _{n}^{(R)}}\cdot e^{j2\pi f_0\tau _s\left( \varphi_j ,\theta_j \right)}\cdot e^{-j2\pi \left( s-1 \right) \varDelta f\tau _j}
\nonumber\\
&\times \frac{\sin \pi T_p\left[ \kappa \left( t-\tau _j \right) +\left( q-1 \right) \varDelta f' \right]}{\pi T_p\left[ \kappa \left( t-\tau _j \right) +\left( q-1 \right) \varDelta f' \right]}
\label{eq.A8a}
\\
&{F}_{n,s,q}^{\left( \mathrm{AF} \right)}\left( t \right) 
\nonumber\\
=&\tilde{E}_t\cdot\rho _q\cdot T_p\cdot e^{-j2\pi f_0\tau _{n}^{(R)}}\cdot e^{j2\pi f_0\tau _s\left( \varphi_j ,\theta_j \right)}\cdot e^{-j2\pi \left( s-1 \right) \varDelta f\tau _j}
\nonumber\\
&\times\frac{\sin \pi T_p\left[ \kappa \left( t-\tau _j \right) +\left( q-1 \right) \varDelta f' +\tau_q f_0\right]}{\pi T_p\left[ \kappa \left( t-\tau _j \right) +\left( q-1 \right) \varDelta f'+\tau_q f_0\right]}
\label{eq.A8b}
\end{align}
\end{subequations}

For a target signal received by the n-th element for FDA-MIMO radar as mentioned in \eqref{eq.12}, the result of the convolution integral in the s-th channel can be expressed as 
\begin{align}
&\int_{T_p}{y_{n}^{\left( \mathrm{F} \right)}\left( \tau \right) \left[ G(\frac{t}{T_p})e^{-j\pi \left[ \kappa (t-\tau )^2+2(s-1)\Delta f(t-\tau ) \right]} \right] \mathrm{d}\tau}
\nonumber\\
=&\xi _t\cdot \sqrt{M_S}\cdot T_p\cdot e^{-j2\pi f_0\tau _{n}^{(R)}}\cdot e^{j2\pi f_0\tau _s\left( \varphi _t,\theta _t \right)}
\nonumber\\
&\times e^{-j2\pi \left( s-1 \right) \varDelta f\tau _t} \cdot \frac{\sin \pi T_p\kappa \left( t-\tau _t \right)}{\pi T_p\kappa \left( t-\tau _t \right)}
\label{eq.A9}
\end{align}

From \eqref{eq.A4a} and \eqref{eq.A8a}, the sampling peaks of the SF jamming after MF for PA and FDA-MIMO radar should both be at
\begin{align}
\tau _{q}^{\left( \mathrm{SF} \right)}=&\tau _j-\frac{T_p\varDelta f'}{B}\left( q-1 \right) \label{eq.A10}
\end{align}
From \eqref{eq.A4b} and \eqref{eq.A8b}, the sampling peaks of the AF jamming after MF for PA and FDA-MIMO radar should both be at
\begin{align}
\tau _{q}^{\left( \mathrm{AF} \right)}=&\tau _j-\frac{T_p f_0}{B}\tau _q-\frac{T_p\varDelta f'}{B}\left( q-1 \right) \label{eq.A11}
\end{align}
Therefore, for two types of the FDA jammer, the range-dimensional MF output of the q-th jamming signal appear at 
\begin{subequations}
\begin{align}
R _{q}^{\left( \mathrm{SF} \right)}=&R _j-T_p\varDelta f'\varDelta R\left( q-1 \right) \label{eq.A12a}
\\
R _{q}^{\left( \mathrm{AF} \right)}=&R _j-\frac{(q-1)T_p}{4B}\cos{\theta_j}\cos{\varphi_j}-T_p\varDelta f'\varDelta R\left( q-1 \right)\label{eq.A12b}
\end{align}
\end{subequations}
where $\varDelta R=c/2B$ and $R_j=c\tau_j/2$ denote the range resolution of radar and the jammer range. Note that \eqref{eq.A12b} uses $\tau _q={d_j}\left( q-1 \right) \cos \varphi _j\cos \theta _j/{c}$, $d_j={\lambda_0}/{2}$ and $\lambda_0=c/f_0$ to simplify.

From \eqref{eq.A12a} and \eqref{eq.A12b}, the jamming frequency offset should satisfy $\varDelta f'\geqslant \frac{1}{T_p(Q-1)}$ in order to generate the jamming peaks at different range resolution cells. Moreover, the maximum frequency shift of the FDA jamming cannot exceed the signal bandwidth, thus the constraint of the jamming frequency offset should be $\varDelta f'\leqslant \frac{B}{Q-1}$ against PA radar, while it should be $\varDelta f'\leqslant \frac{\varDelta f}{Q-1}$ against FDA-MIMO radar since it has a inherent frequency shift. In summary, the conditions of the jamming frequency offset for MF process can be written as
\begin{equation}
\frac{1}{T_p(Q-1)}\leqslant \varDelta f'\leqslant \frac{B}{Q-1}\leqslant \frac{\varDelta f}{Q-1}
\label{eq.A13}
\end{equation}
Furthermore, to ensure that the power of the false targets generated by each FDA jammer antenna are not less than the target power, compared with \eqref{eq.A5} for PA radar, and \eqref{eq.A9} for FDA-MIMO radar, the q-th jamming signal power should satisfy
\begin{subequations}
\begin{align}
\left( \rho _{q}^{\left( \mathrm{P} \right)} \right) ^2&\geqslant M\frac{\sigma _{t}^{2}}{\left( E_t \right) ^2}
 \label{eq.A14a}
\\
\left( \rho _{q}^{\left( \mathrm{F} \right)} \right) ^2&\geqslant M_S\frac{\sigma _{t}^{2}}{\left( \tilde{E}_t \right) ^2}
\label{eq.A14b}
\end{align}
\end{subequations}
where $\sigma _{t}^{2}=\mathrm{E}\left\{ \left| \xi _t \right|^2 \right\} $. $\rho _{q}^{\left( \mathrm{P} \right)}$ is against PA radar according to \eqref{eq.A4a}, \eqref{eq.A4b}, and \eqref{eq.A5} and $\rho _{q}^{\left( \mathrm{F} \right)}$ is against FDA-MIMO radar according to \eqref{eq.A8a}, \eqref{eq.A8b}, and \eqref{eq.A9}.

\section*{APPENDIX B}

This appendix shows the proof of Proposition 2. 

Firstly, we discuss $\varUpsilon _{\mathrm{P}}^{\left( \mathrm{SF} \right)}$ and $\varUpsilon _{\mathrm{P}}^{\left( \mathrm{AF} \right)}$ against PA radar. Substituting \eqref{eq.23a}, \eqref{eq.23b}, and \eqref{eq.23c} into \eqref{eq.26}, then 
\begin{subequations}
\begin{align}
\varUpsilon _{\mathrm{P}}^{\left( \mathrm{SF} \right)}=&E_t\sum_{q=1}^Q{\rho _q\int_{T_p}{e^{j2\pi \left( q-1 \right) \varDelta f'\left( t-\tau_j \right)}\left| A\left( t \right) \right|^2\mathrm{d}t}}\label{eq.B1a}
\\
\varUpsilon _{\mathrm{P}}^{\left( \mathrm{AF} \right)}=&E_t\sum_{q=1}^Q{\rho _q\int_{T_p}{e^{j2\pi [f_0 \tau_q+\left( q-1 \right) \varDelta f'\left( t-\tau_j \right)]}\left| A\left( t \right) \right|^2\mathrm{d}t}}\label{eq.B1b}
\end{align}
\end{subequations}
where $\tau _q=(q-1)d_j\cos{\varphi_j}\cos{\theta_j}/c$. Assume that $\left| A\left( t \right) \right|^2$ is rectangular wave function with time interval $\left[ -\frac{T_p}{2},\frac{T_p}{2} \right]$ and its Fourier transform can be expressed as $\mathcal{A} \left( f \right) $, 
\begin{align}
\mathcal{A} \left( f \right)= &\int_{-\infty}^{\infty}{\left| A\left( t \right) \right|^2e^{-j2\pi ft}\mathrm{d}t}
\nonumber\\
=&e^{-j\pi fT_p}\mathrm{sinc}\left( T_pf \right)
\label{eq.B2}
\end{align}
where $\mathrm{sinc}\left( T_pf \right) ={\sin (\pi T_pf)}/{(\pi T_pf)}$. Then \eqref{eq.B1a} and \eqref{eq.B1b} can be written as
\begin{subequations}
\begin{align}
\varUpsilon _{\mathrm{P}}^{\left( \mathrm{SF} \right)}=&E_t\sum_{q=1}^Q{\rho _qe^{-j2\pi \left( q-1 \right) \varDelta f'\tau _j}\mathcal{A} \left[ -\left( q-1 \right) \varDelta f' \right]}
\nonumber\\
\approx &E_t\sum_{q=1}^Q{\rho _q\frac{\sin \left[ \pi T_p\left( q-1 \right) \varDelta f' \right]}{\pi T_p\left( q-1 \right) \varDelta f'}}
\label{eq.B3a}
\\
\varUpsilon _{\mathrm{P}}^{\left( \mathrm{SF} \right)}=&E_t\sum_{q=1}^Q{\rho _qe^{-j2\pi[\left( q-1 \right) \varDelta f'\tau_j -f_0 \tau_q]}\mathcal{A} \left[ -\left( q-1 \right) \varDelta f' \right]}
\nonumber\\
\approx &E_t\sum_{q=1}^Q{\rho _q e^{j2\pi f_0 \tau_q}  \frac{\sin \left[ \pi T_p\left( q-1 \right) \varDelta f' \right]}{\pi T_p\left( q-1 \right) \varDelta f'}}
\label{eq.B3b}
\end{align}
\end{subequations}
where \eqref{eq.B3a} and \eqref{eq.B3b} ignore $e^{-j2\pi \left( q-1 \right) \varDelta f'\tau _j}$ and $e^{-j2\pi \left( q-1 \right) \varDelta f'T_p}$ due to $\varDelta f'\ll f_0$. Compared \eqref{eq.B3a} with \eqref{eq.B3b}, AF jamming considers the transmit array steering $e^{j2\pi f_0 \tau_q}$ and requires more jamming power than the SF jamming since the main beam of its jammer antennas is not orientated to the radar as the SF jammer antennas. Defining the following $Q\times1$ auxiliary vectors,
\begin{subequations}
\begin{align}
\boldsymbol{g}\left( \varphi ,\theta \right) =\left[ \begin{matrix}
  1&    \cdots&   e^{j2\pi \frac{d}{\lambda _0}\left( Q-1 \right) \cos \varphi \cos \theta}\\
\end{matrix} \right] ^{\mathrm{T}}
\label{eq.B4a}
\\
\boldsymbol{\varPhi }\left( \varDelta f' \right) =\left[ \begin{matrix}
  1&    \cdots&   {\mathrm{sinc}\left[ T_p\left( Q-1 \right) \varDelta f' \right]}\\
\end{matrix} \right] ^{\mathrm{T}}
\label{eq.B4b}
\end{align}
\end{subequations}
meanwhile using \eqref{eq.23a}, then \eqref{eq.B3a} and (\ref{eq.B3b}) can be rewritten as
\begin{subequations}
\begin{align}
\varUpsilon _{\mathrm{P}}^{\left( \mathrm{SF} \right)}=&E_t\cdot\boldsymbol{\rho }\cdot \boldsymbol{\varPhi }\left( \varDelta f' \right) 
\label{eq.B5a}
\\
\varUpsilon _{\mathrm{F}}^{\left( \mathrm{SF} \right)}=&E_t\cdot\left[ \boldsymbol{\rho }\odot \boldsymbol{g}^{\mathrm{T}}\left( \varphi _j,\theta _j \right) \right] \cdot \boldsymbol{\varPhi }\left( \varDelta f' \right) 
\label{eq.B5b}
\end{align}
\end{subequations}
Under the condition of $\varDelta f'< \frac{1}{(Q-1){T_p}}$, focusing on the q-th element in $\boldsymbol{\varPhi }\left( \varDelta f' \right)$, since its first null is at $\Delta f'_u=\frac{1}{\left( q-1 \right) T_p} > \Delta f'$ according to the properties of sinc function, $\varUpsilon _{\mathrm{P}}^{\left( \mathrm{SF} \right)}$ and $\varUpsilon _{\mathrm{P}}^{\left( \mathrm{AF} \right)}$ are  monotonically decreasing with respect with $\Delta f'\in \left[ 0,\frac{1}{\left( Q-1 \right) T_p} \right] $.

Secondly, we discuss $\boldsymbol{\varUpsilon} _{\mathrm{F}}^{\left( \mathrm{SF} \right)}$ and $\boldsymbol{\varUpsilon} _{\mathrm{F}}^{\left( \mathrm{AF} \right)}$ against FDA-MIMO radar with $S$ subarrays. Substituting \eqref{eq.23a}, \eqref{eq.23b}, \eqref{eq.23c}, and \eqref{eq.28} into \eqref{eq.32}, the u-th row and v-th column elements ($u,v=1,...,S$) of $\boldsymbol{\varUpsilon} _{\mathrm{F}}^{\left( \mathrm{SF} \right)}$ and $\boldsymbol{\varUpsilon} _{\mathrm{F}}^{\left( \mathrm{AF} \right)}$ can be represented as
\begin{subequations}
\begin{align}
&[ \boldsymbol{\varUpsilon} _{\mathrm{F}}^{\left( \mathrm{SF} \right)} ] _{u,v}
=\tilde{E}_t\sum_{q=1}^Q{\rho _p\mathcal{A}^{(q)} _{u,v}\left( \varDelta f,\varDelta f' \right)  e^{-j2\pi \left( q-1 \right) \varDelta f' \tau_j}}\label{eq.B6a}
\\
&[ \boldsymbol{\varUpsilon} _{\mathrm{F}}^{\left( \mathrm{AF} \right)} ] _{u,v}=\tilde{E}_t \sum_{q=1}^Q{\rho _p\mathcal{A}^{(q)} _{u,v}\left( \varDelta f,\varDelta f' \right)  e^{j2\pi[f_0 \tau_q-\left( q-1 \right) \varDelta f'\tau_j ]}}\label{eq.B6b}
\end{align}
\end{subequations}
where 
\begin{align}
\mathcal{A}^{(q)} _{u,v}\left( \varDelta f,\varDelta f' \right)=& \mathcal{A} \left[ \left( u-v \right) \varDelta f-\left( q-1 \right) \varDelta f' \right]
\label{eq.B7}
\end{align}
When $\varDelta fT_p$ is a large positive integer, according to the relationship between $\varDelta f$, $T_p$ and $\varDelta f'$, $\varDelta f'\leqslant \frac{1}{(Q-1)T_p}\ll \varDelta f$, 
\begin{align}
\mathcal{A}^{(q)} _{u,v}\left( \varDelta f,\varDelta f' \right)\approx e^{-j\pi T_p\left( u-v \right) \varDelta f}\mathcal{A} \left[ -\left( q-1 \right) \varDelta f' \right] 
\label{eq.B8}
\end{align}
The diagonal elements and the remaining elements can be calculated as
\begin{equation}
\mathcal{A}^{(q)} _{u,v}\left( \varDelta f,\varDelta f' \right) =\begin{cases}
  \,\, ~~~~~~~~~~0,              ~~~~~~~~ ~     u\ne v\\
  \mathcal{A} \left[ -\left( q-1 \right) \varDelta f' \right] , u= v\\
\end{cases}
\label{eq.B9}
\end{equation}
Therefore, $\boldsymbol{\varUpsilon} _{\mathrm{F}}^{\left( \mathrm{SF} \right)}$ and $\boldsymbol{\varUpsilon} _{\mathrm{F}}^{\left( \mathrm{AF} \right)}$ can be simplified to the $S\times S$ diagonal matrices with $\varUpsilon _{\mathrm{P}}^{\left( \mathrm{SF} \right)}$ and $\varUpsilon _{\mathrm{P}}^{\left( \mathrm{AF} \right)}$ as the diagonal elements, respectively.
\begin{subequations}
\begin{align}
&\boldsymbol{\varUpsilon }_{\mathrm{F}}^{\left( \mathrm{SF} \right)}\approx \frac{\tilde{E}_t}{E_t}\varUpsilon _{\mathrm{P}}^{\left( \mathrm{SF} \right)}\boldsymbol{I}_S
\label{eq.B10a}
\\
&\boldsymbol{\varUpsilon }_{\mathrm{F}}^{\left( \mathrm{AF} \right)}\approx \frac{\tilde{E}_t}{E_t}\varUpsilon _{\mathrm{P}}^{\left( \mathrm{AF} \right)}\boldsymbol{I}_S
\label{eq.B10b}
\end{align}
\end{subequations}
where $\varUpsilon _{\mathrm{P}}^{\left( \mathrm{SF} \right)}$ and $\varUpsilon _{\mathrm{P}}^{\left( \mathrm{AF} \right)}$ are referred in (\ref{eq.B5a}) and (\ref{eq.B5b}).

\setcounter{TempEqCnt}{\value{equation}} 
\setcounter{equation}{74}
\begin{figure*}[b]
\hrulefill
\begin{align}
 Y_{\varphi _j}\left( \varDelta f' \right) =\frac{1}{\sigma _{n}^{2}}\left[ \boldsymbol{a}_{r}^{\left( \varphi _j \right)} \right] ^{\mathrm{H}}\boldsymbol{a}_{r}^{\left( T \right)}-
\frac{\sum_{q=1}^Q{\rho _{q}^{2}}}{\sigma _{n}^{2}\left( \sigma _{n}^{2}+N\sum_{q=1}^Q{\rho _{q}^{2}} \right)}\left| \varUpsilon _{\mathrm{P}}^{\left( \mathrm{FDA} \right)}\boldsymbol{a}_{r}^{\left( \varphi _j \right)} \right| ^{2}\left[ \boldsymbol{a}_{r}^{\left( \varphi _j \right)} \right] ^{\mathrm{H}}\boldsymbol{a}_{r}^{\left( T \right)}
\label{eq.C3}
\end{align}
\end{figure*}

\setcounter{TempEqCnt}{\value{equation}} 
\setcounter{equation}{82}

\begin{figure*}[b]
\vspace{-0.10in}
\hrulefill
\begin{subequations}
\begin{align}
Y_{\varphi _j}\left( \varDelta f' \right)=&\frac{1}{\sigma _{\mathrm{n}}^{2}}\left[ \boldsymbol{c}^{\left( \varphi_j \right)} \right] ^{\mathrm{H}}\boldsymbol{c}^{\left( T \right)}\left[ \boldsymbol{a}_{r}^{\left( \varphi_j \right)} \right] ^{\mathrm{H}}\boldsymbol{a}_{r}^{\left( T \right)}
\nonumber\\
-&\frac{\sum_{q=1}^Q{\rho _{q}^{2}}}{\sigma _{n}^{2}\left( \sigma _{n}^{2}+S\cdot N\cdot \sum_{q=1}^Q{\rho _{q}^{2}} \right)}\left[ \boldsymbol{c}^{\left( \varphi_j \right)} \right] ^{\mathrm{H}}\left[ \boldsymbol{\varUpsilon }_{\mathrm{F}}^{\left( \mathrm{FDA} \right)}\boldsymbol{c}^{\left( \varphi_j \right)}\left[ \boldsymbol{c}^{\left( \varphi_j \right)} \right] ^{\mathrm{H}}\left[ \boldsymbol{\varUpsilon }_{\mathrm{F}}^{\left( \mathrm{FDA} \right)} \right] ^{\mathrm{H}} \right] \boldsymbol{c}^{\left( T \right)}\left| \boldsymbol{a}_{r}^{\left( \varphi_j \right)} \right|^2\left[ \boldsymbol{a}_{r}^{\left( \varphi_j \right)} \right] ^{\mathrm{H}}\boldsymbol{a}_{r}^{\left( T \right)}
\label{eq.C12a}
\\
Y_{R_j}\left( \varDelta f' \right)=&\frac{1}{\sigma _{\mathrm{n}}^{2}}\left[ \boldsymbol{c}^{\left( R_j \right)} \right] ^{\mathrm{H}}\boldsymbol{c}^{\left( T \right)}\left[ \boldsymbol{a}_{r}^{\left( R_j \right)} \right] ^{\mathrm{H}}\boldsymbol{a}_{r}^{\left( T \right)}
\nonumber\\
-&\frac{\sum_{q=1}^Q{\rho _{q}^{2}}}{\sigma _{n}^{2}\left( \sigma _{n}^{2}+S\cdot N\cdot \sum_{q=1}^Q{\rho _{q}^{2}} \right)}\left[ \boldsymbol{c}^{\left( R_j \right)} \right] ^{\mathrm{H}}\left[ \boldsymbol{\varUpsilon }_{\mathrm{F}}^{\left( \mathrm{FDA} \right)}\boldsymbol{c}^{\left( R_j \right)}\left[ \boldsymbol{c}^{\left( R_j \right)} \right] ^{\mathrm{H}}\left[ \boldsymbol{\varUpsilon }_{\mathrm{F}}^{\left( \mathrm{FDA} \right)} \right] ^{\mathrm{H}} \right] \boldsymbol{c}^{\left( T \right)}\left| \boldsymbol{a}_{r}^{\left( R_j \right)} \right|^2\left[ \boldsymbol{a}_{r}^{\left( R_j \right)} \right] ^{\mathrm{H}}\boldsymbol{a}_{r}^{\left( T \right)}\label{eq.C12b}
\end{align}
\end{subequations}
\end{figure*}
\setcounter{equation}{\value{TempEqCnt}} 
\setcounter{equation}{72}

\section*{APPENDIX C}

This appendix gives the detailed derivations of \eqref{eq.43a} and \eqref{eq.43b} for PA radar, and \eqref{eq.45a}, \eqref{eq.45a} and \eqref{eq.45b} for FDA-MIMO radar. We derive three measurements for two different cases of phased-MIMO radar, PA radar and FDA-MIMO radar.

\subsection*{Case 1: PA radar}

Following the matrix inversion lemma [\ref{cite9}, \ref{cite46}], $\boldsymbol{R}_{u}^{-1}$ for PA radar can be calculated as
\begin{equation}
\boldsymbol{R}_{u}^{-1}\approx\frac{1}{\sigma _{n}^{2}}\left[ \boldsymbol{I}_N-\frac{\sum_{q=1}^Q{\rho _{q}^{2}}}{\sigma _{n}^{2}+N\sum_{q=1}^Q{\rho _{q}^{2}}}\boldsymbol{j}_{\mathrm{P}}\boldsymbol{j}_{\mathrm{P}}^{\mathrm{H}} \right] 
\label{eq.C1}
\end{equation}
where 
\begin{equation}
\boldsymbol{j}_{\mathrm{P}}\boldsymbol{j}_{\mathrm{P}}^{\mathrm{H}}=\left| \varUpsilon _{\mathrm{P}}^{\left( \mathrm{FDA} \right)} \right|^2\boldsymbol{a}_{r}^{\left( \varphi _j \right)}\left[ \boldsymbol{a}_{r}^{\left( \varphi _j \right)} \right] ^{\mathrm{H}}
\label{eq.C2}
\end{equation}
Substituting  \eqref{eq.35a}, \eqref{eq.38a} into \eqref{eq.39a} and using \eqref{eq.C1} and \eqref{eq.C2}, we can get \eqref{eq.C3}. Using the following auxiliary scalars,
\setcounter{equation}{\value{TempEqCnt}} 
\setcounter{equation}{75}
\begin{subequations}
\begin{align}
\mathcal{X} _{r}^{\left( \varphi_j \right)}=&\left[ \boldsymbol{a}_{r}^{\left( \varphi _j \right)} \right] ^{\mathrm{H}}\boldsymbol{a}_{r}^{\left( T \right)}\label{eq.C4a}
\\
\varPsi ^{\left( \varphi _j \right)}_{\mathrm{P}}\left( \varDelta f'\right)=&\left| \varUpsilon _{\mathrm{P}}^{\left( \mathrm{FDA} \right)}\boldsymbol{a}_{r}^{\left( \varphi _j \right)} \right| ^{2}\label{eq.C4b}
\end{align}
\end{subequations}
then \eqref{eq.C3} can be rewritten as
\begin{equation}
\left| Y_{\varphi _j}\left( \varDelta f' \right) \right|=\left| \frac{\mathcal{X} _{r}^{\left( \varphi _j \right)}}{\sigma _{n}^{2}}-\frac{\varPsi ^{\left( \varphi _j \right)}_{\mathrm{P}}\left( \varDelta f' \right) \mathcal{X} _{r}^{\left( \varphi _j \right)}\left( \sum_{q=1}^Q{\rho _{q}^{2}} \right)}{\sigma _{n}^{2}\left( \sigma _{n}^{2}+N\sum_{q=1}^Q{\rho _{q}^{2}} \right)} \right|
\label{eq.C5}
\end{equation}

For output SINR, substituting \eqref{eq.35a} and \eqref{eq.38a} into \eqref{eq.39c} yields
\begin{equation}
\mathrm{SINR}_o\left( \varDelta f' \right) =\frac{M\sigma _{t}^{2}}{\left| \boldsymbol{v}_{\mathrm{P}}^{\mathrm{H}}\boldsymbol{R}_{\mathrm{u}}^{-1}\boldsymbol{x}_{\mathrm{P}} \right|^2}
\label{eq.C6}
\end{equation}
where $\sigma _{t}^{2}=\mathrm{E}\left\{ \left| \xi _t \right|^2 \right\}$. Substituting \eqref{eq.6}, \eqref{eq.33a}, and \eqref{eq.C1} into \eqref{eq.C6}, then
\begin{equation}
\mathrm{SINR}_o\left( \varDelta f' \right) =\frac{M\sigma _{t}^{2}}{\left| \frac{\mathcal{X} _{r}^{\left( J \right)}}{\sigma _{n}^{2}}-\frac{\varPsi _{\mathrm{P}}^{\left( J \right)}\left( \varDelta f' \right) \mathcal{X} _{r}^{\left( J \right)}\left( \sum_{q=1}^Q{\rho _{q}^{2}} \right)}{\sigma _{n}^{2}\left( \sigma _{n}^{2}+N\sum_{q=1}^Q{\rho _{q}^{2}} \right)} \right|^2}
\label{eq.C7}
\end{equation}
where 
\begin{subequations}
\begin{align}
\mathcal{X} _{r}^{\left( J \right)}=&\left[ \boldsymbol{a}_{r}^{\left( J \right)} \right] ^{\mathrm{H}}\boldsymbol{a}_{r}^{\left( T \right)}\label{eq.C8a}
\\
\varPsi ^{\left( J \right)}_{\mathrm{P}}\left( \varDelta f'\right)=&\left| \varUpsilon _{\mathrm{P}}^{\left( \mathrm{FDA} \right)}\boldsymbol{a}_{r}^{\left( J \right)} \right|^{2}\label{eq.C8b}
\end{align}
\end{subequations}

\subsection*{Case 2: FDA-MIMO radar}

Following the matrix inversion lemma [\ref{cite46}], $\boldsymbol{R}_{u}^{-1}$ for FDA-MIMO radar with $S$ subarrays can be calculated as
\begin{equation}
\boldsymbol{R}_{u}^{-1}=\frac{1}{\sigma _{n}^{2}}\left[ \boldsymbol{I}_{SN}-\frac{\sum_{q=1}^Q{\rho _{q}^{2}}}{\sigma _{n}^{2}+SN\sum_{q=1}^Q{\rho _{q}^{2}}}\boldsymbol{j}_{\mathrm{F}}\boldsymbol{j}_{\mathrm{F}}^{\mathrm{H}} \right] 
\label{eq.C9}
\end{equation}
where 
\begin{align}
\boldsymbol{j}_{\mathrm{F}}\boldsymbol{j}_{\mathrm{F}}^{\mathrm{H}}=&\left[ \boldsymbol{\varUpsilon }_{\mathrm{F}}^{\left( \mathrm{FDA} \right)}\boldsymbol{c}^{\left( R_j \right)}\left[ \boldsymbol{c}^{\left( R_j \right)} \right] ^{\mathrm{H}}\left[ \boldsymbol{\varUpsilon }_{\mathrm{F}}^{\left( \mathrm{FDA} \right)} \right] ^{\mathrm{H}} \right] 
\nonumber\\
&\otimes \left[ \boldsymbol{a}_{r}^{\left( R_j \right)}\left[ \boldsymbol{a}_{r}^{\left( R_j \right)} \right] ^{\mathrm{H}} \right] 
\label{eq.C10}
\end{align} 
Substituting \eqref{eq.13} and \eqref{eq.C9} into \eqref{eq.38b} and using \eqref{eq.35b}, we can get $\left|Y_{\varphi _j}\left( \varDelta f' \right)\right|$ and $\left|Y_{R _j}\left( \varDelta f' \right)\right|$ in \eqref{eq.C12a} and \eqref{eq.C12b} for FDA-MIMO radar, respectively. Using \eqref{eq.C4a}, \eqref{eq.C4b}, and the following auxiliary scalars,
\setcounter{equation}{\value{TempEqCnt}} 
\setcounter{equation}{83}
\begin{subequations}
\begin{align}
\mathcal{X} _{t}^{\left( \varphi _j \right)}=&\left[ \boldsymbol{c}^{\left( \varphi _j \right)} \right] ^{\mathrm{H}}\boldsymbol{c}^{\left( T \right)}
\label{eq.C13a}
\\
\mathcal{X} _{t}^{\left( R_j \right)}=&\left[ \boldsymbol{c}^{\left( R_j \right)} \right] ^{\mathrm{H}}\boldsymbol{c}^{\left( T \right)}
\label{eq.C13b}
\\
\mathcal{X}_{r}^{\left( R_j \right)}=&\left[ \boldsymbol{a}_{r}^{\left( R_j \right)} \right] ^{\mathrm{H}}\boldsymbol{a}_{r}^{\left( T \right)}
\label{eq.C13c}
\\
\varPsi _{\mathrm{F}}^{\left( \varphi_j \right)}\left( \varDelta f' \right) =&\left[ \boldsymbol{c}^{\left( \varphi_j \right)} \right] ^{\mathrm{H}} \boldsymbol{\varUpsilon }_{\mathrm{F}}^{\left( \mathrm{FDA} \right)}\boldsymbol{c}^{\left( \varphi_j \right)}\left[ \boldsymbol{c}^{\left( \varphi_j \right)} \right] ^{\mathrm{H}}\left[ \boldsymbol{\varUpsilon }_{\mathrm{F}}^{\left( \mathrm{FDA} \right)} \right] ^{\mathrm{H}}  
\nonumber\\
&\cdot\boldsymbol{c}^{\left(T \right)}\left| \boldsymbol{a}_{r}^{\left( \varphi_j \right)} \right|^2
\label{eq.C13d}
\\
\varPsi _{\mathrm{F}}^{\left( R_j \right)}\left( \varDelta f' \right) =&\left[ \boldsymbol{c}^{\left( R_j \right)} \right] ^{\mathrm{H}} \boldsymbol{\varUpsilon }_{\mathrm{F}}^{\left( \mathrm{FDA} \right)}\boldsymbol{c}^{\left( R_j \right)}\left[ \boldsymbol{c}^{\left( R_j \right)} \right] ^{\mathrm{H}}\left[ \boldsymbol{\varUpsilon }_{\mathrm{F}}^{\left( \mathrm{FDA} \right)} \right] ^{\mathrm{H}} \nonumber\\
&\cdot \boldsymbol{c}^{\left( T \right)}\left| \boldsymbol{a}_{r}^{\left( R_j \right)} \right|^2
\label{eq.C13e}
\end{align}
\end{subequations}
then \eqref{eq.C12a} and \eqref{eq.C12b} can be rewritten as 
\begin{subequations}
\begin{align}
\left| Y_{\varphi _j}\left( \varDelta f' \right) \right|=&\frac{\mathcal{X} _{t}^{\left( \varphi _j \right)}\mathcal{X} _{r}^{\left( \varphi _j \right)}}{\sigma _{\mathrm{n}}^{2}}-\frac{\varPsi _{\mathrm{F}}^{\left( \varphi _j \right)}\left( \varDelta f' \right) \mathcal{X} _{r}^{\left( \varphi _j \right)}(\sum_{q=1}^Q{\rho _{q}^{2}})}{\sigma _{n}^{2}\left( \sigma _{n}^{2}+S N(\sum_{q=1}^Q{\rho _{q}^{2}}) \right)}
\label{eq.C14a}
\\
\left| Y_{R _j}\left( \varDelta f' \right) \right|=&\frac{\mathcal{X} _{t}^{\left( R _j \right)}\mathcal{X} _{r}^{\left( R _j \right)}}{\sigma _{\mathrm{n}}^{2}}-\frac{\varPsi _{\mathrm{F}}^{\left( R _j \right)}\left( \varDelta f' \right) \mathcal{X} _{r}^{\left( R _j \right)}(\sum_{q=1}^Q{\rho _{q}^{2}})}{\sigma _{n}^{2}\left( \sigma _{n}^{2}+S N (\sum_{q=1}^Q{\rho _{q}^{2}}) \right)}\label{eq.C14b}
\end{align}
\end{subequations}

For output SINR, substituting \eqref{eq.35b} and \eqref{eq.38b} into \eqref{eq.39c} yields
\begin{equation}
\mathrm{SINR}_o\left( \varDelta f' \right) =\frac{M_S\sigma _{t}^{2}}{\left| \boldsymbol{v}_{\mathrm{F}}^{\mathrm{H}}\boldsymbol{R}_{\mathrm{u}}^{-1}\boldsymbol{x}_{\mathrm{F}} \right|^2}
\label{eq.C15}
\end{equation}
Substituting \eqref{eq.13}, \eqref{eq.35b}, \eqref{eq.C9} into \eqref{eq.C15}, then
\begin{equation}
\mathrm{SINR}_o\left( \varDelta f' \right) =\frac{M_S\sigma _{t}^{2}}{\left| \frac{\mathcal{X} _{t}^{\left( J \right)}\mathcal{X} _{r}^{\left( J \right)}}{\sigma _{\mathrm{n}}^{2}}-\frac{\varPsi _{\mathrm{F}}^{\left( J \right)}\left( \varDelta f' \right) \mathcal{X} _{r}^{\left( J \right)}(\sum_{q=1}^Q{\rho _{q}^{2}})}{\sigma _{n}^{2}\left( \sigma _{n}^{2}+S N (\sum_{q=1}^Q{\rho _{q}^{2}}) \right)} \right|^2}
\label{eq.C16}
\end{equation}
where 
\begin{subequations}
\begin{align}
\mathcal{X} _{t}^{\left( J \right)}=&\left[ \boldsymbol{c}^{\left( J \right)} \right] ^{\mathrm{H}}\boldsymbol{c}^{\left( T \right)}\label{eq.C17a}
\\
\varPsi ^{\left( J \right)}_{\mathrm{F}}\left( \varDelta f'\right)=&\left[ \boldsymbol{c}^{\left( J \right)} \right] ^{\mathrm{H}} \boldsymbol{\varUpsilon }_{\mathrm{F}}^{\left( \mathrm{FDA} \right)}\boldsymbol{c}^{\left( J \right)}\left[ \boldsymbol{c}^{\left( J \right)} \right] ^{\mathrm{H}}\left[ \boldsymbol{\varUpsilon }_{\mathrm{F}}^{\left( \mathrm{FDA} \right)} \right] ^{\mathrm{H}} \nonumber\\
&\cdot \boldsymbol{c}^{\left( T \right)}\left| \boldsymbol{a}_{r}^{\left( J \right)} \right|^2\label{eq.C17b}
\end{align}
\end{subequations}

\bibsection*{REFERENCES}{}
\def\refname{}

\end{document}